\documentclass[12pt]{article}
\usepackage[table,dvipsnames]{xcolor}
\usepackage[utf8]{inputenc}
\usepackage{tikz}
\usetikzlibrary{shapes.geometric, snakes}
\usepackage{pgfplots}
\usepackage{graphicx}
\usepackage{imakeidx}
\usepackage{comment}
\usepackage{framed}
\usepackage{subcaption} 
\usepackage{amsmath,amsfonts,amssymb,amsthm}
\usepackage{mathtools}
\usepackage{commath}
\usepackage[sc,osf]{mathpazo}
\usepackage[margin=1in]{geometry}
\usepackage[math]{cellspace}
\usepackage{tocloft}
\usepackage{hyperref}
\hypersetup{colorlinks,citecolor=red,urlcolor=blue,hypertexnames=true}

\usepackage{csquotes}
\usepackage[backend=biber, style=authoryear,citestyle=authoryear,natbib, maxbibnames=999]{biblatex}
\DeclareBibliographyDriver{article}{
  \usebibmacro{bibindex}
  \usebibmacro{begentry}
  \usebibmacro{author/editor+others/translator+others}
  \setunit{\labelnamepunct}\newblock
  \usebibmacro{title}
  \newunit\newblock
  \usebibmacro{journal}
  \newunit
  \printfield{year}
  \newunit
  \printfield{volume}
  \setunit{\bibpagespunct}
  \printfield{pages}
  \usebibmacro{finentry}
}
\DeclareNameAlias{sortname}{family-given}
\AtBeginBibliography{

}
\DeclareFieldFormat[article, inbook, incollection, inproceedings, misc, thesis, unpublished]{title}{#1}  

\newtheorem{assumption}{Assumption}
\newtheorem{proposition}{Proposition}

\newtheorem{definition}{Definition}
\newtheorem{remark}{Remark}

\title{Trade, Growth, and Product Innovation}
\author{\setcounter{footnote}{0}
Carlos G\'oes
\thanks{\emph{cgoes@ucsd.edu}. I am particularly grateful for guidance and supervision from Marc Muendler, Fabian Trottner, Fabian Eckert, Valerie Ramey, Kyle Handley, and Johannes Wieland. I benefited from helpful discussions with  Jim Hamilton, Titan Alon, Ana Maria Santacreu, Thomas Sampson, Ezra Oberfield, Costas Arkolakis, Andr\'es Rodriguez-Claire, Treb Allen, Mayara F\'elix, Ernest Liu, Fernando Parro, Erhan Artuc, Chris Tonetti,  Juan Herreño, Rafael Dix-Carneiro, AV Ludovice, and Raymond Riezman as well as with seminar participants at multiple conferences. All errors and omissions are solely my responsibility. } \\[-3pt] \textit{\small UC San Diego}}
\date{This version: June 3, 2024 \\ First version: September 29, 2023 \\ For the most recent version \href{https://github.com/omercadopopular/omercadopopular.github.io/blob/master/files/GoesC-JMP.pdf?raw=true}{click here}.}

\begin{document}

\maketitle

\begin{abstract}
Can trade integration induce product innovation? I document that countries that joined the European Union (EU) started producing more product varieties, investing more in R\&D, and trading more compared to candidate countries that did not join at a given horizon. Additionally, I show that a plausibly exogenous increase in market access increases the probability of a given country starting production of and exporting a given product. To rationalize this reduced-form evidence, I propose a new quantitative framework that integrates the forces of specialization and market size. This is a dynamic general equilibrium model of frictional trade and endogenous growth with arbitrarily many asymmetric countries that nests the Eaton-Kortum model of trade and the Romer growth model as special cases. The key result is an analytical expression to decompose gains from trade into dynamic and static components. In this framework, the product innovation growth rate increases with higher market access.  Finally, a quantitative version of the model suggests that: (a) the EU enlargement increased its long-run yearly growth rate by about 0.10pp; and (b) dynamic gains can account for between 65-90\% of total welfare gains from trade. \\


\end{abstract}

\newpage

\section{Introduction}

Over the last decades, the trade literature converged to a broad consensus regarding how to summarize the static gains from trade. But there is no similar consensus on how to measure dynamic gains from trade\footnote{For a comprehensive review of the literature and the different mechanisms that link trade, growth, and innovation, see the paper by \textcite{melitz_trade_2021}}. In this paper, I address this topic by examining the mechanisms through which trade integration can induce product innovation. Economic theory presents conflicting viewpoints regarding this question. \textit{Canonical trade theory} typically suggests that increased economic integration should cause countries to produce a \textit{smaller range} of produced goods\footnote{In the class of Ricardian models, this follows naturally: as a country opens up to trade, it specializes in a smaller set of goods. But this also happens in the class of Melitz models. As a country opens up to trade, due to the selection effect, the least productive firms of each country exit the market, which results in a smaller range of firms (or, equivalently, goods) in either market. This result holds with asymmetric populations and symmetric productivity distributions or even with asymmetric productivity distributions, as long as the countries are not too dissimilar \textemdash for the latter see \textcite{demidova_productivity_2008}.}. Models that emphasize growth and innovation, such as those common in \textit{macroeconomics}, often emphasize the role of market size for having an incentive to innovate and produce a \textit{large range} of goods\footnote{This is true of a very large class of endogenous growth models in macroeconomics, both with and without scale effects. See, for instance, Chapter 13 of \textcite{acemoglu_introduction_2008}.}.

This paper integrates these two traditions by conceiving a global marketplace in which the economic forces of \textit{specialization} and \textit{market access} are jointly operating and developing tractable and intuitive ways of modeling them in a dynamic framework fit for policy evaluation. First, I show that after large events of trade integration \textemdash the expansion waves of the European Union (EU) \textemdash the countries that joined the EU started producing more product varieties, investing more in research and development (R\&D), and trading more compared to candidate countries that did not join at a given horizon. Additionally, I show that a plausibly exogenous increase in market access increases the probability of a given country starting production of and exporting a given product. These facts are all suggestive of a \textit{dynamic market access effect}. Second, to rationalize this reduced-form evidence, I propose a new dynamic general equilibrium model of frictional trade and endogenous growth with arbitrarily many asymmetric countries that nests the \textcite{eaton_technology_2002} model of trade and the \textcite{romer_endogenous_1990} growth model as special cases. Third, I provide analytical expressions decomposing gains from trade into dynamic and static components; growth and welfare into ``Romer'' and ``Eaton-Kortum'' parts; and show analytically that the product innovation and R\&D growth rates increase with higher market access, which is consistent the facts I used as motivation for the model. Lastly, I use a numerical version of the model to estimate the welfare effects of 2004 enlargement of the EU, in this framework: (a) the enlargement increased its long-run yearly growth rate by about 0.10pp; and (b) dynamic gains can account for between 65-90\% of total welfare gains from trade.

My focus on \textit{product innovation} stems from two key reasons, one theoretical and one empirical. From a theoretical standpoint, the new product margin can have large welfare implications. Empirically, around trade liberalization episodes, the bulk of trade creation comes from the extensive margin\footnote{For the former, \textcite{romer_new_1994} has shown that in a simple trade model, adding extensive margin can make welfare costs of a 10\% tariff increase from 1\% to 20\%. For the latter, \textcite{kehoe_how_2013} provide an extensive documentation of the empirical facts.}.    

The paper starts by documenting a set of facts related to the Eastwards enlargement of the EU. Compared to countries that selected into being candidates of the EU but were not yet members, New Member States (NMS) started: (a) producing more product varieties; (b) spending more on private R\&D per capita; and (c) having larger trading values. 

Later, in order to go beyond correlational analysis, I exploit the fact that, once NMS join the EU, they not only have preferential access to the European market, but they also have to adhere to the Common Commercial Policy of the European Union. NMS have immediate preferential access to third-party markets via pre-existing trade agreements between the EU and these third-party markets.

Importantly, the NMS did not get to negotiate the tariff variation that they face \textemdash these were only a byproduct of the EU accession process. In this context, through an event-study design, a plausibly exogenous increase in market access leads to a higher probability of initiating production and exporting a given product \textemdash i.e, leads to product innovation in the extensive margin.

I develop a dynamic general equilibrium model that is consistent with both the stylized facts and the market access mechanism to rationalize this reduced-form evidence. Like much of the trade and growth literature, the model presented in this paper incorporates forward-looking dynamics. However, unlike much of the literature, it shies away from stylized simplifications, such as symmetric countries or two-country cases. It encompasses an arbitrary number of asymmetric countries, costly trade, and is fit for counterfactual quantitative exercises. Therefore, it fits neatly into the tradition of quantitative trade models in international trade or policy counterfactuals using dynamic stochastic general equilibrium models in macroeconomics.\footnote{In the trade literature, this is the modern world of ``trade theory with numbers'' \parencite{costinot_trade_2014}. In the macroeconomics literature, this is the use of macro models as ``the	leading	tool''	for assessing the effect of policy changes in	``an	open	and	transparent	manner.'' \parencite{christiano_dsge_2018}.}  

In the model, in each source country, there are producers of final goods varieties that combine labor and intermediate goods using a constant returns to scale technology. They source differentiated intermediate varieties from foreign countries. These countries differ in their product spaces: some countries have a measure of intermediate goods that are larger than others.

Intermediate goods are \textit{non-rival} in the same spirit as in the endogenous growth literature. As new varieties are invented, they can be simultaneously and immediately sourced by final goods producers everywhere, inducing increasing returns.

In this framework, international trade induces substitutability across non-rival goods. Trade also implies that the measure of intermediate varieties that effectively diffuses to each country will be a price-weighted average of the measure of varieties imported from all trade partners. If there are no trade costs, all countries will share the same effective measure of varieties. Conversely, in autarky, each country will only take advantage of its own varieties. 

At each destination, final good varieties from every source are aggregated into a final composite good with some probability. Those actually sourced for aggregation will be only the lowest-cost varieties at each destination. Prices depend on productivities, taken to be the realization of a random variable.

The final composite good is used for household consumption and as an input for the production of intermediate varieties and research \& development (R\&D). Once a new blueprint is invented, each intermediate goods producer has perpetual rights over the production of its variety. They produce under monopolistic competition and set prices optimal prices accordingly through market-specific price discrimination. Since this model embeds an input-output structure, the optimal monopolist prices will depend on the price of the final composite good at the intermediate source countries.

Forward-looking households use equity markets to invest their savings in the R\&D of new goods. For each unit of the final good invested in a new R\&D project, there is a risky return on investment with probability determined by a \textit{Poisson} process. At an aggregate level: domestic households hold a balanced portfolio of infinitely many small firms, such that they face no idiosyncratic risk; savings equal investment; investment flows determine the growth of varieties; and a non-arbitrage condition connects the real interest rate (the asset market) to real returns on R\&D (the equity market).

Over the balanced growth path (BGP), I prove that the equilibrium will be characterized by a stable distribution of income and measures of varieties, the real interest rates will equalize across countries, and all countries will grow at the same rate. Even though there are no international capital markets, trade acts as a vehicle that will integrate R\&D stocks and returns. Countries with larger labor forces will have larger equilibrium measures of varieties, which is a fact also observed in the cross-section of countries in the data.

Exploiting the linearity of income in the measure of varieties, I derive an analytical decomposition for the BGP growth rate across labor and capital income shares of GDP, whose elements can be further interpreted as ``Romer'' and ``Eaton-Kortum'' components, giving intuitive meaning to the results. The Eaton-Kortum component of growth is very much Ricardian, i.e., related to technology, while the Romerian is related to market access, both domestically and internationally.

Another contribution of the paper is to provide a formula for welfare gains from trade that decomposes welfare into static and dynamic components. The welfare formula subsumes the static results of \textcite{arkolakis_new_2012} into a dynamic framework. Like the growth formula, the static component of welfare also has Ricardian and Romerian margins, with the Romerian margin augmenting the Ricardian one through an extensive margin. One of the technical contributions of the model is a tractable way of integrating a new product margin into the Eaton-Kortum framework, which is one of the workhorse models in the international trade literature and lacks such a margin.

By comparing the static and dynamic components of welfare, the model clarifies that they work through different mechanisms, rationalizing the two forces of market access and specialization. The reason is that the former operates on households as consumers and the latter on households as producers and investors

An additional theoretical insight lies in accounting for market access as an avenue for growth and product innovation. Increased market access is related to a higher steady-state equilibrium product innovation growth rate. This finding highlights the positive impact of trade integration on fostering product innovation and is consistent with the reduced-form evidence presented in the beginning of the paper.

The final contribution is to set up and calibrate a quantitative version of the model that solves for the endogenous balanced growth path of the model with an experiment of asymmetric country groups and costly trade. I then use this framework and apply trade cost shocks to replicate the policy scenario of the 2004 Eastwards enlargement of the European Union.

The outcome of the numerical exercise is a set of results and decompositions of both static and dynamic welfare as a result of the EU enlargement. This toolkit suggests that: (a) the EU enlargement increased its long-run yearly growth rate by about 0.10pp; (b) the share of ``Eaton-Kortum'' share in static gains from trade can vary widely across countries, being as large as 90\% for some countries and as small as 10\% for some other countries; (c) dynamic gains can account for a large share of total welfare gains from trade; and (d) the share of dynamic gains also varies across countries, ranging from 65-90\% of total welfare gains from trade.

\paragraph{Related Literature} This paper adds to the theoretical literature on trade and growth \textemdash and in particular to trade and product innovation. The literature can be traced back to the seminal paper by \textcite{romer_endogenous_1990}. While Romer does not develop a full model, he mentions in the paper that a natural extension of his model ``pertain to its implications for growth, trade, and research.''\footnote{This is in section VII of \textcite{romer_endogenous_1990}.} Extensions of the Romer model of endogenous growth of product innovation to a two-country framework were later done by \textcite{rivera-batiz_economic_1991} and \textcite{rivera-batiz_international_1991} as well as \textcite{grossman_comparative_1990}, in a very similar framework. I extend the Romer growth model to a multiple asymmetric country framework and combine it with a modern quantitative Ricardian trade model of \textcite{eaton_technology_2002}. 

The model is also related to the work by \textcite{acemoglu_world_2002}, who proposed a model with Armington trade that features an AK-model of trade and growth with a stable distribution of income over the balanced growth path. While groundbreaking, they restrict their analysis to the costless trade case, while in this paper trade costs can be positive with much more heterogeneity across countries.

Since modeling the complete state space of dynamics and countries is nontrivial, most of the trade and growth literature has to make compromises. Part of the literature simplifies by assuming a world of symmetric countries (\cite{perla_equilibrium_2015}; \cite{sampson_dynamic_2016}) or a two-country world (\cite{eaton_innovation_2006}; \cite{hsu_innovation_2019}; \cite{helpman_foreign_2023}). Another part, while adding the heterogeneity to the cross-section, rules out forward-looking dynamics and models growth as some external diffusion process (\cite{buera_global_2020}, \cite{cai_knowledge_2022}). My model departs from most of the literature by having both asymmetric countries and forward-looking dynamics in a theoretical and quantitative framework. 

As will be clear in the next section, it is a ``true macro model'' combined with a ``true trade model.'' In this sense, it is more similar to the very recent models of \textcite{sampson_technology_2023} and \textcite{kleinman_neoclassical_2023}. However, unlike mine, the latter is a model of convergence rather than a model of long-run growth and the former is a model of firm-productivity growth rather than product innovation.

My paper makes two sets of contributions to the empirical literature. First, it documents a collection of facts using production-and-trade data around the enlargement episodes of the European Union. This first part of the analysis is more akin to papers like \textcite{hummels_variety_2005}, \textcite{bernard_margins_2009}, \textcite{kehoe_how_2013}, and \textcite{arkolakis_extensive_2020}, which provide noncausal documentation of novel stylized facts regarding the extensive margin. But the paper also goes beyond that, using plausibly exogenous variation in an event-study design using a very detailed source-destination-product-year dataset. In doing so, it relates more papers like \textcite{goldberg_imported_2010}, \textcite{bas_input-trade_2012}, \textcite{argente_patents_2020}, and \textcite{rachapalli_learning_2021}, which estimate well-identified empirical effects regarding product innovation.

The paper is organized as follows. Section \ref{section: empirical} introduces the empirical evidence that motivates the work, first summarizing some stylizing facts and then providing some causal evidence on the relationship between market access and product innovation. Later, Section \ref{section: theory} lays down the theory, introduces the model, defines the equilibrium, and states the main results regarding the existence of the balanced growth path and the equilibrium growth rate being decreasing in trade costs. Afterward, Section \ref{section: quatification} uses a numerical version of the model to estimate the dynamic welfare effects of the 2004 enlargement of the European Union and decomposes them using the key results of the previous section. Finally, I conclude by trying to relate the main takeaways to the general literature and where the main advances were.

\section{Empirical Evidence}\label{section: empirical}

This section describes the evidence related to international trade and product innovation in the context of the different enlargement waves of the European Union (EU). First, it describes the data. Then it presents some stylized facts comparing new member states (NMS) of the EU relative to candidate countries. Finally, it uses an event-study approach to isolate some plausibly exogenous variation of trade costs on the probability of initiating production of a new product.

\paragraph{Data sources} Production data comes from Eurostat's Prodcom (\textit{Production Communautaire}), which is an annual full coverage survey of the European mining, quarry and manufacturing sectors, reporting the value of production of 4,000+ different product-lines of EU members and candidate countries. Data are really high-quality and coverage error is estimated to be below 10\%. These data allows one to create a time-series of product counts for products actually produced in each member state and candidate countries of the European Union.    

Bilateral tariff data come from WITS (World Integrated Trade Solution Trade Stats). It consolidates tariff data from the UNCTAD's Trade Analysis Information System (TRAINS) as well as from the WTO. Bilateral trade flow data comes from UNCOMTRADE. 

I matched all of these to the production data using Eurostat's concordance between Prodcom product-codes, which are finer than Harmonized System (HS) 6-digit level, to a HS-6 digit level. WITS and UNCOMTRADE data come natively at an HS-6 digit level. Combined, these constitute a novel production-and-trade matched database.

I also collected data on (a) the dates of accession of new member states to the European Union; (b) trade agreements existent and entered into force between the European Union and third parties before 2004; and (c) expenditure in private research \& development expenditures per capita. The first two come from hand collecting documents and tables from the European Commission's official websites while the latter comes from Eurostat.

Further details on the data used, data matching, and data construction are on Appendix \ref{appendix: data}.

\begin{figure}[htp!]
    \centering
     \includegraphics[width=0.9\linewidth]{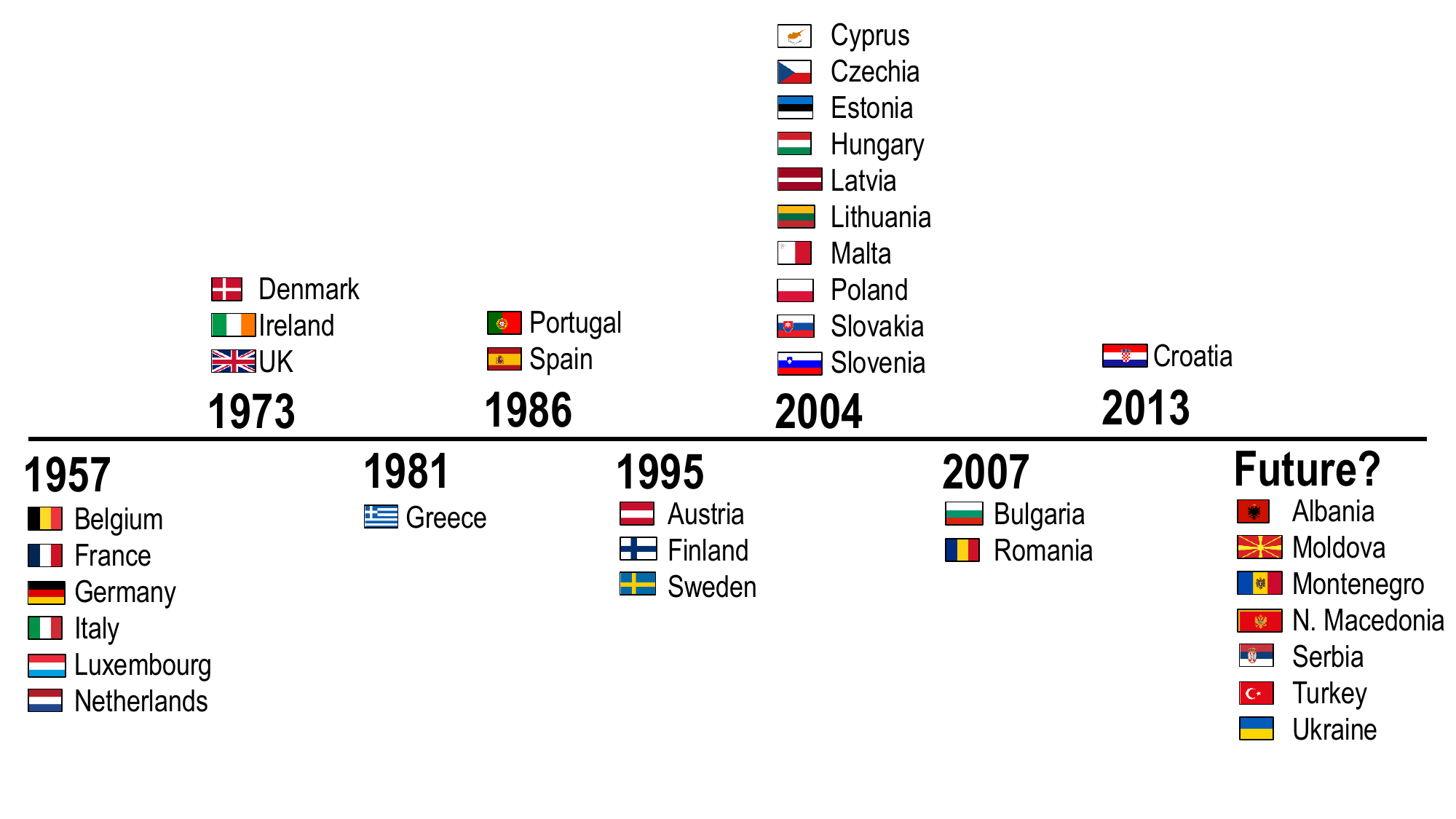}

    \caption{\textbf{Institutional Context: Timeline of European Union Enlargement.} The EU enlargement comes in waves. The future cohorts serve as comparison groups, for some time, to previous cohorts. The largest expansion wave is 2004.}\label{fig: institutional-context}
\end{figure}

\paragraph{Institutional Context} Throughout this paper, the institutional setting will be expansion of the European Union \textemdash or its \textit{enlargement}, how it is typically called in EU language. The enlargement happened in different waves as the EU included more members from the six original members that created the group in 1957, as shown in Figure \ref{fig: institutional-context}.

I use this setting in three ways. First, I exploit the staggered nature of the EU expansion to summarize some stylized facts regarding some key statistics of New Member States (NMS) of the EU relative to candidate countries who are not yet members of the EU at a given horizon.

Then, focusing on the largest expansion wave in 2004, I show that that the adoption of the EU's Common Commercial Policy induced a plausibly exogenous variation in market access between 9 NMS that joined the EU in 2004\footnote{While ten countries joined the EU in 2004, I do not have product-level PRODCOM data for Malta.} and 12 third-party countries with previously existing trade agreements with the EU. Finally, in the numerical quantification exercise, I will also use the enlargement waves as natural country groups to estimate the welfare effects of the 2004 enlargement.

\paragraph{Stylized Facts}

 My stylized facts compare two groups: countries that became new members of the European Union \textit{relative to countries that self-selected into becoming candidates for EU membership but were not yet members at a given time horizon}, exploiting the staggered nature of the enlargement. Here, one can think of countries that became EU members as individual members of a ``treatment group'' and candidate countries that applied for EU membership but had not yet become members by that time as individual members of a ``control group.'' Of course, since treatment assignment, in this case, is not random, this is not actually a true experiment.

In this paper, to avoid the potential biases of the Two-Way Fixed Effects (TWFE) estimator in summarizing the data, I adopt the Callaway-Sant'Anna (CS) estimator\footnote{\textcite{goodman-bacon_difference--differences_2021} shows that the TWFE estimator is a weighted average of all possible two period-two group comparisons and that, as emphasized by \textcite{borusyak_revisiting_2022}, it is biased if treatment effects are heterogeneous. \textcite{sun_estimating_2021} proposed a new estimator that accommodates treatment effect heterogeneity, which was later generalized by  \textcite{callaway_difference--differences_2021}.}. In a nutshell, CS calculates group-specific treatment effects by: (a) comparing the treated group with either the not-yet-treated groups or the never-treated groups; and then (b) aggregating them into an average treatment on the treated given a specific set of weights. This estimator is consistent even if true treatment effects are heterogeneous.

Therefore, even if the objective is to simply summarize the data rather than to make causal claims, one would still want to avoid making ``forbidden comparisons.'' The CS estimator, in this case, will simply recover the average difference in outcomes for NMS relative to countries that are candidate countries but are not yet members, at different horizons around EU enlargement events. In Appendix \ref{appendix: data}, I formal description of the CS estimator used here. 

Given the weighting scheme of the CS estimator, in the estimates reported below, the event that will have the largest weight will be the 2004 EU expansion, which enlarged membership by ten countries, but only nine are observed in PRODCOM data. The other episodes of expansion -- Bulgaria and Romania, in 2007; and Croatia, in 2013 -- influence the estimates with proportional weights for the horizons in which data is available. It is important to highlight that throughout the sample, there is readily available data for candidate countries that never became EU members, which serve as a natural comparison group.
 
 Here, I run the staggered difference-in-differences event study regressions for a set of variables, using similar models. First, using the \textit{measure of produced varieties} as the dependent variable. Then with \textit{log of real private research and development expenditures} and the  \textit{log of real value of yearly trade} as dependent variables.
 
 The frame of reference is to take these variables as aggregate \textit{macro moments}. Relative to a candidate country that did not fully integrate its economy with the European Union and did not have preferential access to the trade partners of the EU, what happens, \textit{on average}, to these variables in New Member States?

\begin{figure}[htp!]
    \centering
     \includegraphics[width=0.8\linewidth]{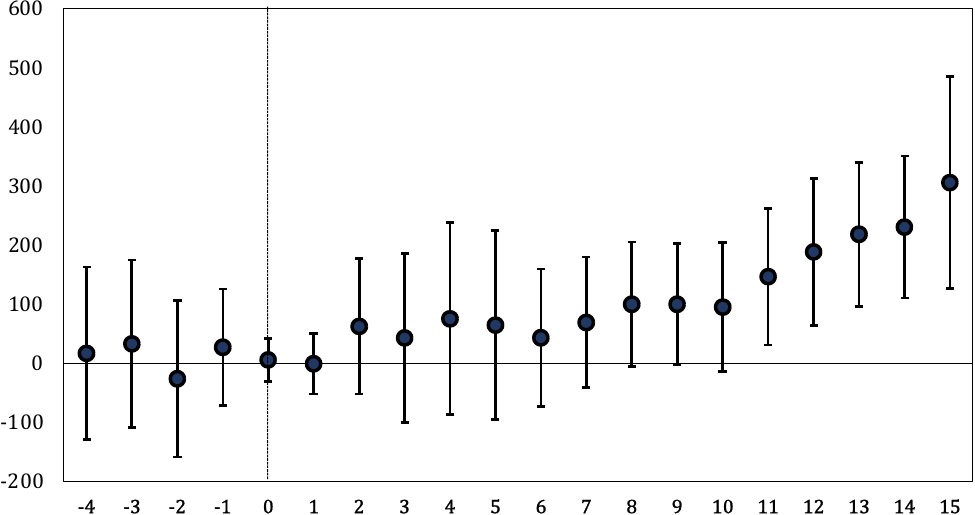}

    \caption{\textbf{Staggered difference-in-differences: Measure of Varieties. \textit{X-axis: years around EU enlargement event. Y-axis: in number of produced varieties.}} This plot shows the estimated coefficients $\theta(t)$ time-specific average treatment on the treated coefficient described by equation \eqref{eq: cs-theta} at the and aggregate level. The bars around the red line denote 95\% bootstrapped standard errors.}
    \label{fig:cs-estimate-measure}
\end{figure}

As shown in Figure \ref{fig:cs-estimate-measure}, fifteen years after membership, the expected differential increase in varieties is $306$, or about $17\%$ relative to the year of membership\footnote{The average treatment on the treated is $306.23$ and the conditional average number of produced varieties in treatment year zero is $1804.6$.}. The differential effect seems to cumulatively increase after the year of membership.

\begin{figure}[htp]
    \centering
     \includegraphics[width=0.8\linewidth]{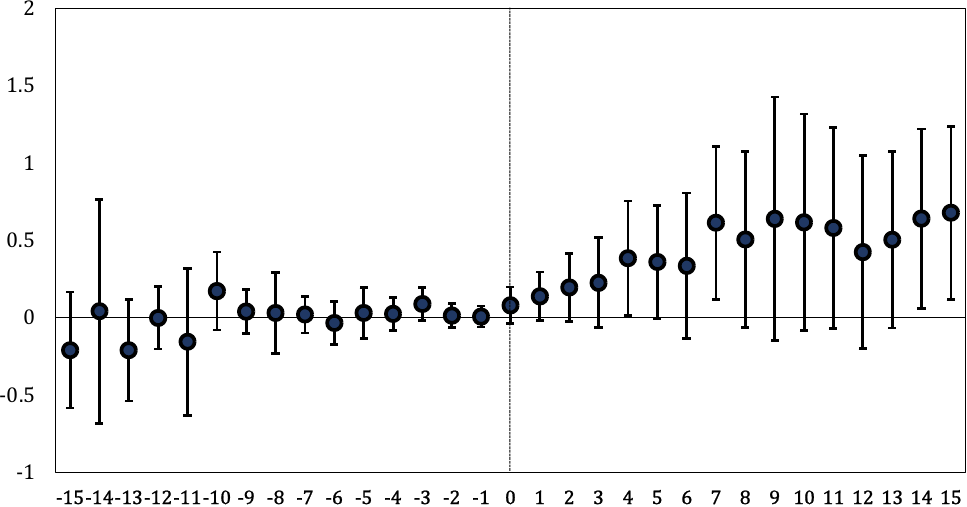}

    \caption{\textbf{Staggered difference-in-differences: Log of Private Research and Development Expenditures Per Capita. \textit{X-axis: years around EU enlargement event. Y-axis: in log value private yearly R\&D expenditures per capita (thousand euro).}} This plot shows the estimated coefficients $\theta(t)$ time-specific average treatment on the treated coefficient described by equation \eqref{eq: cs-theta} at the aggregate level. The bars around the red line denote 95\% bootstrapped standard errors.}
    \label{fig:cs-estimate-log-rd}
\end{figure}

The effects regarding private R\&D, shown in Figure \ref{fig:cs-estimate-log-rd} show a clear break in trend in the differential averages around the year of membership. Fifteen years after the expected differential growth in private R\&D expenditures is about 60\%.

\begin{figure}[htp]
    \centering
     \includegraphics[width=0.8\linewidth]{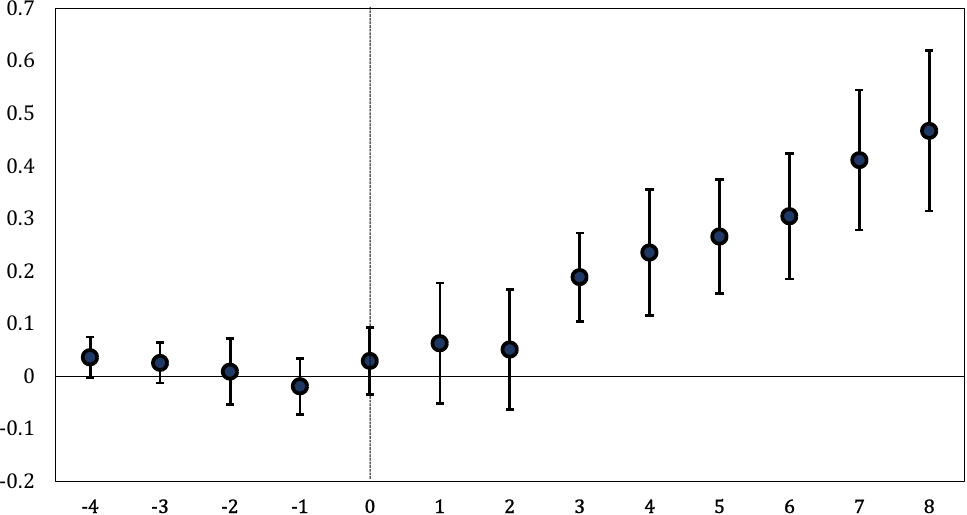}

    \caption{\textbf{Staggered difference-in-differences: Log of Real Value of Yearly Trade. \textit{X-axis: years around EU enlargement event. Y-axis: in log real value of yearly trade (thousand US Dollars).}} This plot shows the estimated coefficients $\theta(t)$ time-specific average treatment on the treated coefficient described by equation \eqref{eq: cs-theta} at the aggregate level. The bars around the red line denote 95\% bootstrapped standard errors.}
    \label{fig:cs-estimate-trade}
\end{figure}

Finally, the results relative to trade also show a differential growth, as illustrated by Figure  \ref{fig:cs-estimate-trade}. There are no signs of pre-treatment trends and, seven years after membership,  the expected differential growth in the value of yearly trade is about 50\%.

\newpage

\paragraph{Causal Evidence}

Once NMS joins the EU, they not only have preferential access to the European market, but they also have to adhere to the Common Commercial Policy of the European Union. This means that these countries have immediate preferential access to third-party markets via previously existing trade agreements between the EU and these third parties. Furthermore, since these trade agreements previously existed, while the NMS had immediate access to them, they did not get to negotiate the tariff variation that they faced \textemdash these were only a byproduct of the EU accession process. 

\begin{figure}[htp]
    \centering
    \begin{tikzpicture}[snake=zigzag, line before snake = 5mm, line after snake = 5mm, line width=0.25mm]
    \draw[red] (0,0) -- (.25,0);
    \draw[red,snake] (.25,0) -- (3,0);
    \draw[red] (3,0) -- (4,0);
    \draw[red] (4,0) -- (11,0);
    \draw[blue,->] (11,0) -- (12,0);

    \draw[red] (0,.95) -- (1,.95);
    \draw[red,snake] (1,.95) -- (3,.95);
    \draw[red] (3,.95) -- (4,.95);
    \draw[blue] (4,.95) -- (12,.95);
    \draw[blue,->] (11,.95) -- (12,.95);

    \foreach \x in {11}
      \draw (\x cm,.205) -- (\x cm,-.205);
    \foreach \x in {4}
      \draw (\x cm,{.95+(.205)}) -- (\x cm,{.95-(.205)});
    \foreach \x in {0}
      \draw (\x cm,{.95+(.205)}) -- (\x cm,{0-(.205)});

    \draw (8,0.95) node[below=.105] {} node[above=.105] {} node[above=.105] {\textcolor{blue}{Preferential Tariffs}};
    \draw (0,0) node[below=.105] {}  node[left=.105] {\textcolor{red}{NMS}};
    \draw (0,0.95) node[below=.105] {} node[above=.105] {} node[left=.105] {\textcolor{blue}{EU}};
    \draw (4,.95) node[below=.5] {} node[above=.105] {} node[below=4] {2000: FTA EU-MX enters into force};
    \draw (11,0) node[below=.105] {} node[above=.105] {} node[below=4] {2004};
    \draw (11,-.3) node[below=.105] {} node[above=.105] {} node[below=4] {$\Delta \tau_{sdip,2004}$};
    \draw (8,0) node[below=.105] {} node[above=.105] {} node[below=.105] {\textcolor{red}{MFN Tariffs}};
  \end{tikzpicture}
    \caption{\textbf{Event Study Design, Constructing the Trade Shock}: \footnotesize{I use the fact that when the NMS joined the EU in 2004, they had immediate preferential access to third-party markets via previously signed EU trade agreements which the NMS did not get to negotiate. The product-level bilateral tariff variation $\Delta \tau_{sdip,2004}$ which was a by-product of the EU accession process is my measure of the market accession shock. In the example above, the EU joined a FTA with Mexico in 2000, but the NMS only joined the EU in 2004, so in 2004 they immediately adhered to these previously negotiated tariff schedule.} }
    \label{fig:event-institutional-setting}
\end{figure}
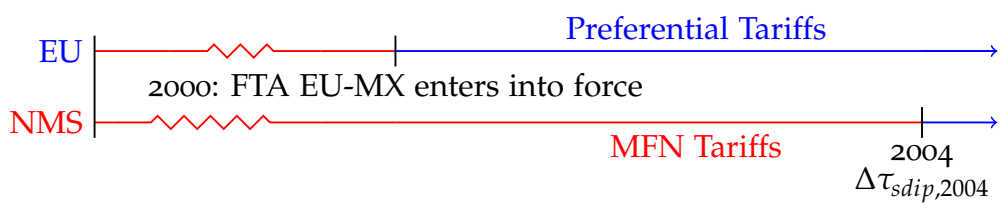

Figure \ref{fig:event-institutional-setting} illustrates how this happened in a specific example: the Free Trade Agreement between the EU and Mexico. The EU joined a FTA with Mexico in 2000, but the NMS only joined the EU in 2004, so in 2004 the latter immediately adhered to these previously negotiated tariff schedule. The product-level bilateral tariff variation $\Delta \tau_{sdip,2004}$ which was a by-product of the EU accession process is my measure of the market accession shock.

I focus on the largest wave of enlargement was in 2004. The source of variation is at the source-country $\times$ destination-country $\times$ HS-6-code product level. The metric of the tariff shock change is simply $\Delta \tau_{sdip,2004} \equiv -(\tau_{sdip,2004} - \tau_{sdip,2003})$, which is the change in the level of effectively applied bilateral tariffs at the product level between 2003 and 2004. In each year, there are about 300 thousand observations.\footnote{Leading up to 2004 there were no large changes in bilateral tariffs between NMS and third parties but between 2003 and 2004 there was a median drop of about 2.5 percentage points. In the years after the enlargement, there was also not a large change in the distribution of bilateral tariff rates, which is shown in Figure \ref{fig:event-tariff-distribution} in Appendix \ref{appendix: data}. Figure \ref{fig:event-delta-tau} in Appendix \ref{appendix: data} plots the distribution of $\Delta \tau_{sdip,2004}$, excluding the zero-valued observations.}

I estimate a sequence of cross-sectional local-projection linear probability models, which estimate what is the marginal effect of a \textit{decrease} in the tariffs on exports of a given product $p$, \textit{conditional on that country $s$ not producing that particular product before joining the EU in 2003}. The fact the data is highly granular permits me to exploit within $industry \times source \times destination \times horizon$ (across
product) variation. 

Formally, I estimate the following equation:
\begin{eqnarray}\label{eq: entry-regressions}
    P\left(X_{sdip,h} > 0 \middle| Y_{s\cdot ip,2003} = 0 \right) &=& 
    \alpha_h + \beta_h \cdot \Delta \tau_{sdip,2004} + \gamma_{sdi,h} + \nu_{sdip,h} \\
    & & \text{for } h \in \{2000, \cdots, 2010\} \nonumber
\end{eqnarray}

\noindent where $X_{sdip,h}$ is the market value of exports between country $s$ and country $d$ of product $p$ of industry $i$ at horizon $h$; $Y_{s\cdot ip,2003}$ is the market value of production in country $s$ of product $p$ of industry $i$ in $2003$; $\alpha_h$ are horizon (time) fixed-effects; $\gamma_{sdi,h}$ are $source \times destination \times industry$ interactions fixed-effects for each $h$. These types of cross-sectional event studies with local projections can be interpreted as differences in differences with continuous treatments\footnote{See \textcite{chodorow-reich_geographic_2019} and \textcite{dube_local_2023}.}. 

This strategy takes seriously the assertion in \textcite{baier_free_2007} that countries engage endogenously in free trade agreements (FTAs). The identification assumption is that conditional on the very saturated fixed effects that this model includes, the unobserved components $\nu_{sdip,h}$ are uncorrelated with the change in tariffs $\Delta \tau_{sdip,2004}$. Intuitively, the identification is robust to a NMS (say, Poland's) policymakers endogenously targeting EU accession to have preferential access to a third-party's (say, Mexico's) car industry (relative to other industries and countries), but not if they want to have preferential access to compact cars relative to SUVs in Mexico. For further details on the methodology, see Appendix \ref{appendix: data}.

As shown in Figure \ref{fig:event-treatment-entry}, an increase in market access by 1 percentage point increases the probability of starting to produce and export a given product by about 1 percent by 2010. To benchmark this result, it is about one-third of the conditional mean $\mathbb{E}[X_{sdip,h} > 0 | X_{s\cdot ip,h} > 0, h > 2003] = 2.9\%$. There are no signs of a pre-existing trend before 2004: both the magnitude of the coefficients and the standard errors are very small before the treatment date.

\begin{figure}[htp!]
    \centering
    \includegraphics[width=0.95\textwidth]{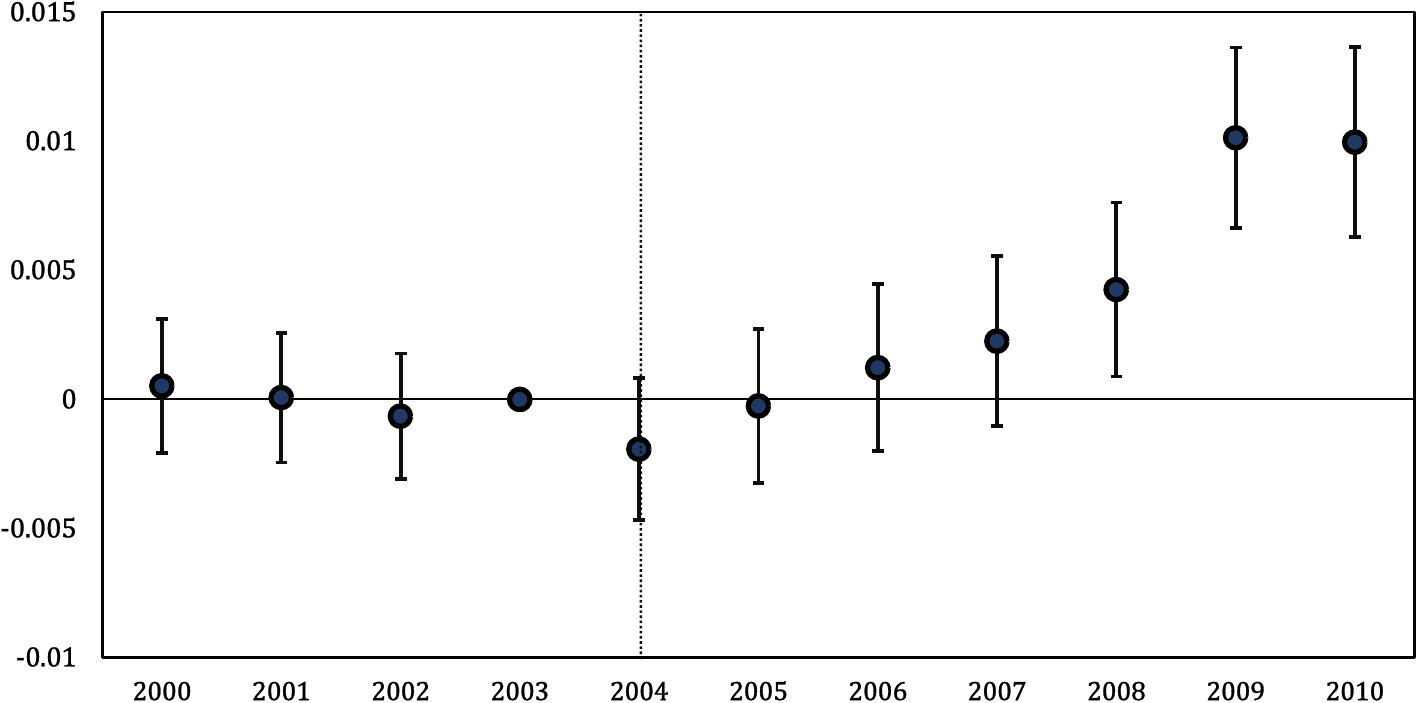}
    \caption{\textbf{Entry Regressions.} \footnotesize{This plot shows the coefficients $\beta_h$ of the local projection linear probability models specified in equation \eqref{eq: entry-regressions}. Each year is a different cross-sectional regression with approximately 300 thousand observations. The whiskers show 95\% confidence intervals with robust standard errors clustered at the source-destination-industry level.}}
    \label{fig:event-treatment-entry}
\end{figure}

\paragraph{Summary of Empirical Evidence} I have documented the following novel facts: as \textit{New Member States} go through a large trade integration event with the European Union, they \textit{start producing new product varieties; investing more in research and development; and trading more in real values relative to candidate countries that are not yet members} at a given horizon. Furthermore, I have shown that as they are exposed to a \textit{plausibly exogenous new market accession shock}, as it happened due to the idiosyncrasies of the Common Commercial Policy of the EU, \textit{they increase their probability of starting production of a new product variety}, which suggests that market access increases the rate of product innovation.

To rationalize this reduced-form evidence, I develop a dynamic general equilibrium model that is consistent with both the stylized facts and the market access mechanism. This is what I do in the following section.

\newpage

\section{Theory}\label{section: theory}

Here I present a dynamic multi-country model of the world economy with intertemporal optimization, investment in research and development, and trade in final and intermediate goods. In this economy, time is continuous with $t \in \mathcal{T} \equiv [0,\infty )$ and countries indexed by $s \in \boldsymbol{K} \equiv \{ 1,\ \cdots,N \}$.

Every country has the ability to produce final goods $\omega \in [0,1]$. However, they differ in their ability to produce non-rival intermediate inputs $\nu \in [0,M_{s}(t)]$, where the upper bound of the interval $M_{s}(t)$ defines the product space of a particular country. Intermediate goods are \textit{non-rival} in the same spirit as in the endogenous growth literature: new blueprints can be simultaneously used by multiple producers at the same time, inducing increasing returns to scale\footnote{See \textcite{jones_growth_2005} and \textcite{jones_paul_2019} for extensive reviews.}.  

As intermediate goods are invented, trade acts as a mechanism that diffuses new blueprints: producers expand their production function by sourcing newly minted inputs from around the world. Exporters are monopolists in their intermediate varieties and therefore have the incentive to invest in the development of new varieties, thereby propelling growth. Therefore, international trade will work as a vehicle that integrates global research and development stocks and induces growth-rate convergence over the balanced growth path.

My goal is to make this model easily accessible and recognizable for someone who is familiar with either modern trade theory or modern growth theory. This model will recover, as special cases, the \textcite{eaton_technology_2002} model of trade and the \textcite{romer_endogenous_1990} model of growth. Some functional form assumptions will be such that this nesting is clear.

\subsection{Demand}

In each country $s \in \boldsymbol{K}$, there is a representative household the maximizes its lifetime utility according to:
\begin{eqnarray*}
    \max_{   C_{s}(t) } & & \int_0^{\infty} \exp\{-\rho t\} \log \left( C_{s}(t)\right) dt \\
 s.t. \quad  P_s(t) I_s(t) + P_{s}(t) C_{s}(t)  &=& r_s(t) A_s(t) + w_s(t) L_{s} 
\end{eqnarray*}

\noindent where $P_{s}(t) C_{s}(t) $ are aggregate consumption good prices and quantities in country $s$; $I_s(t)$ are instantaneous investment flows; and $w_{s}(t), r_{s}(t)$ are wages and interest rates. At any instant, the state of asset holdings is simply the cumulative investment flows: $A_s(t) \equiv \int_0^t I(s) ds$\footnote{This, of course, implies that one can write investments as $I_s(t) = \dot{A}_s(t)$, which clarifies the optimal control problem at hand.}. 

Households choose a sequence of consumption quantities for the aggregate good, satisfying the Euler Equation:

\begin{equation}\label{eq: euler-equation}
    \frac{\dot{C}_s(t)}{C_s(t)} = \frac{r_s(t)}{P_s(t)}  - \rho
\end{equation}

\subsection{Production and Trade in Varieties}

There are three kinds of producers in each country: \textit{those who produce varieties of the final good}, \textit{those who produce varieties of intermediate goods}, and \textit{those who invest in research and development}. This section will focus on the two first ones.

\paragraph{Final Goods Producers.} In each country, a local assembler for the final composite good $Y_d(t)$ who operates under perfect competition uses the least expensive variety  $\omega \in [0,1]$ available at $d \in \boldsymbol{K}$ with the following technology:

\begin{equation*}
    Y_d(t) = \Big[ \int_0^1 y_{d}(t,\omega)^{\frac{\sigma-1}{\sigma}}  d\omega \Big] ^{\frac{\sigma}{\sigma-1}}
\end{equation*}
\noindent where $\sigma > 1$ is a constant elasticity of substitution across sourced varieties $\omega$. Under these assumptions, the ideal price index of the final good satisfies $P_{s}(t)$ satisfies:

\begin{equation*}
    P_{s}(t) = \Big[ \int_0^1 p_{s}(t, \omega)^{1-\sigma}  d\omega \Big] ^{\frac{1}{1-\sigma}}
\end{equation*}

A producer of each variety $\omega \in [0,1]$ of the final good is endowed with a constant returns to scale technology that combines labor and intermediate inputs $\nu \in [0,M_{s}(t)]$ coming from multiple countries $k \in \boldsymbol{K}$:

\begin{equation}\label{eq:production-function}
    y_{s}(t, \omega) = z_{s}(t, \omega)  [\ell_{s}(t, \omega)]^{1-\alpha} \left( \frac{1}{\alpha} \sum_{k \in \boldsymbol{K}}  \int_{0}^{M_{k}(t)} [x_{ks}(t, \omega,\nu)]^{\alpha} d\nu  \right)
\end{equation}

\noindent where $z_{s}(t, \omega)$ is total factor productivity; $\ell_{s}(t, \omega)$ is factor demand for labor for variety $\omega \in [0,1]$ located in country $s$; and $x_{ks}(t, \omega,\nu)$ is the demand for a intermediate good of variety $\nu \in [0,M_{k}(t)]$ sourced from country $k$ for production as an input of a final good in country $s$.

Non-rival intermediate goods varieties are differentiated across countries: an input $\nu \in [0,M_{k}(t)]$ is different from $\nu \in [0,M_{n}(t)]$, even if it is indexed by the same symbol. For instance, the first one may be a twelve-core computer chip from Estonia while the second one may be a large language model from Malta. Additionally, note that countries differ in their ability to produce intermediate goods, which is denoted by the upper bound of the integral $M_{k}(t)$. Optimal demand for an intermediate good satisfies:

    \begin{equation}\label{eq: demand-intermediate}
            x_{ks}(t,\omega, \nu) = \left[ \frac{p_{ks}(t,\omega, \nu)}{p_{ss}(t, \omega)} \right] ^{- \frac{1}{1-\alpha}} \cdot \ell_{s}(t, \omega) \cdot z_{s}(t, \omega)^{\frac{1}{1-\alpha}}
    \end{equation}

\paragraph{Intermediate Goods Producers.} Each intermediate goods producer in country $s$ has perpetual rights over the production of each variety $\nu \in [0,M_{s}(t)]$. They are endowed with a linear technology that transforms one unit of the final good into one unit of the intermediate good. 

\begin{assumption}[Trade Costs] \label{ass: trade-costs} Trade is subject to iceberg trade costs, which implies that shipping a final or intermediate good variety from source region $s$ to a consumer in region $d$ requires producing $\tau_{sd} \ge 1$, where $\tau_{dd} = 1$ and $\tau_{sd}=\tau_{ds}$ for all $s,d \in \boldsymbol{K}$.
\end{assumption}

Given assumption \eqref{ass: trade-costs}, intermediate goods producers face heterogeneous marginal costs and set optimal prices accordingly through market-specific price discrimination. They take marginal costs and demand curves as given and choose optimal prices to maximize profits, with the optimal price being a mark-up over marginal costs for every variety $\nu$ and $\omega$:

\begin{equation*}
    p^M_{ks}(t) = \frac{\tau_{ks} P_{k}(t) }{\alpha} \qquad \forall \omega \in [0,1], \quad \nu \in  [0, M_{k}(t)]
\end{equation*}

Note that this is the standard result of profit maximization under monopolistic competition with two variations. First, as in most trade models, prices are differentiated by destination and are inclusive of trade costs $\tau_{ks}$. Second, since intermediate goods use one unit of the final good at the origin country $k$ to produce one unit of the intermediate good, its marginal cost is $P_k(t)$. Optimal monopolist prices being independent of variety $\nu$ imply that demand for is symmetric:

\begin{equation*}
    \bar{x}_{ks}(t, \omega) \equiv \left[ z_s(\omega) \cdot \frac{p_{ss}(t, \omega)}{p^M_{ks}(t)} \right]^{\frac{1}{1-\alpha}}   \cdot \ell_{s}(t, \omega) \qquad \forall \nu \in [0, M_k(t)]
\end{equation*}

\noindent Given the result above, rewrite the final goods firm maximization problem in the following way:

\begin{equation}\label{eq: production-function-main}
 \max_{\ell_{s}(t, \omega)} \frac{1}{\alpha} [ p_{ss}(t, \omega) \cdot z_{s}(t, \omega)]^{\frac{1}{1-\alpha}} \cdot \tilde{M}_{s}(t) \cdot \ell_{s}(t, \omega)   - \ell_{s}(t, \omega) w_{s}(t) 
\end{equation}

\noindent which comes from substituting for $\bar{x}_{ks}(t)$ and defining:

\begin{equation}
    \underbrace{\tilde{M}_{s}(t)}_{\substack{\text{effective}\\ \text{measure of}\\ \text{input varieties}}} \equiv  \sum_{k \in \boldsymbol{K}} \underbrace{M_{k}(t)}_{\substack{\text{measure of} \\ \text{varieties} \\
    \text{in each } k}} \cdot  \left( \underbrace{p^M_{ks}(t)}_{\substack{\text{optimal monopolist} \\ \text{price from } k \text{ to } s} } \right)^{1-\eta}  \qquad \text{where } \eta = \frac{1}{1-\alpha}
\end{equation}

The effective measure of input varieties is a key object in this model that captures the diffusion of non-rival intermediate goods to country $s$. It measures input varieties sourced from each country weighted by marginal cost. The first term on the right-hand side captures heterogeneity in the source-country measure of varieties since final goods producers are sourcing intermediate varieties internationally. The second term captures the substitutability across intermediate goods, controlled by the elasticity of substitution $\eta$. The exponent $1-\eta < 0$ down-weights the relative importance of intermediate goods coming from source countries $k$ with relatively more expensive intermediate inputs.

This object also makes explicit how the model nests both the Eaton-Kortum model of trade and the Romer growth model. If $\alpha \to 0$, then there is no intermediate sector. The technology $\eqref{eq: production-function-main}$ collapses into a linear production function as in \textcite{eaton_technology_2002}. If the world is in autarky  \textemdash i.e., if $\tau_{sd} \to \infty$ for all $d \neq s$ \textemdash then, after setting $P_s(t) =1$ as the num\'eraire of the home economy, $\tilde{M}_{s}(t) = \alpha^{\frac{\alpha}{1-\alpha}} M_s$ and the final goods technology becomes linear in labor with an extensive margin $M_s$, as in \textcite{romer_endogenous_1990}\footnote{For a more detailed description of the nesting, see Appendix \ref{appendix: nesting}.}.

Even outside the limiting cases expressed above, one should observe that the final goods producer's technology in this model is related to those in both the Eaton-Kortum and the Romer models. It is equivalent to a simple Eaton-Kortum model that uses the final good as an intermediate input with an added extensive margin shifter $\tilde{M}_{s}(t)$. It also is related to the technology of the Romer model, which is linear in labor (or human capital), except that in this model the measure of varieties component is a weighted average of inputs coming from domestic and international suppliers.

Note that with a slight redefinition, one can also interpret a transformation of $\tilde{M}_s(t)$ as the price of a composite basket of intermediate goods, as it is standard in many models that resort to a constant elasticity of substitution:

\begin{equation}
    P^M_{s}(t) \equiv    \tilde{M}_{s}(t) ^{\frac{1}{1-\eta}} =  \left( \sum_{k \in \boldsymbol{K}} M_{k}(t) \cdot  p^M_{ks}(t)   ^{1-\eta} \right)^{\frac{1}{1-\eta}}
\end{equation}

\noindent which is how, due to notational convenience, this object will appear throughout the rest of the paper. Define the value in final goods in country $d \in \boldsymbol{K}$ to be $P_d(t) Y_d(t)$. Then, using the definitions above and the properties of C.E.S. and Cobb-Douglas, intermediate goods sales by country $s$ in country $d$ equal:

\begin{eqnarray*}
    M_s(t) p^M_{sd}(t) x_{sd}(t) = \alpha \cdot M_{s} \left( \frac{ p^M_{sd}(t) }{P^M_d(t)} \right) ^{1-\eta}  \cdot P_d(t) Y_d(t)  =  \alpha \cdot \lambda_{sd}^M(t) \cdot P_d(t) Y_d(t) 
\end{eqnarray*}

The last equation follows from defining intermediate trade shares $\lambda_{sd}^M(t)  \equiv \frac{ M_{s} p_{sd}^M(t)^{1-\eta} }{\sum_{k \in \boldsymbol{K}}  M_{k} p_{kd}^M(t)^{1-\eta} }$.

In the standard \textcite{romer_endogenous_1990} model, assemblers source intermediate goods exclusively from domestic suppliers. One important implication of that assumption is symmetry: the price of all intermediate goods will be the same. Conversely, in this framework, when sourcing intermediate goods from multiple countries $k \in \boldsymbol{K}$, the prices of these goods will no longer be necessarily the same. 

Economically, it is this lack of symmetry in prices that will induce substitutability across varieties sourced from different countries, which is reflected in the composite price of intermediate goods above. The more dissimilar countries are in terms of their relative unit costs, the more substitution across intermediate goods will occur. Hence, accommodating asymmetric countries in a dynamic framework will be an important feature of this model. 

\paragraph{Trade in final goods.} The factory gate price $p_{ss}(t, \omega)$ for a variety has three components: the unit production cost $w_{s}(t)$, the price of intermediate goods $P^M_{s}(t)$, and a producer-specific productivity $z_{s}(t, \omega)$. Destination prices also include iceberg trade costs. Under perfect competition, consumers in country $d$ choose the lowest price variety $\omega$ available at the domestic market:

\begin{equation}\label{eq: landed-prices}
    p_{d}(t, \omega) = \min_{s \in \boldsymbol{K}} \left\{ p_{sd}(t, \omega) \right\} = \min_{s \in \boldsymbol{K}} \left\{ \tau_{sd} p_{ss}(t, \omega) \right\} = \min_{s \in \boldsymbol{K}} \left\{  \frac{ P^M_s(t)^{\alpha} w_{s}(t)^{1-\alpha}  \tau_{sd}}{ z_{s}(t, \omega)} \right\}
\end{equation}

\begin{assumption}[Productivity draws]\label{ass: productivity} 
Following \textcite{eaton_technology_2002}, assume that $z_{s}(t, \omega)$ is an iid random variable drawn from a market-specific Fr\'echet distribution

\begin{equation*}
    F_{s}(t)(z) = \exp\!\left\{ - T_{s} \, z^{-\theta} \right\}.
\end{equation*}

\noindent where $T_{s}$ is the the \textit{scale parameter} and $\theta$ is the shape parameter.
\end{assumption}

Given assumption ~\eqref{ass: productivity}, both prices and demanded quantities (which are functions of productivity draws) are also random variables. By the law of large numbers, the share of varieties sourced from $s$ to $d$ equals\footnote{Since there are infinitely many varieties $\omega$ and productivities are iid random variables, by the law of large numbers, the share of varieties sourced from $s$ to $d$ converges almost surely to the probability of sourcing a specific variety from $s$ to $d$.}:

\begin{equation}\label{eq: gravity}
    \lambda^F_{sd}(t) \equiv \frac{E^F_{sd}(t)}{E^F_{d}(t)} = \frac{T_{s}  (P^M_s(t)^{\alpha} w_{s}(t)^{1-\alpha} \tau_{sd} )^{-\theta}}{\sum_{n \in \boldsymbol{K}}{T_{n}  (P^M_n(t)^{\alpha} w_{n}(t)^{1-\alpha} \tau_{nd} )^{-\theta}}}
\end{equation}

\noindent where $E^F_{sd}(t)$ denotes the expenditure on final goods going from country $s$ to country $d$; $E^F_{d}(t)$ denotes total expenditure on final goods in country $d$.

\subsection{Research and Development}

The \textit{research sector} creates new varieties of the intermediate good. One can think of this sector as investing in the invention of new machines, which result in new blueprints. These firms use $\psi$ units of the final good as inputs to research and development (R\&D), but success is not guaranteed.

\begin{assumption}[Research and Development Process] The success rate of R\&D follows a \textit{Poisson} process with flow arrival rate equal to $\psi I_{s}(t) dt$, where $I_{s}(t)$ is the research input, measured in units of the final good per time unit. 
\end{assumption}

Once researching firms invent a new machine, they hold perpetual monopoly rights over the new variety $\nu$. They can either set up their own shop to produce and enjoy the profits of producing such variety at the market or, alternatively, they can sell the rights to this patent to an intermediate variety producer. In either case, domestic households, that finance the invention of new varieties through capital markets, will collect the profits. 

The economic value of a new variety is the present value of producing the new varieties and selling them as intermediate inputs to final goods producers, which is, at period $t$:
    
\begin{equation}\label{eq: value-firm}
    V_{s}(t, \nu) = \int_{t}^{\infty} \exp \left\{ -\int_{t}^{\tau} \frac{r_{s}(k)}{P_s(k)} dk  \right\} \pi_s(\tau, \nu) d\tau
\end{equation}
    
where $\pi_s(\tau, \nu)$ is the flow profit per variety per unit of time. Research firms will only invest if the expected return of their investment is positive, that is $ \psi V_{s}(t, \nu) I_{s}(t, \nu) - P_{s}(t) I_{s}(t, \nu) \ge 0$. With free entry, this condition holds with equality and in equilibrium it pins down the value of each variety: $V_{s}(t, \nu) = P_{s}(t) / \psi$.

Since the only asset market in this economy is the domestic equity market, domestic households save by funding investments in new varieties through a balanced portfolio of infinitely many small firms, such that they face no idiosyncratic risk. At the aggregate level, then, $\dot{M}_{s}(t) = \psi I_{s}(t)$, where $I_{k}(t)$ is the level of aggregate investment in the domestic economy measured in units of the final good.  The value of aggregate assets is simply the value of all invented varieties $P_s(t) A_s(t) = M_s(t) V_{s}(t)$ and, since the arrival rate of ideas is constant, the total stock of assets is at any instant a function of the total measure of varieties $A_s(t) = M_s(t) / \psi$.

Taking the derivative of both sides of \eqref{eq: value-firm} with respect to time and noting that both $V_{s}(t, \nu)$ and $\pi_s(\tau, \nu)$ are independent of $\nu$ pins down the real interest rate in this economy. The result is a non-arbitrage condition relating returns on assets to returns on R\&D:


\begin{equation}\label{eq: rents}
    \frac{r_s(t)}{P_s(t)} = \underbrace{\frac{\psi \cdot \pi_s(t, \nu)}{P_{s}(t)}}_{\text{flow dividend rate}} +  \underbrace{\frac{\dot{P}_s(t)}{P_{s}(t)}}_{\text{capital gains}}
\end{equation}

\subsection{Market Clearing and Equilibrium}\label{sec: equilibrium-bgp}

\paragraph{Factor Market Clearing} Let $Y_d(t)$ denote the total output of the final good and $X_d(t), I_d(t)$ denote the use of the final good as inputs for the production of intermediate inputs and R\&D, respectively. Then total output in the final good for a given country must satisfy:

\begin{equation}
    Y_d(t) = C_d(t) + I_d(t) + X_d(t)
\end{equation}

\noindent where $I_d(t)$ and $C_d(t)$ are pinned down by the dynamic problem, described below, and $X_d(t)$ can be expressed as a function of aggregate demand in all destinations.

\paragraph{Expenditure Determination}  Flow aggregate profits $\Pi_{s}(t) \equiv \int_0^{M_s(t)} \pi(t,\nu) d\nu$ are a constant fraction of revenue:

\begin{equation}\label{eq: profits}
    \Pi_{s}(t) = \frac{\alpha}{\eta} \cdot \sum_{d \in \boldsymbol{K}} \lambda_{sd}^M(t)  \cdot P_d(t) Y_d(t) 
\end{equation}

On the expenditure side, GDP of each destination country $s \in \boldsymbol{K}$ country will be exhausted as the combination of the total expenditures of labor and capital income:

\begin{equation}
    P_s(t) Y_s(t) = w_s(t) L_s + \Pi_{s}(t)
\end{equation}

From the income side, nominal GDP must equal the sum of total flow payments received domestically and from the rest of the world:

\begin{equation}\label{eq: expenditure-market-clearing}
     P_s(t) Y_s(t) = \sum_{d \in \boldsymbol{K}}  \left[  (1-\alpha) \lambda_{sd}^F(t) + \frac{\alpha }{\eta } \lambda_{sd}^M(t) \right]   P_d(t) Y_d(t)
\end{equation}

\paragraph{Trade Balance} Since savings equals investment and there is no access to international capital markets in this economy, GDP accounting requires that trade balances in each country and net exports are equal to zero at any instant:

\begin{eqnarray}\label{eq: trade-balance}
    & & \sum_{ d \neq s \in \boldsymbol{K}} \lambda_{sd}^F(t) P_d(t) Y_d(t) + \alpha \sum_{ d \neq s \in \boldsymbol{K}} \lambda_{sd}^M(t) \left[ \sum_{ k' \in \boldsymbol{K}} \lambda_{dk'}^F(t) P_{k'}(t)Y_{k'}(t) \right] =  \nonumber \\
    & & [1 - \lambda_{ss}^F(t) ]  P_{s}(t)Y_{s}(t) + \alpha   [1 - \lambda_{ss}^M(t) ]  \left[ \sum_{ k' \in \boldsymbol{K}} \lambda_{dk'}^F(t)  P_{k'}(t)Y_{k'}(t) \right] 
\end{eqnarray}

\paragraph{Dynamic Equilibrium} The dynamics in each of the countries of this world economy are governed by the following system of differential equations:

\begin{eqnarray}\label{eq: dynamical-system}
    \dot{C}_s(t) &=&  \left[ \frac{r_s(t)}{P_s(t)} - \rho \right] C_s(t) \\
    \dot{M}_s(t) &=& \frac{r_s(t)}{P_s(t)} M_s(t) + \psi \frac{w_s(t)}{P_s(t)} L_s - \psi C_s(t) \nonumber
\end{eqnarray}

As it is clear from the system above, the dynamics of the model are essentially neoclassical. However, since openness to trade impacts the cross-sectional distribution of wages and prices, it will also impact the path of consumption product measures over time.

The first equation \textemdash the Euler Equation \textemdash states that the household in a country $s \in \boldsymbol{K}$ will choose an upward-sloping consumption path if the real interest rate is greater than the rate of time preference. The higher this gap, the more a household will be willing to defer current consumption and take advantage of higher returns in the asset and R\&D markets.

The second equation is less obvious to interpret in its current form, but it states that the growth in the product measure in each country is proportional to the net investment rate. Since expected profits of new varieties are always positive, the net investment rate is also always positive, which means that new varieties are always created, inducing growth in this model.

A more explicit way to observe the net investment rate is by writing the second equation in its equivalent asset representation. Since $M_s(t) = \psi A_s(t)$, then:

\begin{equation*}
    \dot{A}_s(t) = I_s(t) = \underbrace{ \frac{r_s(t)}{P_s(t)} A_s(t)}_{\text{real capital income}} +  \underbrace{\frac{w_s(t)}{P_s(t)} L_s}_{\text{real labor income}} -  \underbrace{C_s(t)}_{\text{real consumption}}
\end{equation*}

\noindent which, along with the discussion regarding the non-arbitrage condition in the previous section, helps clarify that asset markets and varieties markets are two sides of the same coin.

The assumption of log preferences substantially simplifies the dynamic problem. In Appendix \ref{appendix: solution-dynamic-problem}, I show that instantaneous consumption is always well-defined as a constant fraction of  lifetime wealth:

\begin{equation*}
    C_s(t) = \rho \left[ \underbrace{A_s(t)}_{\text{wealth at }t}  +  \underbrace{\int_t^\infty  \frac{w_s(\tau)}{P_s(\tau)} L_s \cdot\exp \left\{  - \bar{r}_s(\tau) \cdot \tau \right\} d\tau}_{\text{PV of future labor income}}  \right]  \nonumber
\end{equation*}

\noindent where $\bar{r}_s(\tau) \equiv \frac{1}{\tau} \int_t^\tau  \frac{r_s(\nu)}{P_s(\nu)}  d\nu$ is the average real interest rate between periods $t$ and $\tau$. Once there is an explicit  solution for consumption at every $t$, the differential equation for $\dot{M}_s(t)$ becomes autonomous, and also has an explicit solution as a function of the path of prices.

It also shows that there is a unique initial choice of consumption that is consistent with the optimal choices described by \eqref{eq: dynamical-system} and the transversality condition. Since the other conditions to satisfy the Maximum Principle are satisfied, this is equivalent to showing that the solution to the dynamic problem is unique.

\begin{definition}[Dynamic Equilibrium] The dynamic equilibrium of the world economy is defined by a collection of paths of consumption quantities, assets stocks, and profit flows $[C_s(t), A_s(t), \Pi_s(t)]$; paths of final goods varieties output quantities $[y_s(t,\omega)]$; paths of intermediate goods varieties output quantities $[x_{ks}(t,\omega,\nu)]$; paths of prices $[w_s(t), r_s(t), P_s(t)$, $p_{ss}(t,\omega), p_{sk}(t,\omega,\nu)]$; and a vector of fundamentals $(\theta, \sigma, \boldsymbol{T}, \boldsymbol{\tau})'$ where $\boldsymbol{T} \equiv \{T_s\}$ is a collection of location parameters of the Fr\'echet distribution and $\boldsymbol{\tau} \equiv [\tau_{sd}]$ is a matrix of trade costs, such that: (a) households maximize utility given the path for prices; (b) final goods firms maximize profits given the path for prices; (c) intermediate goods firms choose prices to maximize profits given demand functions and final goods prices; (d) trade balances; and (e) factors and goods markets clear.   
\end{definition}

\paragraph{Homogeneity of Income in Equilibrium} One of the key properties of this model is that real income, real wages, and real profits are a function of the measure of varieties $M_s(t)$ and of terms that are homogeneous of degree zero in the distribution of the measure of varieties $\{ M_k(t) \}_{k \in \textbf{K}}$. Note that for real aggregate labor and aggregate capital income, respectively, can be expressed as:

\begin{eqnarray*}
    \frac{w_s(t)  L_s }{P_s(t)} &=& M_s(t)  \cdot  \left(  \frac{ T_{s} }{   \lambda^F_{ss}(t) } \right) ^{\frac{1}{\theta(1-\alpha)}} \cdot  \left( \lambda^M_{ss}(t) \right)^{-1} \cdot  L_s \equiv M_s(t) \times \mathcal{R}^w_s(t)  \\
    \frac{\Pi_s(t)}{P_s(t)} &=& M_s(t) \cdot \frac{\alpha}{\eta} \sum_{d \in \boldsymbol{K}} \frac{ \left( \tau_{sd} P_s(t) \right)^{1-\eta}}{\sum_{k' \in \boldsymbol{K}} M_{k'} \left( \tau_{{k'}d} P_{k'}(t) \right)^{1-\eta} } \frac{P_d(t) Y_d(t)}{P_s(t)} \equiv M_s(t) \times \mathcal{R}_s^\pi(t) 
\end{eqnarray*}

\noindent which, of course, means that Real GDP is also a function of $M_s(t)$ times a term that is homogeneous of degree zero in the distribution of the measure of varieties:

\begin{equation}\label{eq: gdp-linear}
    Y_s(t) = \frac{w_s(t)  L_s }{P_s(t)} + \frac{\Pi_s(t)}{P_s(t)} = M_s(t) \times \left[ \mathcal{R}^w_s(t) +  \mathcal{R}^\pi_s(t) \right] \equiv  M_s(t) \times \mathcal{R}_s(t) 
\end{equation}

This property is important because it is the mechanism that induces increasing returns to scale in this model. It will also be important to characterize the existence of the Balanced Growth Path (BGP).

Intuitively, even though $\mathcal{R}_s(t)$ is a complicated function, the fact that real income is a function of $M_s(t)$ times a term that is homogeneous of degree zero in $\{ M_k(t) \}_{k \in \textbf{K}}$ means that this model falls within the broader class of AK-models in macroeconomics, a property that is inherited from the Romer side of production. Many of the well-behaved properties of AK-models about long-run growth will carry through to this model.

Furthermore, as it will become clear below, trade will influence both the returns to idea creation ($\mathcal{R}_s$) as well as the stock of ideas ($M_s$). But it is important to bear in mind that some of the mechanisms behind growth in this economy are increasing returns to scale exemplified by the linearity of the production function.

\subsection{Balanced Growth Path}

\paragraph{Autarky} Under autarky, which is a special case in which trade costs are prohibitively high such that countries are isolated as single-country economies, the BGP exists and is unique for each individual economy.  

\begin{proposition}[Growth rates under autarky]\label{prop: BGP-autarky} 
    If $\tau_{sd} \to \infty$ for all $s \neq d$, then growth rates in real consumption $g_s^{\text{autarky}}$ in every country $s \in \boldsymbol{K}$ are proportional to domestic market size:

    \begin{equation*}\label{eq: BGP-autarky}
        g_s^{\text{autarky}} =    \frac{\alpha \cdot \psi }{\eta}  \cdot  \frac{Y_s(t^*)}{M_s(t^*)}   - \rho 
    \end{equation*}

\end{proposition}

\begin{proof}
    Appendix \ref{appendix: bgp}.
\end{proof}

Intuitively, Proposition \eqref{prop: BGP-autarky} characterizes the BGP in a collection of closed \textit{AK} economies with expanding varieties each of them as in the original \textcite{romer_endogenous_1990} model. Growth happens endogenously in each of the countries as households invest in the equity market to fund new intermediate varieties. However, the mass of non-rival goods available for production will be completely different across different countries, since final good producers only have access to domestic intermediate inputs and are therefore less productive than they would be if they were trading internationally. Similarly, in general, BGPs will be characterized by different growth rates. Note that growth rates $g_s^{\text{autarky}}$ are indeed constant because $\frac{Y_s(t^*)}{M_s(t^*)} = \frac{M_s(t^*) \times \mathcal{R}_s(t^*)}{M_s(t^*)}$ is homogeneous of degree zero in $M_s(t^*)$ for each $s \in \boldsymbol{K}$.

\paragraph{Zero gravity} Now move on to characterize the equilibrium growth rates under the polar opposite case: zero gravity. This is one in which trade is costless and even geographical barriers are nonexistent. The term comes from \textcite{eaton_technology_2002}.

\begin{proposition}[BGP under zero gravity]\label{prop: bgp-zero-gravity}
    If $\tau_{sd} = 1$ for all $(s,d)$, then there is a unique world equilibrium growth rate $g^{\text{zero gravity}}$ that satisfies:

    \begin{equation}\label{eq: BGP-zero-gravity}
    g^{\text{zero gravity}} =  \frac{\alpha \cdot \psi }{\eta} \cdot  \frac{\sum_{d \in \boldsymbol{K}} Y_d(t^*)}{\sum_{d \in \boldsymbol{K}} M_d(t^*)}     - \rho 
    \end{equation}

\end{proposition}

\begin{proof}
    Appendix \ref{appendix: bgp}
\end{proof}

By comparing $g^{\text{zero gravity}}$ and $g_{s}^{autarky}$, it is immediately clear that while the latter is proportional to \textit{domestic} value added per variety $Y_{s}(t^*) / M_{s}(t^*) $, the former is proportional to \textit{global} value added per variety $ \sum_{d \in \boldsymbol{K}} Y_d(t^*) / \sum_{d \in \boldsymbol{K}} M_d(t^*) $. Intuitively, under zero gravity, growth happens as if the world were a single integrated Romer economy. 

It is clear that the growth rate must be common under zero gravity because the expression on the right-hand side of \eqref{eq: BGP-zero-gravity} is the same for each country. Since $g^{\text{zero gravity}}$ must be a constant, each element in the sum $\frac{Y_d(t^*)}{\sum_{d \in \boldsymbol{K}} M_d(t^*)} = \frac{M_d(t^*) \times \mathcal{R}_d(t^*)}{\sum_{d \in \boldsymbol{K}} M_d(t^*)}$ must be homogeneous of degree zero in $[M_k(t^*)]_{k \in \boldsymbol{K}}$. This, in turn, implies that returns equalize and $\mathcal{R}_d(t^*) = \mathcal{R}(t^*)$.

In the absence of trade costs, the world economy is fully integrated in terms of final goods varieties suppliers and the law of one price holds in the final good. As the final good serves as an input for intermediate varieties, the price of intermediate varieties equalizes globally. A corollary is that the effective measure of input varieties $\tilde{M}_s(t^*)$ also equalizes globally, indicating that non-rival inputs fully diffuse across the world. 

Note, however, that income levels need not be the same in this world economy. In fact, those countries that have a higher relative wage at the start of the BGP will have a higher wage relative forever. Therefore, under zero gravity, this model features a \textit{stable global distribution of income} as in the model of Armington trade and capital accumulation-driven growth of \textcite{acemoglu_world_2002}.

Using the linearity of income in equilibrium, the growth rate over the BGP be decomposed into the following expression, which relates how growth affects labor and capital income, respectively:

    \begin{equation*}
    \small
    g^{\text{zero gravity}} =  \psi \rho   \left[ T_s^{\frac{1}{\theta(1-\alpha)}}  \cdot \left( \frac{w_s(t^*)^{\theta(1-\alpha)}}{\sum_{k \in \boldsymbol{K}} w_k(t^*)^{\theta(1-\alpha)}} \right)^{-\frac{1}{\theta(1-\alpha)}} \cdot L_s \cdot  \left( \frac{M_s(t^*)}{\sum_{k \in \boldsymbol{K}}M_k(t^*)} \right)^{-1} + \frac{\alpha}{\eta} \frac{\sum_{d \in \boldsymbol{K}} Y_d(t^*)}{\sum_{d \in \boldsymbol{K}} M_d(t^*)}   \right]
    \end{equation*}

The second term within the square brackets, which is related to capital income, shows that profits per variety equalize under zero gravity, with every country having the same level of market access due to the absence of trade frictions and equalization of global prices. The first term within the brackets relates to labor income and shows how wages $w_s(t^*)$ and the measure of varieties $M_s(t^*)$, which are endogenous objects, interact with parameters such as the labor force size $L_s$ and technology $T_s$.

\begin{figure}[htp!]
    \centering
     \includegraphics[width=0.8\linewidth]{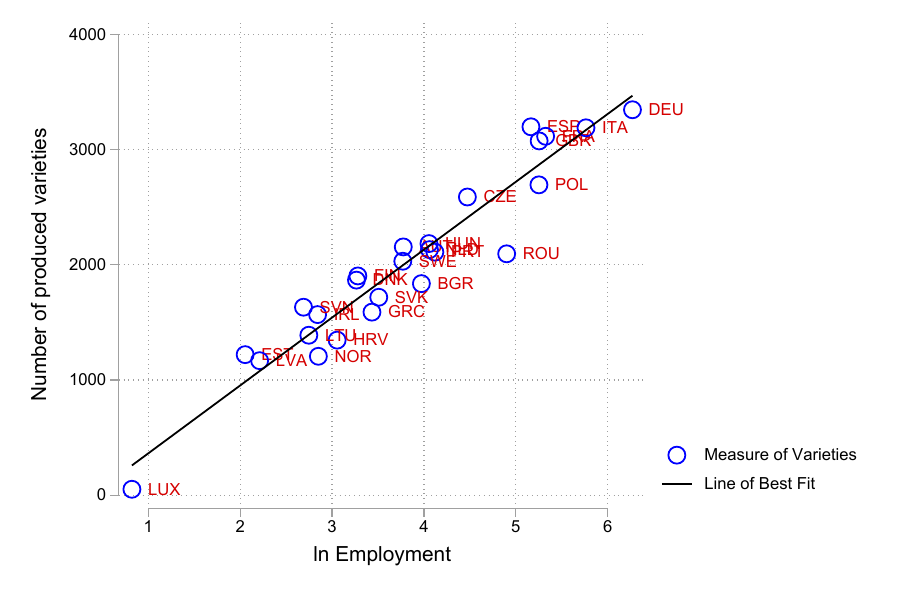} 

    \caption{\textbf{Measures of Variety and Labor Force.} Long Run Averages between 2000-2020 for the cross-country correlation between the Size of the Labor Force ($ln$ of Employment) and the Measure of Produced Varieties. Data come from the Penn World Tables 10.10 and Prodcom, respectively.}
    \label{fig: measure-employment}
\end{figure}

Since growth rates must be equal for all countries, those countries with better technology and higher labor forces will have a proportionately higher real wage and a higher share in the global measure of varieties. The relationship between the size of a country's labor force and the measure of varieties can also be observed in the data, as seen in Figure \ref{fig: measure-employment}.

\paragraph{Costly but finite trade} I now arrive at the more realistic case of a BGP of positive but finite trade costs.  

\begin{proposition}[Balanced growth with costly trade]\label{prop: bgp-general-case} Given a vector of fundamentals $(\theta, \sigma, \boldsymbol{T}, \boldsymbol{\tau})$, if $\tau_{sd} \in (1,  \infty)$ for all $s \neq d$, there \textit{exists a balanced growth path world equilibrium growth rate} satisfying:

    \begin{equation}\label{eq: growth-rate-general}
        g_{s} = \psi \rho \left[  \underbrace{\left(  \frac{ T_{s} }{   \lambda^F_{ss}(t^*) } \right) ^{\frac{1}{\theta(1-\alpha)}}  }_{\text{Eaton-Kortum}} \times \underbrace{\alpha^{1-\eta}\cdot \frac{L_s}{\lambda^M_{ss}(t^*)} }_{\substack{\text{Romer} \\ \text{Domestic}}} + \underbrace{\frac{\alpha}{\eta} \cdot  \sum_{d \in \boldsymbol{K}} \lambda_{sd}^M(t^*) \cdot  \frac{P_d(t^*) Y_d(t^*)}{P_s(t^*) M_s(t^*)}}_{\substack{\text{Romer} \\ \text{Global}}}  \right]
     \end{equation}

and where $g_s = g_{s'} (\forall s, s' \in \boldsymbol{K})$. Furthermore, the growth rate can be decomposed into ``Eaton-Kortum'' and ``Romer'' components.
\end{proposition}

\begin{proof}
    Appendix \ref{appendix: bgp}.
\end{proof}

Like in the previous subsection, equation \eqref{eq: growth-rate-general} shows how growth affects both components of GDP. One can further interpret these components and relate them to the two canonical models that are the building blocks of this framework.

I termed the capital income part, which is the second term within the square brackets, the \textit{Romer Global} component. The reason is that this component is increasing in each country's market access \textemdash i.e., proportional to each country's share of the global intermediate goods market $(\lambda_{sd}^M)$. Intuitively, profits will be related to a firm's sales and to markets all over the world and, therefore, to its market share in each of those destination markets. This component is also decreasing in the price of the final good $(P_d)$ in the source country since due to the input-output structure embedded in the lab-equipment version of the Romer model the price of the final good is the marginal cost of R\&D investment.

Real labor income can be partitioned into two components: an \textit{Eaton-Kortum component} and a \textit{Romer Domestic} component. The \textit{Eaton-Kortum component} is very much Ricardian: real labor income improves with technological improvements ($T_s$) and as the domestic trade share in final ($\lambda^F_{ss}$)  decreases, consistent with the Ricardian intuition that specialization leads to gains from trade.

The \textit{Romer Domestic} component incorporates domestic market size effects, by integrating the size of the domestic labor force $L_s$. But it also adjusts for the diffusion of differentiated intermediate goods, which is embedded in the summary statistic of domestic intermediate trade share  ($\lambda^M_{ss}$)\footnote{To see that, note that: $\lambda^M_{ss}(t^*) = M_s(t^*) \left( \frac{ p_{ss}^M (t^*) }{ P_{s}^M (t^*) } \right)^{1-\eta} =  \frac{ M_s(t^*) }{ \tilde{M}_{s} (t^*) } p_{ss}^M (t^*)^{1-\eta} $.}.

The different components of \eqref{eq: growth-rate-general} make it clear why a BGP requires common growth rates. Both $\lambda_{ss}^M(t^*)$ and $\lambda_{ss}^F(t^*)$ are homogeneous of degree zero in $[M_k(t^*)]_{k \in \boldsymbol{K}}$ and prices $P_s(t^*)$, $P_d(t^*)$ must be constant along a BGP. For a BGP, $g_{s}$ must be a constant, and therefore $\frac{Y_d(t^*)}{M_s(t^*)}=\frac{M_d(t^*) \times \mathcal{R}_d(t)}{M_s(t^*)}$ must be homogeneous of degree zero in $[M_k(t^*)]_{k \in \boldsymbol{K}}$.  This can only happen if $\mathcal{R}_d(t^*)$ is homogeneous of degree zero in each $d$, which implies that $\mathcal{R}_d(t^*) = \mathcal{R}(t^*)$.

Intuitively, along the BGP, the real interest rate equalizes globally\footnote{One way to see that is through the Euler equation. Since the Euler Equation governs the growth rate in consumption and over the BGP $g_c = g_M$ for all countries, a corollary is that the real interest rate must equalize globally.}. Even though there are no international equity markets, the fact that households can invest in new varieties through equity markets and earn expected profits that are linked to exports means that trade acts as a vehicle to integrate international R\&D and equity markets. In a balanced growth equilibrium, then, prices and the endogenous distribution of the measure of varieties $[M_s(t^*)]_{s \in \boldsymbol{K}}$ will adjust to make sure that returns and, therefore, growth rates equalize.

\paragraph{How are growth rates related to market access?} After characterizing the existence of the BGP, one can turn to the discussion of what happens to the equilibrium growth rates after there is a change in trade costs. Here, these trade costs are directly related to market access, since the mechanism that propels growth is the incentive to have equity claims in the profits of variety exporters. The growth rate is a general equilibrium object that depends on the whole distribution of prices across countries and periods. Therefore, characterizing changes to it is not a trivial task.

Nonetheless, in order to connect the theory to the empirical and quantitative analysis, to gain some intuition, first consider what happens to the \textit{long-run} equilibrium growth rate after a permanent change in trade costs in a world of symmetric countries. While that is an important restriction, it allows for a closed-formed intuitive solution: in that case, $g^*$ can be shown to unambiguously increase in the long run after an episode of trade liberalization.

\begin{proposition}[Effects of changes in trade costs over the long run in symmetric economies]\label{prop: comparative-statics-long-run} Suppose there exist a collection of symmetric economies that grow over the BGP with costly trade with trade costs $\tau > 1$. Then $\frac{\partial g^*}{\partial \tau} < 0$.
\end{proposition}

\begin{proof}
    Appendix \ref{appendix: bgp}
\end{proof}

In this model, the long-run growth rate will change after a permanent change in trade costs if there is a change in the effective market size, represented by how much of the global market exporters can tap into \textemdash that is why foreign aggregate demand $\sum_{k} P_k(t^*) Y_k(t^*)$ is modulated by intermediate trade share $\sum_{k} \lambda^M_{sd}(t^*)$ in the profit formula.

\begin{figure}[htp!]
    \centering
     \includegraphics[width=0.8\linewidth]{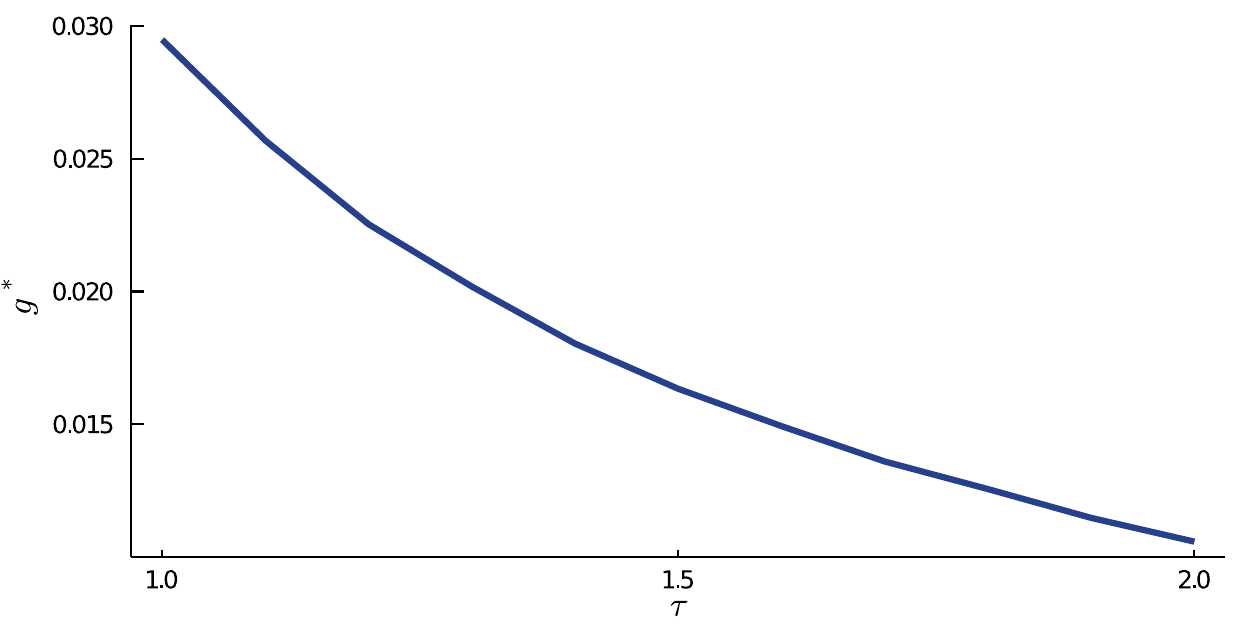} 

    \caption{\textbf{Long-run growth in two asymmetric economies as a function of changing trade costs.} Results from a numerical simulation of the equilibrium growth rate $g^*$ of two asymmetric economies that differ in their populations but are otherwise equal. Parameters are the following:  $L = [1,1.03]$, $\sigma = 0.77, \theta = 2.12, \alpha = 1/3, \rho = 0.03, T = [1,1], \psi = 2.46$.}
    \label{fig: growth-assymetric-two-countries}
\end{figure}

In a symmetric world, it can be shown in closed form that real profits increase when trade costs go down. The intuition translates to numerical exercises with asymmetric countries. As an example plot in Figure \ref{fig: growth-assymetric-two-countries} shows how these results look like in a numerical exercise with two asymmetric countries. 

\newpage

\subsection{Welfare}

   With log preferences, at any moment, consumption over the BGP is a fraction of assets plus real labor income. Since such consumption flow grows at a constant rate $g^*$ and the measure of products is simply a linear transformation of assets, as shown in Appendix \ref{appendix: welfare}, welfare along the BGP can be decomposed between a product measure component, a real income component, and a growth component.  

    \begin{equation}\label{eq: welfare}
        { \footnotesize \int_{t^*}^\infty \exp \{-\rho (t-t^*)\}   \log \left(\exp \{g^* t\} C_s(t^*) \right)  dt = \underbrace{\underbrace{\log \left( \frac{1}{\psi} M_s(t^*) \right)}_{\text{product measure}}}_{\text{transitional}} + \underbrace{\underbrace{ \frac{1}{\rho} \log \left( \frac{w_s(t^*)L_s}{P_s(t^*)} \right)}_{\text{real income}} }_{\text{static}} + \underbrace{ \underbrace{\frac{g^*}{\rho^2}}_{\text{growth}}}_{\text{dynamic} }}
    \end{equation}

     Consider what happens to welfare after a change in trade costs from $\boldsymbol{\tau}$ to $\boldsymbol{\tau} + d\boldsymbol{\tau}$, as in \textcite{arkolakis_new_2012} \textemdash hereinafter ACR. In this dynamic setting, to make a comparison to the static framework, I need to compare what happens across the two BGPs, comparing the preserved value of discounted lifetime utility across the beginning of the two initial equilibria. For that, let me introduce some notation: suppose $t^*$ is the initial period of the original BGP; $t^{**}$ is the first period of the final BGP and let $\hat{x} \equiv x(t^{**})/x(t^*)$.
     
     Then, relative level changes in the first component of welfare across two BGPs can be expressed as $\log \left( \widehat{M}_s \right)$. Changes in the equilibrium product measure will depend on whether the measure of varieties in country $s$ expands or contracts, \textit{relative to the distribution of varieties across countries}, across BGPs. There is no general prediction in the model regarding the direction of this effect.
    
    Countries that have started with a measure of varieties above optimal (relative to other countries) will see a shift in exports (and therefore R\&D expenditures) towards other countries and will see their measures of varieties shrink. The opposite is true for countries that started with a measure of varieties below optimal.
    
    Importantly, however, this first component will not compound over time, as highlighted by the fact is not multiplied by the factor $\rho^{-1}$ or $\rho^{-2}$. This means that it will only change the (relative) income level that a given country arrives with at the BGP and it will have no impact going forward. Therefore, this is a \textit{transitional effect} of welfare. For most reasonable calibrations of $\rho$, the transitional effect will have a very small weight on total welfare changes.

    The second component will be familiar to most trade economists. It looks like the traditional \textit{static welfare formula in ACR}. In the same spirit as ACR, one can also write the static welfare component in changes: 

    \begin{equation}\label{eq: static-welfare}
        \frac{1}{\rho} \log \left( \widehat{\frac{w_s }{P_s }} \right) =  \frac{1}{\rho}  \log \underbrace{\left( \widehat{\lambda_{dd}^F} ^{-\frac{1}{(1-\alpha)\theta}} \right)}_{\text{Eaton-Kortum}} + \frac{1}{\rho\eta} \log \underbrace{\left(\sum_{k \in \boldsymbol{K}}  \mu_{k}   \cdot \widehat{M}_{k} \cdot  \left( \frac{\widehat{p}^M_{kd} }{ \widehat{P}_{s} } \right)^{1-\eta}  \right)}_{\text{Romer}} 
    \end{equation}
    
    \noindent where $\mu_{k} \equiv \frac{ M_{k}(t^*) \cdot  \left( \frac{ p^M_{ks}(t^*) }{P_{s}(t^*) } \right)^{1-\eta} }{\sum_{k \in \boldsymbol{K}} M_{k}(t^*) \cdot  \left( \frac{ p^M_{ks}(t^*) }{P_{s}(t^*) } \right)^{1-\eta}}$.

    This component preserves the standard feature that changes in consumer welfare are decreasing in changes in domestic trade share $\widehat{\lambda_{ss}^F}$\footnote{The elasticity of this effect is $-\frac{1}{(1-\alpha)\theta}$ rather than the standard $-\frac{1}{\theta}$ due to the input-output structure of the model.}. This captures the Ricardian intuition of the model: at the margin, there are static gains from specialization in this model.

    Like the growth formula, the static component of welfare also has Eaton-Kortum and Romer components. Here, the Romer component impacts welfare by augmenting Ricardian gains through an extensive margin. It is represented by the weighted change in the measure of varieties, accounting for previous weights $\mu_{k}$, changes in the measures of varieties in each country $k \in \boldsymbol{K}$ across equilibria $\widehat{M}_{k}$, and changes in the prices of foreign intermediate goods relative to the domestic consumer price index $ \left( \frac{ p^M_{ks}(t^*) }{P_{s}(t^*) } \right)$ at the domestic market.

    Note that this welfare impact from product innovation resembles how the change in the measure of varieties shows up in the ACR formula in Melitz-type models. This highlights that the nested structure of production featured in this model effectively adds an extensive margin to the Eaton-Kortum framework.

    While they do not compound over time, both of these effects have an impact in every period over the BGP as it is made clear by it being multiplied by the factor $\rho^{-1}$. For that reason, these can be understood as a \textit{level effect} or \textit{static effect} of welfare.

    The third and last component is the common growth rate $g^*$. Importantly, since it compounds the BGP level of consumption, it is multiplied by a factor $\rho^{-2}$ rather than $\rho^{-1}$ and it will in general have a larger weight on welfare. This is a metric of \textit{dynamic gains from trade}, which is a \textit{growth effect} of welfare.
    
    Changes in the growth component of welfare will be defined as the change in the growth rage: $\frac{g^{**} - g^{*}}{\rho^2}.$ Since growth rates equalize along the BGP, changes in the growth component of welfare will also be shared across all countries. However, since the other components will differ, the share of the dynamic component of welfare in total gains from trade will therefore be different across countries. 
    
    The discount rate $\rho$ will have an important role in attributing weights across the dynamic, static components, and transitional components of welfare. Intuitively, the lower the $\rho$, the more patient the agent is, and the more relevant the dynamic component of welfare will become. 

    \paragraph{Welfare as buyers and as sellers} Comparing the dynamic and static components of welfare yields important insights regarding the economic mechanisms behind this model. In fact, the forces of specialization and innovation are reflected in these two components. 
    
    To see that, note that, as made clear by \eqref{eq: static-welfare}, since $1-\eta < 1$ country $s$'s \textit{static welfare} is decreasing in the price of foreign intermediate goods. The static welfare formula captures the effect of $s$ as final producers and consumers. As $s$ purchases more foreign intermediate varieties for a cheaper price, it becomes more productive by increasing its effective measure of varieties $\tilde{M}_s(t)$. The other side of the coin is that it decreases the local price index $P_s(t)$, which directly benefits consumers and increases welfare.
    
    By contrast, the growth component of welfare is \textit{increasing} in the price of foreign intermediate goods coming from $k$, in each destination markets $d$, relative to the price of intermediate goods from the source country $s$ at those same destination market:
    
    \begin{eqnarray*}
        g_s \propto \frac{\alpha}{\eta}  \cdot  \sum_{d \in \boldsymbol{K}}  P_d(t)Y_d(t)    \left[ \sum_{k 
        \in \boldsymbol{K}}  M_{k}(t)  \left(\frac{p_{kd}^M(t) }{p_{sd}^M(t)} \right)^{1-\eta} \right]^{-1}          
    \end{eqnarray*}

     The intuition for this contrast is quite straightforward and underscores the different underlying economic mechanisms of the model. The Romerian part of growth captures the effect of $d$ as forward-looking investors in the R\&D market and intermediate good producers. Since the intermediate goods are substitutes, all else equal, demand for intermediate goods from $s$ and maximized profits are higher when the price of intermediate goods of foreign competitors from third-party countries $k$ relative to domestic producers from $s$ at each destination market $d$ is higher.
    
    Intuitively, the \textit{growth effect} captures that, from a \textit{seller's perspective}, the domestically produced and exported intermediate variety $s$ is more attractive and competitive when foreign varieties $k$ are more expensive. Conversely, the \textit{static effect} captures that, from a \textit{buyer's perspective}, when foreign varieties $k$ relative to one's domestic purchasing power at $d$ are more expensive, the domestic consumer is worse off. Both channels are economically sensible and the model captures both mechanisms.

    Along the BGP, prices and measures of varieties will adjust to make sure that growth rates equalize such that $g_s = g_{s'}$ for every $s, s' \in \boldsymbol{K}$. While the economic mechanisms are still operating under the hood and take over if there is any shock that drives the system off the BGP, these different effects will wash out once differences in prices, measures of varieties and wages endogenously adjust towards a BGP.

\section{Quantification and Policy Exercise} \label{section: quatification}

This section describes a numerical quantification of the model, which solves for three endogenous objects along the Balanced Growth Path: (a) the distribution of wages; (b) the distribution of Measures of Varieties; and (c) the common equilibrium growth rate. I calibrate the model to EU-15 countries and the New Member States (NMS) that joined in the 2004 expansion.

To simplify the exercise, I group these countries into six sets: corresponding to the six waves of the expansion of the European Union of to 2004\footnote{This simplification is just a matter of computational tractability. As described in Appendix \ref{appendix: computational}, each guess of my solution algorithm solves for a static version of an Eaton-Kortum model with input-output linkages, which itself has multiple steps for solutions. So the problem grows quite fast in complexity in the number of countries. Improving the solution algorithms for this new class of dynamic models is a fruitful avenue of future research.}. The country groups are asymmetric both in terms of labor force and productivity. The groups are: 

\begin{enumerate}
    \item \textbf{1957}: Belgium, France, Germany, Italy, Luxembourg, and the Netherlands \textemdash the original members;
    \item \textbf{1973}: Denmark, Ireland, and the United Kingdom;
    \item \textbf{1981}: Greece;
    \item \textbf{1986}: Portugal and Spain;
    \item \textbf{1995}: Austria, Finland, and Sweden;
    \item \textbf{2004}: Czech Republic, Estonia, Hungary, Latvia, Lithuania, Poland, Slovakia, and Slovia \textemdash the New Member States (NMS).
\end{enumerate}

The solution method is straightforward. I calibrate the model to a baseline scenario and then change iceberg trade costs to induce a trade liberalization shock. By comparing the endogenous equilibria along the Balanced Growth Path of these two scenarios, which include distributions of the measures of varieties $[M_s(t^*)]_{s \in \boldsymbol{K}}$ and wages $[w_s(t^*)]_{s \in \boldsymbol{K}}$ as well as a common equilibrium growth rate $g^*$, I can infer the welfare consequences of a change in this parameter along the BGP.

\paragraph{Model Calibration} My estimates of the short-term ($\sigma$) and long-term ($\theta$) elasticities of trade come from \textcite{boehm_long_2020}, which are $\sigma=0.76$ and $\theta=2.12$, respectively. The results are not very sensitive to using a $\theta=4.0$. The vector of labor force $\{L_s\}$ comes from Penn World Tables. The share of intermediate goods $\alpha = 0.36$ is set to equal the average share of intermediate goods in the sample of countries between 2000-2003 from the World Input-Output Database.

I use observed trade flows to infer trade costs. The strategy goes back to \textcite{head_increasing_2001}. According to the handbook chapter by \textcite{head_gravity_2014}, the index is called the Head-Ries Index (HRI) since 2011 (when the working paper version of \textcite{eaton_trade_2016} was published). As shown in Appendix \ref{appendix: computational}, I can write trade costs as:

\begin{equation}\label{eq: head-ries}
    \tau_{sd} = \left(  \frac{E^F_{sd}(t^*)}{E^F_{dd}(t^*)} \cdot \frac{E^F_{ds}(t^*)}{E^F_{ss}(t^*)}  \right)^{- \frac{1}{2 \theta (1-\alpha)}}
\end{equation}

\noindent where each flow $E^F_{sd}(t^*)$ defined to be an average between $2000-2003$. The data on bilateral expenditure values $E^F_{sd}(t^*)$ comes from the World Input-Output Database.

\begin{figure}[htp!]
    \centering
    \includegraphics[width=0.9\textwidth]{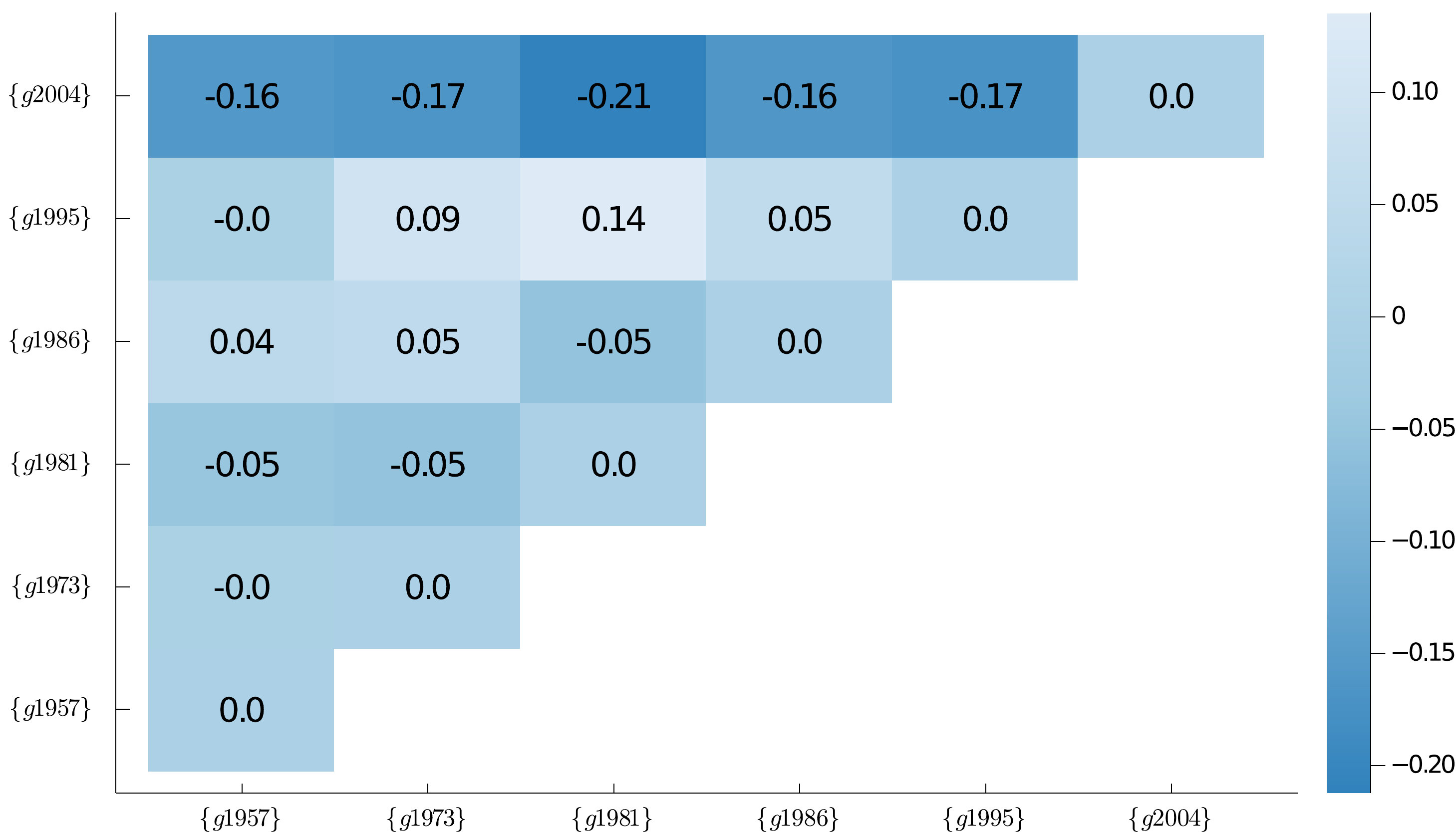}
    \caption{\textbf{Changes in Trade Costs Before and After 2004 EU Enlargement (in percentage terms).} \footnotesize{This matrix shows the bilateral changes in trade costs, calculated using the method inferred from equation \eqref{eq: head-ries}, before and after the 2004 EU Enlargement. The before period is an average for the years 2000-2003 and the after period is an average for the years 2004-2007. Underlying data comes from the World Input-Output Database.}}
    \label{fig:matrix-trade-costs}
\end{figure}

Figure \ref{fig:matrix-trade-costs} plots the change in trade costs before and after the 2004 enlargement of the European Union, calculated from an average of for the years 2000-2003 for the immediate "before" period and an average for the years 2004-2007 for the immediate "after" period. Calculations confirm that there were large reductions ($- 15-20\%$) in trade costs between NMS and the Western European countries during this period, which is consistent with bilateral tariff data between the NMS and Western Europe from TRAINS and the WTO\footnote{See the Appendices from \textcite{caliendo_goods_2021} for a detailed description of the data.}. Changes in trade costs across the other groups have been comparatively small except for one calculated \textit{increase} in trade costs between Greece (\textit{g1981}) and Austria, Finland, and Sweden (\textit{g1995}).

This is important because these changes in trade costs will act as the main shock across calibrations of BGPs in my numerical exercise.  It is relevant that the key driver of changes across equilibria is the enlargement of the EU.

The location parameter of the Fr\'echet distribution $\{T_s\}$ and $\psi$ are free parameters that I vary to match the distribution of wages and the average growth rate of the EU-15 countries in the $1989-2003$ period \textemdash i.e., fifteen years prior to the 2004 expansion of the European Union. The rationale is that I am calibrating this model to BGP growth rate and the EU-15 countries were very likely closer to the BGP than the transition economies of Eastern Europe, so it is reasonable to match the model to their growth rate.

\paragraph{Model Validation} To validate the model, there are some untargeted moments one can look at. First, compare the relative change in real wages across the two BGPs. The predicted changes in the distribution of wages across equilibria in the model can be compared with the relative income growth of each country group around the EU enlargement.  

Since the wages distribution is only pinned down up up to the distribution, if real wages of a given group take a larger share of the distribution in the later BGP, an implication is that it must have grown faster than average between those periods. To compare the data with the model, a natural comparison is to use GDP per capita growth rate net of the average of the EU, which yields a income that is normalized for the periods of 1998-2003 and 2005-2010. The way to interpret the data is to see whether or not each group's income per capita grew faster (slower) than average across these periods.

\begin{figure}[htp]
    \centering
    \includegraphics[width=\textwidth]{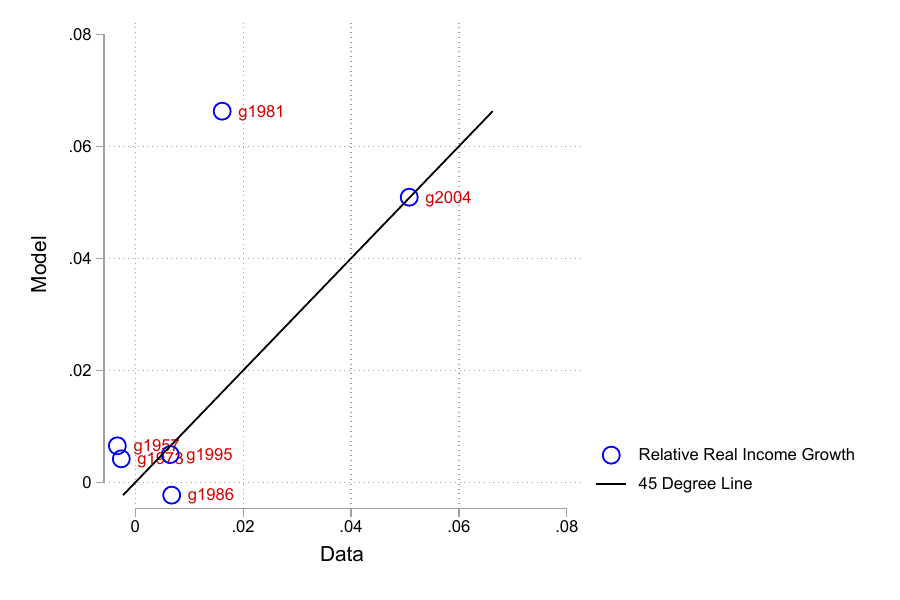}
    \caption{\textbf{Model Validation: Changes in Real Wages, Relative to the Average.} \footnotesize{In the model, the distribution of wages $\lambda_w \cdot [w_s(t^*)]_{s \in \boldsymbol{K}}$ is normalized with a choice of $\lambda_w$ such that $\sum_{s \in \boldsymbol{K}} w_s(t^*) L_s = 1$. What is shown in the chart is the percentage change across equilibria $\frac{w_s(t^{**})/P_s(t^{**})}{w_s(t^{*})/P_s(t^{*})} -1$, where $L_s$ is assumed to be fixed. In the data, for consistency, I calculated annual GDP per capita then subtracted it from the average of the group for the periods of 1998-2003 and 2005-2010. I then calculated changes and plotted the data. Data comes from the Penn World Tables 10.01.}}
    \label{fig:change-real-wages}
\end{figure}

Here, one can see that the model in fact matches the data quite correctly. It predicts relative a catch-up of the New Member States (\textit{g2004}). The model predicts that real wages in NMS would grow about $5.1\%$ faster than the average of the Western European countries, which is very close to observed in the data ($5\%$). As seen in Figure \ref{fig:change-real-wages}, most of the other country groups also fall very close to the 45-degree line, suggesting that the model's predictions are reasonable.

One exception is the 1981 wave, for which the model substantially over-predicts relative real income growth. The reason being that such a wave consists of a single country: Greece. And the aftermath of the EU enlargement $2005-2010$ includes the first years of the Greek debt crisis. Naturally, the model cannot anticipate the negative shocks of the deep recession of the late 2010s in Greece.

Second, compare the (endogenous) distribution of the number of produced varieties in the model to the distribution of the number of produced varieties in the data for the 2000-2003 period. Once again, the observations fall mostly along the 45-degree line, suggesting that the model does a good job in replicating the empirical distribution. As one exception, now the model \textit{underestimates} the actual share of total produced varieties in the NMS (\textit{g2004}).

\begin{figure}[htp!]
    \centering
    \includegraphics[width=\textwidth]{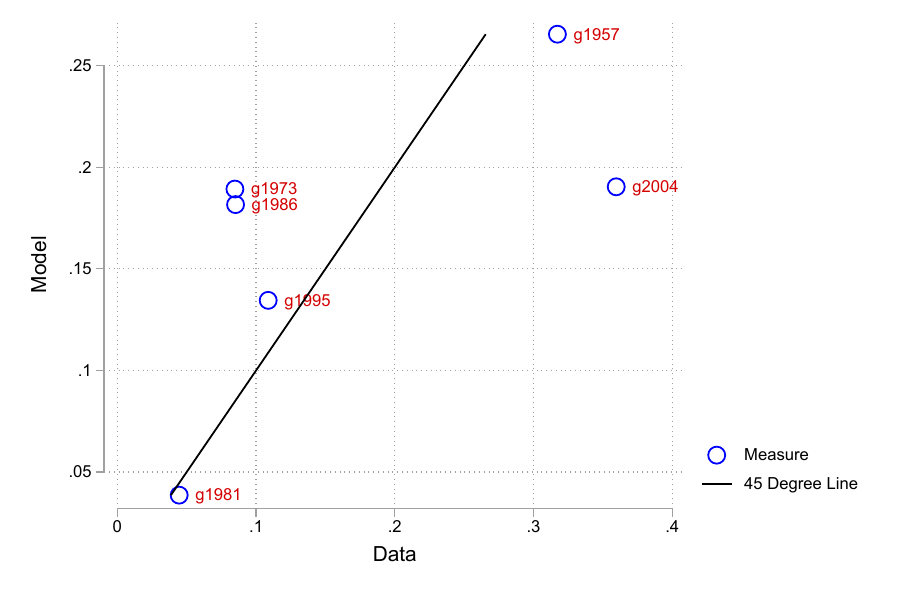}
    \caption{\textbf{Model Validation: Distribution in the Number of Produced Varieties Across Regions.} \footnotesize{In the model, the distribution of measures of varieties $\lambda_M \cdot [M_s(t^*)]_{s \in \boldsymbol{K}}$ is normalized with a choice of $\lambda_M$ such that $\sum_{s \in \boldsymbol{K}} M_s(t^*) = 1$. For consistency in the comparison, what I show in the data bars are the relative shares of each country group in the total universe of the product measure, or: $M_s(t) / \sum_{s' \in \boldsymbol{K}}M_{s'}(t) $}. This assumes, as in the model, that product varieties in the data are differentiated across countries, so the global product space is $\sum_{s' \in \boldsymbol{K}}M_{s'}(t)$. Data comes from Prodcom (Eurostat) and are averages for the 2000-2003 period.}
    \label{fig:measures-model}
\end{figure}

\begin{figure}[htp!]
    \centering
    \includegraphics[width=\textwidth]{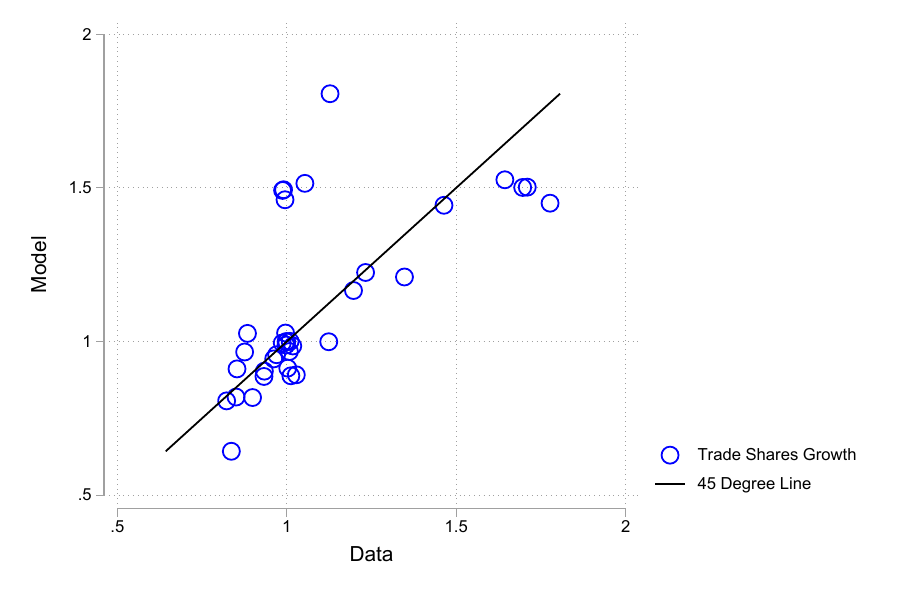}
    \caption{\textbf{Model Validation: Changes in Trade Shares.} \footnotesize{The model object is plotted is $\hat{\lambda}^F_{sd}$: the change in the final sector trade share. In the data, this is total trade shares renormalized to account for the fact that there is no rest-of-the-world in the sample. The before and after periods are 1998-2003 and 2005-2010, respectively.}}
    \label{fig:lambdas-model}
\end{figure}

Finally, compare the changes in trade shares across equilibria. The model captures changes in trade shares really well, as shown in \eqref{fig:lambdas-model}. Trade expands particularly in exports from NMS to Western European countries, which is captured by the upper quadrant observations that lie close to the 45-degree line. Here, the exceptions are the trade flows from Western Europe towards the NMS. The model predicts a symmetrical response in terms of trade expansion, while in reality, the gains were much more a relative market access from the NMS into the EU market than the other way around.

\paragraph{Results} The main result of this exercise relates to the theoretical welfare decomposition in equation \eqref{eq: welfare}. One can compare two paths of consumption along the BGP and decompose them into:

        \begin{eqnarray*}
         & & \int_{\tau}^\infty \exp \{-\rho (t-\tau)\} \left[  \log \left(\exp \{g^{**} t\} C_s(t^{**},\tau) \right) - \log \left(\exp \{g^* t\} C_s(t^{*},\tau) \right) \right]  dt = \\
         & & \underbrace{\log \left(\hat{M}_s \right)}_{\text{transitional}} + \underbrace{ \frac{1}{\rho} \log \left( \widehat{\frac{w_s}{P_s}} \right)}_{\text{static}} + \underbrace{ \frac{g^{**} - g^*}{\rho^2}}_{\text{dynamic} }
    \end{eqnarray*}

where $C_s(t^{*},\tau), C_s(t^{**},\tau)$ are the paths of consumption along the original and new BGPs, respectively.

For all countries, the transitional component is negligible. They never contribute with more than $0.03\%$ of total absolute value of welfare, in the largest case.

Static gains from trade can be as large as 5-6\% of domestic income in the case of NMS (\textit{g2004}) or Greece (\textit{g1981}) or even \textit{negative} or close to zero in the case of the Western European countries such as Portugal and Spain (\textit{g1986}). In the case of Greece and the NMS, they account for 38\% and 32\% of total welfare gains from trade, respectively.

\begin{figure}[htp!]
    \centering
    \includegraphics[width=0.8\textwidth]{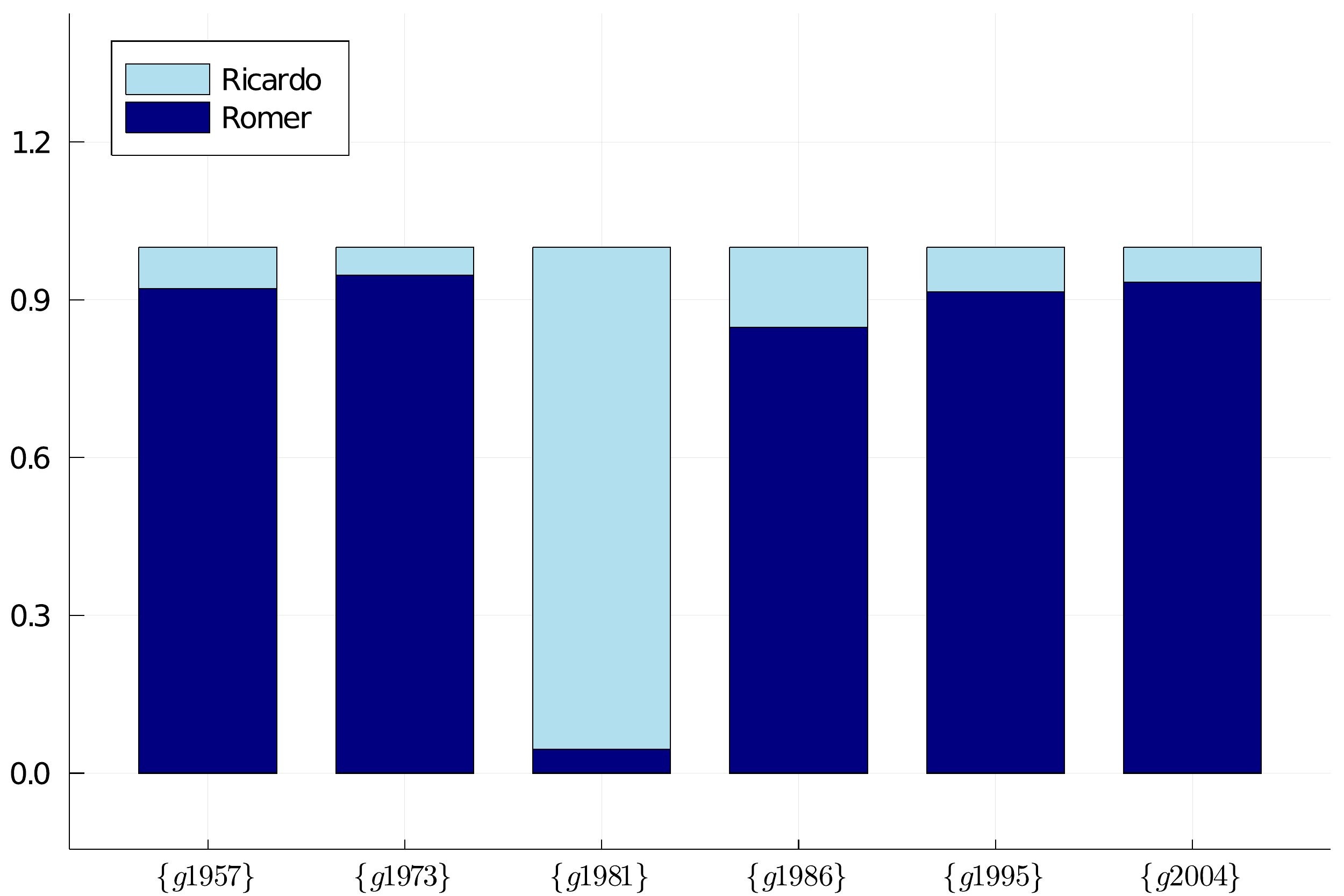}
    \caption{\textbf{Static Welfare Decomposition.} \footnotesize{Static Welfare Decomposition Across its Eaton-Kortum and Romer Components, according to equation \eqref{eq: static-welfare}.}}
    \label{fig:decomposition-static-welfare}
\end{figure}

These changes can be further decomposed into ``Eaton-Kortum'' and ''Romer'' parts of static welfare using equation \eqref{eq: static-welfare}. Results in Figure \ref{fig:decomposition-static-welfare} show a wide variation of the Ricardian component share in total changes in static welfare. While for most country groups that share is about 10\% of total changes in static welfare, for Portugal and Spain (\textit{g1986}) it accounts for nearly 25\% of static welfare changes while for Greece (\textit{g1981}) the Ricardian share accounts for more than 90\% of changes in total welfare.

Finally, the main numerical outcome of the exercise is the differences in growth rates across BGPs $g^{**} - g^*$. In the current calibration, \textit{the trade liberalization embedded in the 2004 enlargement of the European Union induced the EU long-run yearly growth rate to increase $0.10pp$}. One implication is that the dynamic part of welfare accounts for the most of gains from trade for all countries. Therefore, not accounting for this channel ignores the majority of gains from trade.

However, the share of total welfare gains it accounts for varies across country groups. According to this model, in the current parametrization, the share of dynamic gains in total welfare gains is between $65\%$ and $90\%$. This is in line with estimates from \textcite{hsu_innovation_2019} ($78\%$) and \textcite{perla_equilibrium_2015} ($85\%$).

\begin{figure}[htp!]
    \centering
    \includegraphics[width=0.8\textwidth]{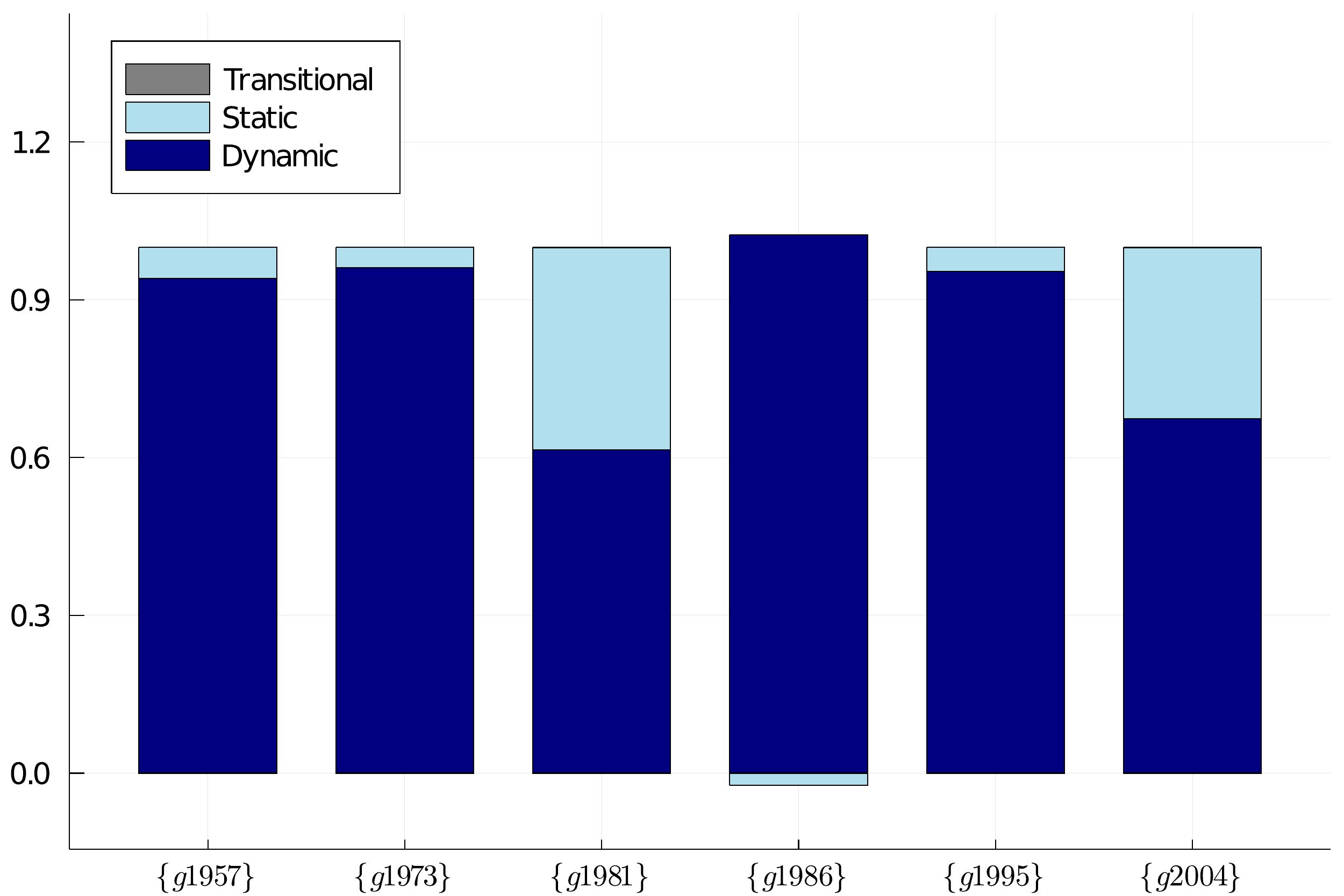}
    \caption{\textbf{Total Welfare Decomposition.} \footnotesize{Welfare Decomposition Across its Transitional Static and Dynamic Components, according to equation \eqref{eq: welfare}.}}
    \label{fig:decomposition-welfare}
\end{figure}

However, in this model, the change in the equilibrium measure of varieties (and hence the real wage) between one BGP and the other can actually decrease, which implies a negative static welfare share. Therefore, for some countries, such as Portugal and Spain (\textit{g1986}) the share of dynamic welfare in total welfare is \textit{larger than 100\%}. These decompositions are in Figure \ref{fig:decomposition-welfare}.

In monetary terms, a back-of-the-envelope calculation suggests an additional $0.10\%$ yearly growth rate to the aggregate GDP of the Western European plus the New Member States since the year of accession \textemdash that is, between 2004 and 2023 \textemdash would have induced an additional current production level of approximately $\$332$ billion in the continent, which accounts for $2.0\%$ of the total level of production of the European Union.

\section{Conclusion}

I focus on the long-lasting question of the relationship between trade and growth and, in particular, trade and product innovation. I make several contributions: theoretical, empirical, and quantitative.

On the theoretical front, my main contribution is a new framework that reconciles the forces of specialization and market size, rationalizes foreign market access as a rationale for growth in a dynamic framework, and provides an analytical formula for dynamic gains from trade. In all of those points, I maintain active dialogues with the literature, such as nesting the Eaton-Kortum model of trade and Romer growth model as special cases of my model and subsuming the ACR static welfare formula in my dynamic welfare formula.

In my empirical work, I rely on the eastward expansion of the European Union and document several new facts that are consistent with the mechanisms of my model. Compared to countries that selected into becoming candidates but had not joined at given horizon, countries started producing more product varieties, investing more in R\&D, and trading more.

I go beyond these facts and exploit plausibly exogenous variation to show that a plausibly exogenous increase in market access leads to a higher probability of initiating production and exporting a given product, which is consistent with the main mechanism of the theoretical model.

Finally, I solve for a quantitative model and replicate the 2004 expansion of the European Union in the computer. The results of the simulation imply that: (a) the EU expansion increased its long-run yearly growth rate by about 0.10pp; and (b) dynamic gains from trade account for somewhere between 65-90\% of total welfare gains from trade.

This paper points to the fact that dynamic gains from trade are likely too large to be ignored. The big generalizable takeaway is that the previous literature has largely underestimated gains from trade, perhaps by as much as one order of magnitude. Advancing on this agenda, perhaps by understanding the transition dynamics, is a fruitful avenue of future research.

\newpage

\printbibliography

@article{cai_knowledge_2022,
	title = {Knowledge {Diffusion}, {Trade}, and {Innovation} across {Countries} and {Sectors}},
	volume = {14},
	issn = {1945-7707},
	url = {https://www.aeaweb.org/articles?id=10.1257/mac.20200084},
	doi = {10.1257/mac.20200084},
	language = {en},
	number = {1},
	urldate = {2023-11-28},
	journal = {American Economic Journal: Macroeconomics},
	author = {Cai, Jie and Li, Nan and Santacreu, Ana Maria},
	month = jan,
	year = {2022},
	pages = {104--145},
}

@misc{helpman_foreign_2023,
	type = {Working {Paper}},
	series = {Working {Paper} {Series}},
	title = {Foreign {Competition} and {Innovation}},
	url = {https://www.nber.org/papers/w31840},
	doi = {10.3386/w31840},
	urldate = {2023-11-08},
	publisher = {National Bureau of Economic Research},
	author = {Helpman, Elhanan},
	month = nov,
	year = {2023},
	doi = {10.3386/w31840},
}

@techreport{hsu_innovation_2019,
	address = {Cambridge, MA},
	title = {Innovation, {Growth}, and {Dynamic} {Gains} from {Trade}},
	url = {http://www.nber.org/papers/w26470.pdf},
	language = {en},
	number = {w26470},
	urldate = {2023-10-18},
	institution = {National Bureau of Economic Research},
	author = {Hsu, Wen-Tai and Riezman, Raymond and Wang, Ping},
	month = nov,
	year = {2019},
	doi = {10.3386/w26470},
	pages = {w26470},
}

@incollection{head_gravity_2014,
	title = {Gravity {Equations}: {Workhorse},{Toolkit}, and {Cookbook}},
	volume = {4},
	isbn = {978-0-444-54314-1},
	shorttitle = {Gravity {Equations}},
	url = {https://linkinghub.elsevier.com/retrieve/pii/B9780444543141000033},
	language = {en},
	urldate = {2023-09-26},
	booktitle = {Handbook of {International} {Economics}},
	publisher = {Elsevier},
	author = {Head, Keith and Mayer, Thierry},
	year = {2014},
	doi = {10.1016/B978-0-444-54314-1.00003-3},
	pages = {131--195},
}

@article{head_increasing_2001,
	title = {Increasing {Returns} versus {National} {Product} {Differentiation} as an {Explanation} for the {Pattern} of {U}.{S}.-{Canada} {Trade}},
	volume = {91},
	issn = {0002-8282},
	url = {https://www.jstor.org/stable/2677816},
	number = {4},
	urldate = {2023-09-26},
	journal = {The American Economic Review},
	author = {Head, Keith and Ries, John},
	year = {2001},
	note = {Publisher: American Economic Association},
	pages = {858--876},
}

@article{eaton_trade_2016,
	title = {Trade and the {Global} {Recession}},
	volume = {106},
	issn = {0002-8282},
	url = {https://www.aeaweb.org/articles?id=10.1257/aer.20101557},
	doi = {10.1257/aer.20101557},
	language = {en},
	number = {11},
	urldate = {2023-09-26},
	journal = {American Economic Review},
	author = {Eaton, Jonathan and Kortum, Samuel and Neiman, Brent and Romalis, John},
	month = nov,
	year = {2016},
	pages = {3401--3438},
}

@book{acemoglu_introduction_2008,
	title = {Introduction to {Modern} {Economic} {Growth}},
	isbn = {978-1-4008-3577-5},
	language = {en},
	publisher = {Princeton University Press},
	author = {Acemoglu, Daron},
	month = dec,
	year = {2008},
	note = {Google-Books-ID: DsPH5fWNdrsC},
}

@article{baier_free_2007,
	title = {Do free trade agreements actually increase members' international trade?},
	volume = {71},
	issn = {0022-1996},
	url = {https://www.sciencedirect.com/science/article/pii/S0022199606000596},
	doi = {10.1016/j.jinteco.2006.02.005},
	number = {1},
	urldate = {2023-09-19},
	journal = {Journal of International Economics},
	author = {Baier, Scott L. and Bergstrand, Jeffrey H.},
	month = mar,
	year = {2007},
	pages = {72--95},
}

@article{bernard_margins_2009,
	title = {The {Margins} of {US} {Trade}},
	volume = {99},
	issn = {0002-8282},
	url = {https://pubs.aeaweb.org/doi/10.1257/aer.99.2.487},
	doi = {10.1257/aer.99.2.487},
	language = {en},
	number = {2},
	urldate = {2023-07-10},
	journal = {American Economic Review},
	author = {Bernard, Andrew B and Jensen, J. Bradford and Redding, Stephen J and Schott, Peter K},
	month = apr,
	year = {2009},
	pages = {487--493},
}

@article{sampson_technology_2023,
	title = {Technology {Gaps}, {Trade} and {Income}},
	volume = {VOL. 113},
	language = {en},
	number = {2},
	journal = {American Economic Review},
	author = {Sampson, Thomas},
	year = {2023},
	pages = {66},
}

@article{christiano_dsge_2018,
	title = {On {DSGE} {Models}},
	volume = {32},
	issn = {0895-3309},
	url = {https://www.aeaweb.org/articles?id=10.1257/jep.32.3.113},
	doi = {10.1257/jep.32.3.113},
	language = {en},
	number = {3},
	urldate = {2023-07-09},
	journal = {Journal of Economic Perspectives},
	author = {Christiano, Lawrence J. and Eichenbaum, Martin S. and Trabandt, Mathias},
	month = aug,
	year = {2018},
	pages = {113--140},
}

@incollection{costinot_trade_2014,
	title = {Trade {Theory} with {Numbers}: {Quantifying} the {Consequences} of {Globalization}},
	volume = {4},
	isbn = {978-0-444-54314-1},
	shorttitle = {Trade {Theory} with {Numbers}},
	url = {https://linkinghub.elsevier.com/retrieve/pii/B9780444543141000045},
	language = {en},
	urldate = {2023-07-09},
	booktitle = {Handbook of {International} {Economics}},
	publisher = {Elsevier},
	author = {Costinot, Arnaud and Rodríguez-Clare, Andrés},
	year = {2014},
	doi = {10.1016/B978-0-444-54314-1.00004-5},
	pages = {197--261},
}

@misc{dube_local_2023,
	type = {Working {Paper}},
	series = {Working {Paper} {Series}},
	title = {A {Local} {Projections} {Approach} to {Difference}-in-{Differences} {Event} {Studies}},
	url = {https://www.nber.org/papers/w31184},
	doi = {10.3386/w31184},
	urldate = {2023-07-02},
	publisher = {National Bureau of Economic Research},
	author = {Dube, Arindrajit and Girardi, Daniele and Jordà, Òscar and Taylor, Alan M.},
	month = apr,
	year = {2023},
	doi = {10.3386/w31184},
}

@article{kleinman_neoclassical_2023,
	title = {Neoclassical {Growth} in an {Interdependent} {World}},
	language = {en},
	author = {Kleinman, Benny and Liu, Ernest and Redding, Stephen J and Yogo, Motohiro},
	year = {2023},
}

@article{rachapalli_learning_2021,
	title = {Learning between {Buyers} and {Sellers} {Along} the {Global} {Value} {Chain}},
	url = {https://cowles.yale.edu/sites/default/files/2022-11/rachapalli-learningbetweenbuyersandsellersalongtheglobalvaluechain-learning-between-buyers-and.pdf},
	language = {en},
	author = {Rachapalli, Swapnika},
	year = {2021},
	pages = {81},
}

@misc{kehoe_how_2013,
	title = {How {Important} {Is} the {New} {Goods} {Margin} in {International} {Trade}?},
	url = {https://www.journals.uchicago.edu/doi/epdf/10.1086/670272},
	language = {en},
	urldate = {2023-05-29},
	author = {Kehoe, Timothy and Ruhl, Kim J.},
	year = {2013},
	doi = {10.1086/670272},
}

@article{demidova_productivity_2008,
	title = {Productivity {Improvements} and {Falling} {Trade} {Costs}: {Boon} or {Bane}?},
	volume = {49},
	issn = {0020-6598},
	shorttitle = {Productivity {Improvements} and {Falling} {Trade} {Costs}},
	url = {https://www.jstor.org/stable/20486843},
	number = {4},
	urldate = {2023-04-17},
	journal = {International Economic Review},
	author = {Demidova, Svetlana},
	year = {2008},
	note = {Publisher: [Economics Department of the University of Pennsylvania, Wiley, Institute of Social and Economic Research, Osaka University]},
	pages = {1437--1462},
}

@misc{boehm_long_2020,
	type = {Working {Paper}},
	series = {Working {Paper} {Series}},
	title = {The {Long} and {Short} ({Run}) of {Trade} {Elasticities}},
	url = {https://www.nber.org/papers/w27064},
	doi = {10.3386/w27064},
	urldate = {2023-02-16},
	publisher = {National Bureau of Economic Research},
	author = {Boehm, Christoph E. and Levchenko, Andrei A. and Pandalai-Nayar, Nitya},
	month = apr,
	year = {2020},
	doi = {10.3386/w27064},
}

@incollection{jones_growth_2005,
	title = {Growth and {Ideas}},
	volume = {1},
	isbn = {978-0-444-52043-2},
	url = {https://linkinghub.elsevier.com/retrieve/pii/S1574068405010166},
	language = {en},
	urldate = {2022-12-07},
	booktitle = {Handbook of {Economic} {Growth}},
	publisher = {Elsevier},
	author = {Jones, Charles I.},
	year = {2005},
	doi = {10.1016/S1574-0684(05)01016-6},
	pages = {1063--1111},
}

@article{jones_paul_2019,
	title = {Paul {Romer}: {Ideas}, {Nonrivalry}, and {Endogenous} {Growth}},
	volume = {121},
	issn = {0347-0520, 1467-9442},
	shorttitle = {Paul {Romer}},
	url = {https://onlinelibrary.wiley.com/doi/10.1111/sjoe.12370},
	doi = {10.1111/sjoe.12370},
	language = {en},
	number = {3},
	urldate = {2022-12-07},
	journal = {The Scandinavian Journal of Economics},
	author = {Jones, Charles I.},
	month = jul,
	year = {2019},
	pages = {859--883},
}

@article{caliendo_goods_2021,
	title = {Goods and {Factor} {Market} {Integration}: {A} {Quantitative} {Assessment} of the {EU} {Enlargement}},
	volume = {129},
	issn = {0022-3808},
	shorttitle = {Goods and {Factor} {Market} {Integration}},
	url = {https://www.journals.uchicago.edu/doi/10.1086/716560},
	doi = {10.1086/716560},
	number = {12},
	urldate = {2022-12-05},
	journal = {Journal of Political Economy},
	author = {Caliendo, Lorenzo and Opromolla, Luca David and Parro, Fernando and Sforza, Alessandro},
	month = dec,
	year = {2021},
	note = {Publisher: The University of Chicago Press},
	pages = {3491--3545},
}

@misc{borusyak_revisiting_2022,
	title = {Revisiting {Event} {Study} {Designs}: {Robust} and {Efficient} {Estimation}},
	shorttitle = {Revisiting {Event} {Study} {Designs}},
	url = {http://arxiv.org/abs/2108.12419},
	urldate = {2022-10-18},
	publisher = {arXiv},
	author = {Borusyak, Kirill and Jaravel, Xavier and Spiess, Jann},
	month = apr,
	year = {2022},
	note = {arXiv:2108.12419 [econ]},
}

@article{sun_estimating_2021,
	series = {Themed {Issue}: {Treatment} {Effect} 1},
	title = {Estimating dynamic treatment effects in event studies with heterogeneous treatment effects},
	volume = {225},
	issn = {0304-4076},
	url = {https://www.sciencedirect.com/science/article/pii/S030440762030378X},
	doi = {10.1016/j.jeconom.2020.09.006},
	language = {en},
	number = {2},
	urldate = {2022-10-18},
	journal = {Journal of Econometrics},
	author = {Sun, Liyang and Abraham, Sarah},
	month = dec,
	year = {2021},
	pages = {175--199},
}

@article{goodman-bacon_difference--differences_2021,
	series = {Themed {Issue}: {Treatment} {Effect} 1},
	title = {Difference-in-differences with variation in treatment timing},
	volume = {225},
	issn = {0304-4076},
	url = {https://www.sciencedirect.com/science/article/pii/S0304407621001445},
	doi = {10.1016/j.jeconom.2021.03.014},
	language = {en},
	number = {2},
	urldate = {2022-10-18},
	journal = {Journal of Econometrics},
	author = {Goodman-Bacon, Andrew},
	month = dec,
	year = {2021},
	pages = {254--277},
}

@article{callaway_difference--differences_2021,
	series = {Themed {Issue}: {Treatment} {Effect} 1},
	title = {Difference-in-{Differences} with multiple time periods},
	volume = {225},
	issn = {0304-4076},
	url = {https://www.sciencedirect.com/science/article/pii/S0304407620303948},
	doi = {10.1016/j.jeconom.2020.12.001},
	language = {en},
	number = {2},
	urldate = {2022-10-18},
	journal = {Journal of Econometrics},
	author = {Callaway, Brantly and Sant’Anna, Pedro H. C.},
	month = dec,
	year = {2021},
	pages = {200--230},
}

@article{acemoglu_world_2002,
	title = {The {World} {Income} {Distribution}*},
	volume = {117},
	issn = {0033-5533},
	url = {https://doi.org/10.1162/003355302753650355},
	doi = {10.1162/003355302753650355},
	number = {2},
	urldate = {2022-05-03},
	journal = {The Quarterly Journal of Economics},
	author = {Acemoglu, Daron and Ventura, Jaume},
	month = may,
	year = {2002},
	pages = {659--694},
}

@article{sampson_dynamic_2016,
	title = {Dynamic {Selection}: {An} {Idea} {Flows} {Theory} of {Entry}, {Trade}, and {Growth} *},
	volume = {131},
	issn = {0033-5533},
	shorttitle = {Dynamic {Selection}},
	url = {https://doi.org/10.1093/qje/qjv032},
	doi = {10.1093/qje/qjv032},
	number = {1},
	urldate = {2022-04-27},
	journal = {The Quarterly Journal of Economics},
	author = {Sampson, Thomas},
	month = feb,
	year = {2016},
	pages = {315--380},
}

@article{bas_input-trade_2012,
	title = {Input-trade liberalization and firm export decisions: {Evidence} from {Argentina}},
	volume = {97},
	issn = {0304-3878},
	shorttitle = {Input-trade liberalization and firm export decisions},
	url = {https://www.sciencedirect.com/science/article/pii/S0304387811000551},
	doi = {10.1016/j.jdeveco.2011.05.010},
	language = {en},
	number = {2},
	urldate = {2022-04-18},
	journal = {Journal of Development Economics},
	author = {Bas, Maria},
	month = mar,
	year = {2012},
	pages = {481--493},
}

@article{rivera-batiz_economic_1991,
	title = {Economic {Integration} and {Endogenous} {Growth}},
	volume = {106},
	issn = {0033-5533},
	url = {https://www.jstor.org/stable/2937946},
	doi = {10.2307/2937946},
	number = {2},
	urldate = {2022-03-05},
	journal = {The Quarterly Journal of Economics},
	author = {Rivera-Batiz, Luis A. and Romer, Paul M.},
	year = {1991},
	note = {Publisher: Oxford University Press},
	pages = {531--555},
}

@article{alvarez_general_2007,
	title = {General equilibrium analysis of the {Eaton}–{Kortum} model of international trade},
	volume = {54},
	issn = {0304-3932},
	url = {https://www.sciencedirect.com/science/article/pii/S0304393206002169},
	doi = {10.1016/j.jmoneco.2006.07.006},
	language = {en},
	number = {6},
	urldate = {2022-03-04},
	journal = {Journal of Monetary Economics},
	author = {Alvarez, Fernando and Lucas, Robert E., },
	month = sep,
	year = {2007},
	pages = {1726--1768},
}

@article{grossman_comparative_1990,
	title = {Comparative {Advantage} and {Long}-{Run} {Growth}},
	volume = {80},
	issn = {0002-8282},
	url = {https://www.jstor.org/stable/2006708},
	number = {4},
	urldate = {2022-02-21},
	journal = {The American Economic Review},
	author = {Grossman, Gene M. and Helpman, Elhanan},
	year = {1990},
	note = {Publisher: American Economic Association},
	pages = {796--815},
}

@article{rivera-batiz_international_1991,
	title = {International trade with endogenous technological change},
	volume = {35},
	issn = {0014-2921},
	url = {https://www.sciencedirect.com/science/article/pii/001429219190048N},
	doi = {10.1016/0014-2921(91)90048-N},
	language = {en},
	number = {4},
	urldate = {2022-02-21},
	journal = {European Economic Review},
	author = {Rivera-Batiz, Luis A. and Romer, Paul M.},
	month = may,
	year = {1991},
	pages = {971--1001},
}

@article{chodorow-reich_regional_2020,
	series = {St. {Louis} {Fed} -{JEDC}-{SCG}-{SNB}-{UniBern} {Conference}, titled "{Disaggregate} {Data} and {Macroeconomic} {Models}"},
	title = {Regional data in macroeconomics: {Some} advice for practitioners},
	volume = {115},
	issn = {0165-1889},
	shorttitle = {Regional data in macroeconomics},
	url = {https://www.sciencedirect.com/science/article/pii/S0165188920300440},
	doi = {10.1016/j.jedc.2020.103875},
	language = {en},
	urldate = {2021-09-13},
	journal = {Journal of Economic Dynamics and Control},
	author = {Chodorow-Reich, Gabriel},
	month = jun,
	year = {2020},
	pages = {103875},
}

@article{chodorow-reich_geographic_2019,
	title = {Geographic {Cross}-{Sectional} {Fiscal} {Spending} {Multipliers}: {What} {Have} {We} {Learned}?},
	volume = {11},
	issn = {1945-7731, 1945-774X},
	shorttitle = {Geographic {Cross}-{Sectional} {Fiscal} {Spending} {Multipliers}},
	url = {https://pubs.aeaweb.org/doi/10.1257/pol.20160465},
	doi = {10.1257/pol.20160465},
	language = {en},
	number = {2},
	urldate = {2020-09-23},
	journal = {American Economic Journal: Economic Policy},
	author = {Chodorow-Reich, Gabriel},
	month = may,
	year = {2019},
	pages = {1--34},
}

@article{goldberg_imported_2010,
	title = {Imported {Intermediate} {Inputs} and {Domestic} {Product} {Growth}: {Evidence} from {India}},
	volume = {125},
	issn = {0033-5533, 1531-4650},
	shorttitle = {Imported {Intermediate} {Inputs} and {Domestic} {Product} {Growth}},
	url = {https://academic.oup.com/qje/article-lookup/doi/10.1162/qjec.2010.125.4.1727},
	doi = {10.1162/qjec.2010.125.4.1727},
	language = {en},
	number = {4},
	urldate = {2021-09-09},
	journal = {The Quarterly Journal of Economics},
	author = {Goldberg, P. K. and Khandelwal, A. K. and Pavcnik, N. and Topalova, P.},
	month = nov,
	year = {2010},
	pages = {1727--1767},
}

@article{romer_new_1994,
	title = {New goods, old theory, and the welfare costs of trade restrictions},
	volume = {43},
	issn = {0304-3878},
	url = {https://www.sciencedirect.com/science/article/pii/0304387894900213},
	doi = {10.1016/0304-3878(94)90021-3},
	language = {en},
	number = {1},
	urldate = {2021-09-09},
	journal = {Journal of Development Economics},
	author = {Romer, Paul},
	month = feb,
	year = {1994},
	pages = {5--38},
}

@article{bernard_multiproduct_2011,
	title = {Multiproduct {Firms} and {Trade} {Liberalization}},
	volume = {126},
	issn = {0033-5533, 1531-4650},
	url = {https://academic.oup.com/qje/article-lookup/doi/10.1093/qje/qjr021},
	doi = {10.1093/qje/qjr021},
	language = {en},
	number = {3},
	urldate = {2021-09-09},
	journal = {The Quarterly Journal of Economics},
	author = {Bernard, A. B. and Redding, S. J. and Schott, P. K.},
	month = aug,
	year = {2011},
	pages = {1271--1318},
}

@techreport{arkolakis_extensive_2020,
	title = {The {Extensive} {Margin} of {Exporting} {Products}: {A} {Firm}-level {Analysis}},
	shorttitle = {The {Extensive} {Margin} of {Exporting} {Products}},
	url = {https://www.nber.org/papers/w16641},
	language = {en},
	number = {w16641},
	urldate = {2021-01-23},
	institution = {National Bureau of Economic Research},
	author = {Arkolakis, Costas and Ganapati, Sharat and Muendler, Marc-Andreas},
	year = {2020},
	doi = {10.3386/w16641},
}

@article{hummels_variety_2005,
	title = {The {Variety} and {Quality} of a {Nation}'s {Exports}},
	volume = {95},
	issn = {0002-8282},
	url = {https://pubs.aeaweb.org/doi/10.1257/0002828054201396},
	doi = {10.1257/0002828054201396},
	language = {en},
	number = {3},
	urldate = {2021-09-04},
	journal = {American Economic Review},
	author = {Hummels, David and Klenow, Peter J},
	month = may,
	year = {2005},
	pages = {704--723},
}

@techreport{melitz_trade_2021,
	address = {Cambridge, MA},
	title = {Trade and {Innovation}},
	url = {http://www.nber.org/papers/w28945.pdf},
	language = {en},
	number = {w28945},
	urldate = {2021-06-28},
	institution = {National Bureau of Economic Research},
	author = {Melitz, Marc and Redding, Stephen},
	month = jun,
	year = {2021},
	doi = {10.3386/w28945},
	pages = {w28945},
}

@article{buera_global_2020,
	title = {The {Global} {Diffusion} of {Ideas}},
	volume = {88},
	issn = {1468-0262},
	url = {https://onlinelibrary.wiley.com/doi/abs/10.3982/ECTA14044},
	doi = {https://doi.org/10.3982/ECTA14044},
	language = {en},
	number = {1},
	urldate = {2021-04-08},
	journal = {Econometrica},
	author = {Buera, Francisco J. and Oberfield, Ezra},
	year = {2020},
	pages = {83--114},
}

@article{eaton_technology_2002,
	title = {Technology, {Geography}, and {Trade}},
	volume = {70},
	issn = {1468-0262},
	url = {https://onlinelibrary.wiley.com/doi/abs/10.1111/1468-0262.00352},
	doi = {https://doi.org/10.1111/1468-0262.00352},
	language = {en},
	number = {5},
	urldate = {2021-01-19},
	journal = {Econometrica},
	author = {Eaton, Jonathan and Kortum, Samuel},
	year = {2002},
	pages = {1741--1779},
}

@techreport{eaton_innovation_2006,
	address = {Cambridge, MA},
	title = {Innovation, {Diffusion}, and {Trade}},
	url = {http://www.nber.org/papers/w12385.pdf},
	language = {en},
	number = {w12385},
	urldate = {2021-01-23},
	institution = {National Bureau of Economic Research},
	author = {Eaton, Jonathan and Kortum, Samuel},
	month = jul,
	year = {2006},
	doi = {10.3386/w12385},
	pages = {w12385},
}

@article{arkolakis_new_2012,
	title = {New {Trade} {Models}, {Same} {Old} {Gains}?},
	volume = {102},
	issn = {0002-8282},
	url = {https://www.aeaweb.org/articles?id=10.1257/aer.102.1.94},
	doi = {10.1257/aer.102.1.94},
	language = {en},
	number = {1},
	urldate = {2021-01-23},
	journal = {American Economic Review},
	author = {Arkolakis, Costas and Costinot, Arnaud and Rodríguez-Clare, Andrés},
	month = feb,
	year = {2012},
	pages = {94--130},
}

@techreport{perla_equilibrium_2015,
	title = {Equilibrium {Technology} {Diffusion}, {Trade}, and {Growth}},
	url = {https://www.nber.org/papers/w20881},
	language = {en},
	number = {w20881},
	urldate = {2021-01-22},
	institution = {National Bureau of Economic Research},
	author = {Perla, Jesse and Tonetti, Christopher and Waugh, Michael E.},
	month = jan,
	year = {2015},
	doi = {10.3386/w20881},
}

@article{argente_patents_2020,
	title = {Patents to {Products}: {Product} {Innovation} and {Firm} {Dynamics}},
	issn = {1556-5068},
	shorttitle = {Patents to {Products}},
	url = {https://www.ssrn.com/abstract=3577811},
	doi = {10.2139/ssrn.3577811},
	language = {en},
	urldate = {2021-01-21},
	journal = {SSRN Electronic Journal},
	author = {Argente, David and Baslandze, Salome and Hanley, Douglas and Moreira, Sara},
	year = {2020},
}

@article{romer_endogenous_1990,
	title = {Endogenous {Technological} {Change}},
	volume = {98},
	url = {http://www.jstor.org/stable/2937632},
	language = {en},
	number = {5,},
	journal = {The Journal of Political Economy},
	author = {Romer, Paul M.},
	year = {1990},
	pages = {S71-- S102},
}
    

\newpage

\appendix

\section{Timeline of EU Trade Agreements}

\begin{table}[htp]

\centering
\begin{tabular}{|l|r|l|r|}
\hline
\rowcolor[HTML]{EAECF0} 
\multicolumn{1}{|c|}{\cellcolor[HTML]{EAECF0}{\color[HTML]{202122} \textbf{Partner}}} &
  \multicolumn{1}{c|}{\cellcolor[HTML]{EAECF0}{\color[HTML]{202122} \textbf{Signed}}} &
  \multicolumn{1}{c|}{\cellcolor[HTML]{EAECF0}{\color[HTML]{202122} \textbf{\begin{tabular}[c]{@{}c@{}}Provisional\\ application\end{tabular}}}} &
  \multicolumn{1}{c|}{\cellcolor[HTML]{EAECF0}{\color[HTML]{202122} \textbf{Full entry into force}}} \\ \hline
\rowcolor[HTML]{F8F9FA} 
{\color[HTML]{202122} Switzerland} &
  {\color[HTML]{202122} 1972} &
   &
  {\color[HTML]{202122} 1973} \\ \hline
\rowcolor[HTML]{F8F9FA} 
{\color[HTML]{202122} Iceland} &
  {\color[HTML]{202122} 1992} &
   &
  {\color[HTML]{202122} 1994} \\ \hline
\rowcolor[HTML]{F8F9FA} 
{\color[HTML]{202122} Norway} &
  {\color[HTML]{202122} 1992} &
   &
  {\color[HTML]{202122} 1994} \\ \hline
\rowcolor[HTML]{F8F9FA} 
{\color[HTML]{202122} Turkey} &
  {\color[HTML]{202122} 1995} &
   &
  {\color[HTML]{202122} 1995} \\ \hline
\rowcolor[HTML]{F8F9FA} 
{\color[HTML]{202122} Tunisia} &
  {\color[HTML]{202122} 1995} &
   &
  {\color[HTML]{202122} 1998} \\ \hline
\rowcolor[HTML]{F8F9FA} 
{\color[HTML]{202122} Israel} &
  {\color[HTML]{202122} 1995} &
  \multicolumn{1}{r|}{\cellcolor[HTML]{F8F9FA}{\color[HTML]{202122} 1996}} &
  {\color[HTML]{202122} 2000} \\ \hline
\rowcolor[HTML]{F8F9FA} 
{\color[HTML]{202122} Mexico} &
  {\color[HTML]{202122} 1997} &
   &
  {\color[HTML]{202122} 2000} \\ \hline
\rowcolor[HTML]{F8F9FA} 
{\color[HTML]{202122} Morocco} &
  {\color[HTML]{202122} 1996} &
   &
  {\color[HTML]{202122} 2000} \\ \hline
\rowcolor[HTML]{F8F9FA} 
{\color[HTML]{202122} Jordan} &
  {\color[HTML]{202122} 1997} &
   &
  {\color[HTML]{202122} 2002} \\ \hline
\rowcolor[HTML]{F8F9FA} 
{\color[HTML]{202122} Egypt} &
  {\color[HTML]{202122} 2001} &
   &
  {\color[HTML]{202122} 2004} \\ \hline
\rowcolor[HTML]{F8F9FA} 
{\color[HTML]{202122} North Macedonia} &
  {\color[HTML]{202122} 2001} &
  \multicolumn{1}{r|}{\cellcolor[HTML]{F8F9FA}{\color[HTML]{202122} 2001}} &
  {\color[HTML]{202122} 2004} \\ \hline
\rowcolor[HTML]{F8F9FA} 
{\color[HTML]{202122} South Africa} &
  {\color[HTML]{202122} 1999} &
  \multicolumn{1}{r|}{\cellcolor[HTML]{F8F9FA}{\color[HTML]{202122} 2000}} &
  {\color[HTML]{202122} 2004} \\ \hline
\rowcolor[HTML]{F8F9FA} 
{\color[HTML]{202122} Chile} &
  {\color[HTML]{202122} 2002} &
  \multicolumn{1}{r|}{\cellcolor[HTML]{F8F9FA}{\color[HTML]{202122} 2003}} &
  {\color[HTML]{202122} 2005} \\ \hline
\end{tabular}
\end{table}

\newpage

\section{Mathematical derivations}
\renewcommand{\theequation}{B.\arabic{equation}}
\setcounter{equation}{0}

\subsection{Optimal control problem}

In the dynamic optimal control problem, the household chooses an optimal path of $C_{s}(t)$ at every instant, taking as given prices. The problem of choosing varieties $c_{s}(t,\omega)$ is separable and can be solved conditional on a path for $C_{s}(t)$, such that only 
aggregates matter for the dynamic path. Therefore, the current-value Hamiltonian for this problem is:

\begin{equation*}
    H(t, C, L, \mu) =\log \left ( C_s(t) \right ) + \mu_s(t) \left[    \frac{r_s(t)}{P_s(t)} A_s(t) + \frac{ w_s(t) }{P_s(t)} L_{s} - C_{s}(t)   \right]
\end{equation*}

\noindent with optimality conditions satisfying:

\begin{eqnarray*}
    \frac{1}{C_s(t)} &=& \mu_s(t)  \\
    \frac{\dot{\mu}_s(t)}{\mu_s(t)} &=& \rho - \frac{r_s(t)}{P_s(t)}
\end{eqnarray*}

\noindent and a transversality condition:

\begin{equation*}
    \lim_{t \to \infty} \left[ \exp \{ - \int_0^t \frac{r_s(\nu)}{P_s(\nu)} d\nu \} P_s(t) A_s(t) \right] = 0
\end{equation*}

Taking time derivatives of the first optimality condition and then replacing for $\frac{\dot{\mu}(t)}{\mu(t)}$ yields the Euler equation:

\begin{equation*}
    \frac{\dot{C}_s(t)}{C_s(t)} =  \left[ \frac{r_s(t)}{P_s(t)} - \rho \right]
\end{equation*}

\subsection{Solution to the dynamic problem}\label{appendix: solution-dynamic-problem}

Growth in each of the $s \in \boldsymbol{K}$ of the national economies evolve according to the following system of differential equations:

\begin{eqnarray*}
    \dot{C}_s(t) &=&   \left[ \frac{r_s(t)}{P_s(t)} - \rho \right] C_s(t) \\
    \dot{M}_s(t) &=& \frac{r_s(t)}{P_s(t)} M_s(t) + \psi \frac{w_s(t)}{P_s(t)} L_s - \psi C_s(t)
\end{eqnarray*}

In this section, I will first derive this system of equations, then solve it. First, one sees that consumption evolves according to a first-order differential equation. Let $a(t) \equiv  \left[ \frac{r_s(t)}{P_s(t)} - \rho \right]$ and write the Euler equation as:

\begin{equation*}
    \dot{C}_s(t) =  a(t) C_s(t)
\end{equation*}

Multiplying both sides by the integration factor $\exp \{ - \int_0^t a(\tau) d\tau \}$:

\begin{equation*}
    \dot{C}_s(t) \exp \{ - \int_0^t a(\tau) d\tau \} - a(t)  C_s(t) \exp \{ - \int_0^t a(\tau) d\tau \}  =  0
\end{equation*}

Now, using Leibnitz lemma, note that the time derivative of \\
$\exp \{ - \int_0^t a(\tau) d\tau \} C_s(t)$ is $\dot{C}_s(t) \exp \{ - \int_0^t a(\tau) d\tau \} - a(t) C_s(t) \exp \{ - \int_0^t a(\tau) d\tau \} $. Therefore, integrating both sides with respect to time:

\begin{equation*}
    \exp \{ - \int_0^t a(\tau) d\tau \} C_s(t) = C(0)
\end{equation*}

\noindent where $C(0)$ is the constant of integration. Dividing both sides by $\exp \{ - \int_0^t a(s) ds \}$ and replacing for $a(t)$ yields the solution for the consumption path:

\begin{equation*}
    C_s(t) = C(0) \exp \left \{\int_0^t  \left[ \frac{r_s(\tau)}{P_s(\tau)}  - \rho  \right] d\tau  \right\}
\end{equation*}

\noindent which can be rewritten as:

\begin{equation*}
    C_s(t) = C_s(0) \exp \left \{   \left[ \bar{r}_s(t) - \rho  \right] t  \right\}
\end{equation*}

\noindent where $\bar{r}_s(t) \equiv \frac{1}{t} \int_0^t  \frac{r_s(\nu)}{P_s(\nu)}  d\nu$ is the average real interest rate between periods $0$ and $t$. Now recall that the budget constraint is:

\begin{equation}
    P_s(t) I_s(t) + P_s(t) C_s(t) = r_s(t) A_s(t) + w_s(t) L_s
\end{equation}

\noindent and that $\psi I_s(t) = \dot{M}_s(t)$ and $\psi A_s(t) = M_s(t)$. Replacing above and solving for $\dot{M}_s(t)$ results in:

\begin{equation*}
    \dot{M}_s(t) = \frac{r_s(t)}{P_s(t)} M_s(t) +  \psi \frac{w_s(t)}{P_s(t)} L_s - \psi C_s(t)
\end{equation*}

\noindent which, after replacement, yields the following equation:

\begin{equation*}
    \dot{M}_s(t) = \frac{r_s(t)}{P_s(t)} M_s(t) + \psi \frac{w_s(t)}{P_s(t)} L_s - \psi C_s(0) \exp \left \{   \left[ \bar{r}_s(t) - \rho  \right] t  \right\}
\end{equation*}

In turn, this equation has a solution satisfying:

\begin{eqnarray*}
    M_s(t) &=& M_s(0) \cdot \exp \left \{ \int_0^t \frac{r_s(\nu)}{P_s(\nu)} d\nu  \right\} \\
    &+&  \int_0^t \psi \frac{w_s(\xi)}{P_s(\xi)} L_s \cdot \exp \left\{ - \int_0^\xi \frac{r_s(\upsilon)}{P_s(\upsilon)} d\upsilon \right\} d\xi \cdot \exp \left \{ \int_0^t \frac{r_s(\nu)}{P_s(\nu)} d\nu  \right\}  \\
    &-& \int_0^t \psi C_s(0) \exp \left \{   \left[ \bar{r}_s(\xi) - \rho  \right] \xi  \right\} \cdot \exp \left\{ - \int_0^\xi \frac{r_s(\upsilon)}{P_s(\upsilon)} d\upsilon \right\} d\xi \cdot \exp \left \{ \int_0^t \frac{r_s(\nu)}{P_s(\nu)} d\nu  \right\}
\end{eqnarray*}

\noindent which, using the definition of $\bar{r}(t)$, becomes:

\begin{eqnarray*}
    M_s(t) &=& M_s(0) \cdot \exp \left \{ \bar{r}(t) \cdot t  \right\} \\
    &+&  \int_0^t \psi \frac{w_s(\xi)}{P_s(\xi)} L_s \cdot\exp \left\{ - \bar{r}(\xi) \cdot \xi \right\} d\xi \cdot \exp \left \{ \bar{r}(t) \cdot t  \right\}  \\
    &-& \psi C_s(0)  \cdot \int_0^t \exp \left \{   \left[ \bar{r}_s(\xi) - \rho  \right] \xi  \right\} \cdot \exp \left\{ - \bar{r}(\xi) \cdot \xi \right\} d\xi \cdot \exp \left \{ \bar{r}(t) \cdot t  \right\}  
\end{eqnarray*}

\noindent simplifying the last integral:

\begin{eqnarray*}
    M_s(t) &=& M_s(0) \cdot \exp \left \{ \bar{r}(t) \cdot t  \right\} \\
    &+&  \int_0^t \psi \frac{w_s(\xi)}{P_s(\xi)} L_s \cdot\exp \left\{ - \bar{r}(\xi) \cdot \xi \right\} d\xi \cdot \exp \left \{ \bar{r}(t) \cdot t  \right\}  \\
    &-& \psi C_s(0)  \cdot \int_0^t \exp \left \{    - \rho   \xi  \right\}  d\xi \cdot \exp \left \{ \bar{r}(t) \cdot t  \right\} 
\end{eqnarray*}

Finally, note that both $P_s(t)$ and $r_s(t)$ are functions of wages. Therefore, given the initial measure of varieties $M_s(0)$ and the wages for all countries, which are defined at every instance through the trade equilibrium, paths for consumption $C_s(t)$, varieties $M_s(t)$ and assets $A_s(t) = 1/\psi M_s(t)$ follow the equations above.

As a final step, one needs to pin down the starting values. $M_s(0)$ is given and calibrated to reflect the technological level of country $s$. Choice of $C_s(0)$, by contrast, is an endogenous object that guarantees that, given lifetime income and the initial level of assets, consumption as governed by the euler equation will be optimal.  Start from the equation above, multiply both sides by $\exp \left \{ - \bar{r}(t) \cdot t  \right\}$:

\begin{eqnarray*}
    \exp \left \{ - \bar{r}(t) \cdot t  \right\} M_s(t) &=& M_s(0)  +  \int_0^t \psi \frac{w_s(\xi)}{P_s(\xi)} L_s \cdot\exp \left\{ - \bar{r}(\xi) \cdot \xi \right\} d\xi \\
    &-& \psi C_s(0)  \cdot \int_0^t \exp \left \{    - \rho   \xi  \right\}  d\xi
\end{eqnarray*}

Now evaluate this equation taking the limit $t\to\infty$.

\begin{eqnarray*}
    \lim_{t\to\infty} \left( \exp \left \{ - \bar{r}(t) \cdot t  \right\} M_s(t)\right) &=& M_s(0)  +  \int_0^\infty \psi \frac{w_s(t)}{P_s(t)} L_s \cdot\exp \left\{ - \bar{r}(t) \cdot t \right\} dt \\
    &-& \psi C_s(0)  \cdot \int_0^\infty \exp \left \{    - \rho   t  \right\}  dt
\end{eqnarray*}

Recall that the transversality condition is:

\begin{equation*}
    \lim_{t \to \infty} \left[ \exp \{ - \int_0^t \frac{r_s(\nu)}{P_s(\nu)} d\nu \} P_s(t) A_s(t) \right] = 0
\end{equation*}

\noindent which states that the value of assets cannot grow faster than the interest rate, the standard no-Ponzi scheme condition. Using the fact that $\psi A_s(t) = M_s(t)$, noting that prices $P_s(t)$ are always positive and finite, and dividing both sides by $P_s(t) / \psi $, we can rewrite this as:

\begin{equation*}
    \lim_{t \to \infty} \left[ \exp \{ - \bar{r}_s(t) t \} M_s(t) \right] = 0
\end{equation*}

Using the fact that $\lim_{t\to\infty} \left( \exp \left \{ - \bar{r}(t) \cdot t  \right\} M_s(t)\right) = 0$, we can then solve for $C_s(0)$ as:

\begin{equation*}
       C_s(0) = \left[ \frac{1}{\psi} M_s(0)  +  \int_0^\infty  \frac{w_s(t)}{P_s(t)} L_s \cdot\exp \left\{  - \bar{r}_s(t) \cdot t \right\} dt  \right] \cdot \left[  \int_0^\infty \exp \left \{    - \rho   t  \right\}  dt \right]^{-1}
\end{equation*}

Using the fact that $\int_0^\infty \exp \left \{    - \rho   t  \right\}  dt = \frac{1}{\rho}$, then:

\begin{eqnarray}\label{eq: initial-consumption}
       C_s(0) &=& \rho \left[ \frac{1}{\psi} M_s(0)  +  \int_0^\infty  \frac{w_s(t)}{P_s(t)} L_s \cdot\exp \left\{  - \bar{r}_s(t) \cdot t \right\} dt  \right] \\
       &=& 
       \rho \left[ \underbrace{A_s(0)}_{\text{initial wealth}}  +  \underbrace{\int_0^\infty  \frac{w_s(t)}{P_s(t)} L_s \cdot\exp \left\{  - \bar{r}_s(t) \cdot t \right\} dt}_{\text{PV of real labor income}}  \right] \nonumber
       \end{eqnarray}

Therefore, at any instant $t$, consumption is proportional to lifetime wealth:

\begin{eqnarray}\label{eq: instantaneous-consumption}
    C_s(t) &=& \rho \left[ A_s(0)  +  \int_0^\infty  \frac{w_s(\tau)}{P_s(\tau)} L_s \cdot\exp \left\{  - \bar{r}_s(\tau) \cdot \tau \right\} d\tau  \right] \cdot \exp \left \{   \left[ \bar{r}_s(t) - \rho  \right] t  \right\} \\
    &=& \rho \left[ A_s(t)  +  \int_t^\infty  \frac{w_s(\tau)}{P_s(\tau)} L_s \cdot\exp \left\{  - \bar{r}_s(\tau) \cdot \tau \right\} d\tau  \right]  \nonumber
\end{eqnarray}

\subsection{Final varieties producers problem} 

Each final goods producer chooses intermediate inputs and labor to maximize profits according to:

\begin{eqnarray*}
    \max_{\ell_{s}(t, \omega), \{x_{ks}(t, \omega,\nu)\}} & & p_{ss}(t,\omega) z_{s}(t, \omega)  [\ell_{s}(t, \omega)]^{1-\alpha} \left( \frac{1}{\alpha} \sum_{k \in \boldsymbol{K}}  \int_{0}^{M_{k}(t)} [x_{ks}(t, \omega,\nu)]^{\alpha} d\nu  \right) \\
    &-& w_{s}(t)\ell_{s}(t, \omega) - \sum_{k \in \boldsymbol{K}}  \int_{0}^{M_{k}(t)} p_{ks}(t, \nu) x_{ks}(t, \omega,\nu) d\nu
\end{eqnarray*}

There are infinitely many first order conditions for this problem: one for each variety $\nu$ and one for labor. These satisfy:

\begin{eqnarray*}
    w_{s}(t)\ell_{s}(t, \omega) &=& (1-\alpha) \cdot p_{ss}(t,\omega) z_{s}(t, \omega)  [\ell_{s}(t, \omega)]^{1-\alpha} \left( \frac{1}{\alpha} \sum_{k \in \boldsymbol{K}}  \int_{0}^{M_{k}(t)} [x_{ks}(t, \omega,\nu)]^{\alpha} d\nu  \right) \\
    p_{ks}(t, \nu) x_{ks}(t, \omega,\nu) &=& \alpha \cdot  p_{ss}(t,\omega) z_{s}(t, \omega)  [\ell_{s}(t, \omega)]^{1-\alpha} \left( \frac{1}{\alpha}  [x_{ks}(t, \omega,\nu)]^{\alpha} \right)
\end{eqnarray*}

Solving for $x_{ks}(t,\omega, \nu)$ yields equation \eqref{eq: demand-intermediate}:

\begin{equation*}
            x_{ks}(t,\omega, \nu) = \left[ \frac{p_{ks}(t,\omega, \nu)}{p_{ss}(t, \omega)} \right] ^{- \frac{1}{1-\alpha}} \cdot \ell_{s}(t, \omega) \cdot z_{s}(t, \omega)^{\frac{1}{1-\alpha}}
\end{equation*}

\subsection{Intermediate varieties producers problem}\label{appendix: intermediate-varieties}

Each intermediate varieties producer holds perpetual rights over variety $\nu$, which they sell to final goods varieties in every country $d \in \boldsymbol{K}$. For each destination, they take demand as given and choose prices to maximize profits at every moment:

\begin{equation*}
    \max_{p_{ks}(t,\omega,\nu)} \frac{1}{\tau_{ks}} p_{ks}(t,\omega,\nu) x_{ks}(t,\omega,\nu) - P_{k}(t) x_{ks}(t,\omega,\nu)
\end{equation*}

Replacing for $x_{ks}(t,\omega,\nu)$:
\begin{equation*}
    \max_{p_{ks}(t,\omega,\nu)} \left[ p_{ks}(t,\omega,\nu) - \tau_{ks} P_{k}(t) \right] \left[ \frac{p_{ks}(t,\omega, \nu)}{p_{ss}(t, \omega)} \right] ^{- \frac{1}{1-\alpha}} \cdot \ell_{s}(t, \omega) \cdot z_{s}(t, \omega)^{\frac{1}{1-\alpha}}
\end{equation*}

\noindent which, after taking the FOC and solving for $p_{ks}(t,\omega,\nu)$ yields the optimal price as a mark-up over marginal price, which is independent of $\omega$ or $\nu$:

\begin{equation*}
    p_{ks}(t,\omega,\nu) = \frac{\tau_{ks} P_{k}(t)}{\alpha} \qquad \forall \omega \in [0,1], \quad \forall \nu \in [0, M_s(t)]
\end{equation*}

Flow aggregate profits $\Pi_{s}(t) \equiv \int_0^{M_s(t)} \pi(t,\nu) d\nu$ are a constant fraction of revenue:

\begin{eqnarray*}
    \Pi_{s}(t) &=& \frac{\alpha}{\eta} \cdot \sum_{d \in \boldsymbol{K}} \lambda_{sd}^M(t)  \cdot P_d(t) Y_d(t)  \\
            &=& \frac{\alpha}{\eta} \cdot \sum_{d \in \boldsymbol{K}}  \frac{M_s \left( p_{sd}^M \right)^{1-\eta} }{P_{d}^M}  \cdot P_d(t) Y_d(t) \\
            &=& \frac{\alpha}{\eta} \cdot \sum_{d \in \boldsymbol{K}}  \frac{M_s \left( \tau_{sd} P_{sd} \right)^{1-\eta} }{\sum_{k' \in \boldsymbol{K}} M_{k'} \left( \tau_{k'd} P_{k'} \right)^{1-\eta} }  \cdot P_d(t) Y_d(t)
\end{eqnarray*}

Profits per variety $\pi_s(t,\nu) = \frac{1}{M_s(t)} \Pi_{s}(t)$ are independent of $\nu$.

\subsection{Trade in Final Goods} \label{appendix: ek-trade}

\paragraph{Trade shares} In this model, since there are infinitely many varieties in the unit interval, the expenditure share of destination region $d \in \boldsymbol{K}$ on goods coming from source country $s \in \boldsymbol{K}$ converge to their expected values. Let $\lambda_{sd}(t,\omega)$ denote the probability that consumers in region $d \in \mathcal{D}$ source variety $\omega$  from region $s \in \mathcal{D}$. For each each $n$, let $A_{n}^{-1}(t,\omega) \equiv \frac{\tilde{x}_{sd}(t)}{\tilde{x}_{nd}(t)}$, with $x_{sd}(t) \equiv (P_s^M(t))^{\alpha} w_{s}(t)^{1-\alpha} \tau_{sd}$. This probability will satisfy:
    
    \begin{eqnarray}
           \lambda_{sd}(t, \omega) &=& Pr \Big( s \text{ is the lowest cost supplier of } \omega \text{ to } d \Big) \nonumber \\
           &=& Pr \Big( \frac{\tilde{x}_{sd}(t)}{z_{s}(t,\omega)} < \min_{(n \neq s)} \Big\{ \frac{\tilde{x}_{nd}(t)}{z_{n}(t,\omega)}  \Big\}  \Big) \nonumber \\
           &=& \int_{0}^{\infty} Pr(z_{s}(t,\omega) = z) Pr ( z_{n}(t,\omega) < z A_n(t) ) dz  \nonumber \\
           &=& \int_{0}^{\infty} f_{s}(t)(z)  \Pi_{(n\neq s)} F_{n}(t)(A_n z ) dz  \nonumber \\
           &=&  \int_{0}^{\infty} \theta T_{s} z^{-(1+\theta)} \exp \left\{ - \left( \sum_{n \in \boldsymbol{K}} T_{n} A_n(t)^{-\theta} \right)z^{-\theta} \right\}  dz   \nonumber \\
           &=& \frac{T_{s} \left(\tilde{x}_{sd}(t) \right)^{-\theta}}{\sum_{n \in \boldsymbol{K}} T_{n} \left( \tilde{x}_{nd}(t) \right)^{-\theta}}  \nonumber \\
           &=& \frac{T_{s} \left(\tilde{x}_{sd}(t) \right)^{-\theta}}{\sum_{n \in \boldsymbol{K}} T_{n} \left( \tilde{x}_{nd}(t) \right)^{-\theta}}  \nonumber \\
           &=& \frac{T_{s} (w_{s}(t)^{1-\alpha} P^M_{s}(t)^{\alpha} \tau_{sd} )^{-\theta}}{\sum_{n \in \boldsymbol{K}}{T_{n}  (w_{n}(t)^{1-\alpha}  P^M_{n}(t)^{\alpha} \tau_{nd} )^{-\theta}}}
    \end{eqnarray}
    
    Now note that $\lambda_{sd}(t,\omega)$ is independent of $\omega$, so the probability of sourcing each variety from $s$ to $d$ is identical. A corollary is that aggregate expenditure trade shares of final goods from $s$ in $d$ will be equal to the probability of sourcing an arbitrary variety from $s$ in $d$.

\paragraph{Price distributions and ideal price index} Recall that, under the assumption of perfect competition, prices equal their marginal costs, such that the price of a variety $\omega$ produced in country $s$ and shipped to $d$ satisfies $p_{sd}(t, \omega) = \frac{\tau_{sd} P^M_s(t)^{\alpha} w_s(t) ^{1-\alpha}}{z_{s}(t, \omega)}$. 

Since $z_s(t, \omega)$ is a random variable, $p_{sd}(t, \omega)$ is also a random variable. We can derive the distribution of prices through the following steps. First, note  that $z_{s}(t, \omega) = \frac{\tau_{sd} P^M_s(t)^{\alpha} w_s(t)^{1-\alpha}}{ p_{sd}(\omega)}$. Then, note that:

\begin{equation*}
    p_{sd}(t,\omega) < p = \frac{\tau_{sd} P^M_s(t)^{\alpha} w_s^{1-\alpha}}{ z} \iff z_s(t,\omega) > z =  \frac{\tau_{sd} P^M_s(t)^{\alpha} w_s^{1-\alpha}}{ p}
\end{equation*}

Therefore:

\begin{eqnarray*}
    G_{sd}(t, \omega)(p) &=& Pr( p_{sd}(t,\omega) < p ) \\
    &=& Pr \left( z_s(t,\omega) > \frac{\tau_{sd} P^M_s(t)^{\alpha} w_s^{1-\alpha}}{ p} \right) \\
    &=& 1 - \exp \{ - T_s (\tau_{sd} P^M_s(t)^{\alpha} w_s(t) )^{-\theta}  p^{\theta}\} \qquad \forall \omega \in [0,1]
\end{eqnarray*}

\noindent which is the distribution of prices of any variety $\omega$ conditional on $s$ being the lowest cost supplier of such a variety to $d$. To derive the unconditional distribution of prices at $d$, realize that:

\begin{eqnarray*}
       G_n(t,\omega) &\equiv& Pr( p_s(t,\omega) < p ) \\ 
       &=& Pr( (\exists s) \text{ for which } p_sd(t,\omega) < p ) \\
       &=& 1- Pr( (\nexists s) \text{ for which } p_{sd}(t,\omega) < p ) \\
       &=& 1 - \prod_{s \in \boldsymbol{K}} Pr( p_{sd}(t,\omega) > p ) \\
       &=& 1 - \prod_{s \in \boldsymbol{K}}  \exp \{ - T_s (\tau_{sd} P^M_s(t)^{\alpha} w_s(t) )^{-\theta} p^{\theta} \\
       &=& 1 - \exp \{ - \sum_{s \in \boldsymbol{K}}  T_s (\tau_{sd} P^M_s(t)^{\alpha} w_s(t) )^{-\theta} p^{\theta}\}
     \end{eqnarray*}

Recall that the price index is defined as:

\begin{eqnarray*}
       P_{d}(t) &=& \left[ \int_0^1 p_d(t,\omega)^{1-\sigma} d\omega \right]^{\frac{1}{1-\sigma}} \\
       &=& \left[ \int_0^\infty p^{1-\sigma} dG_n(t,p) \right]^{\frac{1}{1-\sigma}} \\
       &=& \left[ \int_0^\infty p^{1-\sigma} \theta p^{\theta-1} \exp \left\{ - \sum_{s \in \boldsymbol{K}} T_{s} (\tau_{sd} P^M_s(t)^{\alpha} w_s
       (t)^{1-\alpha} )^{-\theta}  p^{\theta} \right\} dp  \right]^{\frac{1}{1-\sigma}}
\end{eqnarray*}

Using a change of variables, let $\nu \equiv \sum_{s \in \boldsymbol{K}} T_{s} (\tau_{sd} P^M_s(t)^{\alpha} w_s
       (t)^{1-\alpha} )^{-\theta}  p^{\theta}$ and note that $d\nu = \theta p^{\theta-1} \sum_{s \in \boldsymbol{K}} T_{s} (\tau_{sd} P^M_s(t)^{\alpha} w_s
       (t)^{1-\alpha} )^{-\theta}  p^{\theta} dp$. Then:
       
\begin{eqnarray}\label{eq: appendix-price-level}
       P_{d}(t) &=&  \left[ \int_0^\infty \left( \frac{\nu}{\sum_{s \in \boldsymbol{K}} T_{s} (\tau_{sd} P^M_s(t)^{\alpha} w_s
       (t)^{1-\alpha} )^{-\theta} } \right)^{\frac{1-\sigma}{\theta}} \exp \left\{ - \nu \right\} d\nu  \right]^{\frac{1}{1-\sigma}} \nonumber \\
       &=& \Gamma \left( \frac{\theta + 1 - \sigma}{\theta} \right)^{\frac{1}{1-\sigma}} \left( \sum_{s \in \boldsymbol{K}} T_{s} (\tau_{sd} P^M_s(t)^{\alpha} w_s
       (t)^{1-\alpha} )^{-\theta} \right)^{-\frac{1}{\theta}} \\
       &=& \Gamma \left( \frac{\theta + 1 - \sigma}{\theta} \right)^{\frac{1}{1-\sigma}} \left( \sum_{s \in \boldsymbol{K}} T_{s} (\tau_{sd}  w_s
       (t)^{1-\alpha} )^{-\theta} \left(  \sum_{n \in \boldsymbol{K}} M_n(t) \left[ \frac{\tau_{ns} P_n(t)}{\alpha} \right]^{-\frac{\alpha}{1-\alpha}} \right) ^{(1-\alpha)\theta} \right)^{-\frac{1}{\theta}}  \nonumber
       .
\end{eqnarray}

\noindent which shows that, given parameters $T_{s}. \tau_{sd}$ and the vector of state variables $\boldsymbol{M}_s(t) = [M_1(t), \cdots, M_N(t)]'$, the closed form solution for the ideal price index $P_d(t)$ is a function of the vector of wages $\boldsymbol{w}(t) = [w_1(t), \cdots, w_N(t)]'$.

\subsection{Market Clearing and Trade Balance}\label{appendix: market-clearing-trade-balance}

\paragraph{Market Clearing} Let $Y_d(t)$ denote the total output of the final good and $X_d(t), I_d(t)$ denote the use of the final good as inputs for the production of intermediate inputs and R\&D, respectively. Then total output in the final good for a given country must satisfy:

\begin{equation*}
    Y_d(t) = C_d(t) + I_d(t) + X_d(t)
\end{equation*}

\noindent where $I_d(t)$ and $C_d(t)$ are pinned down by the dynamic problem, described below, and $X_d(t)$ can be expressed as a function of aggregate demand in all destinations:

\begin{equation*}
    X_d(t) \equiv \sum_{k \in \boldsymbol{K}} M_d(t) \cdot \left( \frac{p^M_{dk}(t)}{P_k^M(t)} \right)^{-\eta} \cdot \alpha \cdot \left( \frac{P^M_k(t)}{P_k(t)} \right)^{-1} \cdot Y_k(t)
\end{equation*}

Combining the equations, one can express aggregate output as a function of the state variable $M_d(t)$, parameters, and wages (both $r_d(t)$ and $P_d(t)$ are functions of wages in every country):

\begin{equation*}
    Y_d(t) = I_d(t) + C_d(t)  + \sum_{k \in \boldsymbol{K}} M_d(t) \cdot \left( \frac{p^M_{dk}(t)}{P_k^M(t)} \right)^{-\eta} \cdot \alpha \cdot \left( \frac{P^M_k(t)}{P_k(t)} \right)^{-1} \cdot Y_k(t)
\end{equation*}

\paragraph{Expenditure Determination}  Flow aggregate profits $\Pi_{s}(t) \equiv \int_0^{M_s(t)} \pi(t,\nu) d\nu$ are a constant fraction of revenue:

\begin{equation*}
    \Pi_{s}(t) = \frac{\alpha}{\eta} \cdot \sum_{d \in \boldsymbol{K}} \lambda_{sd}^M(t)  \cdot P_d(t) Y_d(t) 
\end{equation*}

On the expenditure side, GDP of each destination country $s \in \boldsymbol{K}$ country will be exhausted as the combination of the total expenditures of labor and capital income:

\begin{equation*}
    P_s(t) Y_s(t) = w_s(t) L_s + \Pi_{s}(t)
\end{equation*}

From the income side, nominal GDP must equal the sum of total flow payments received domestically and from the rest of the world:

\begin{equation*}
     P_s(t) Y_s(t) = \sum_{d \in \boldsymbol{K}}  \left[  (1-\alpha) \lambda_{sd}^F(t) + \frac{\alpha }{\eta } \lambda_{sd}^M(t) \right]   P_d(t) Y_d(t)
\end{equation*}

\paragraph{Trade Balance}

Total exports are equal to:

\begin{equation*}
    EX_d(t) = \underbrace{\sum_{ d \neq s \in \boldsymbol{K}} \lambda_{sd}^F(t) P_d(t) Y_d(t)}_{\text{exports in final goods}} + \underbrace{\alpha \sum_{ d \neq s \in \boldsymbol{K}} \lambda_{sd}^M(t) \left[ \sum_{ k' \in \boldsymbol{K}} \lambda_{dk'}^F(t) P_{k'}(t)Y_{k'}(t) \right]}_{\text{exports in intermediates}}
\end{equation*}

Total imports are equal to:

\begin{equation*}
    IM_d(t) = \underbrace{ [1 - \lambda_{ss}^F(t) ]  P_{s}(t)Y_{s}(t)}_{\text{imports in final goods}} + \underbrace{\alpha   [1 - \lambda_{ss}^M(t) ]  \left[ \sum_{ k' \in \boldsymbol{K}} \lambda_{dk'}^F(t)  P_{k'}(t)Y_{k'}(t) \right] }_{\text{imports in intermediates}}
\end{equation*}

Since there are no international capital markets in this economy, trade will be balanced at any instant. This means that:

\begin{eqnarray*}
    & & \sum_{ d \neq s \in \boldsymbol{K}} \lambda_{sd}^F(t) P_d(t) Y_d(t) + \alpha \sum_{ d \neq s \in \boldsymbol{K}} \lambda_{sd}^M(t) \left[ \sum_{ k' \in \boldsymbol{K}} \lambda_{dk'}^F(t) P_{k'}(t)Y_{k'}(t) \right] =  \nonumber \\
    & & [1 - \lambda_{ss}^F(t) ]  P_{s}(t)Y_{s}(t) + \alpha   [1 - \lambda_{ss}^M(t) ]  \left[ \sum_{ k' \in \boldsymbol{K}} \lambda_{dk'}^F(t)  P_{k'}(t)Y_{k'}(t) \right] 
\end{eqnarray*}

\subsection{Homogeneity of Income in Equilibrium}

The trade share for final goods is:

\begin{eqnarray*}
    \lambda^F_{sd}(t) &\equiv& \frac{E^F_{sd}(t)}{E^F_{d}(t)} = \frac{T_{s}  (P^M_s(t)^{\alpha} w_{s}(t)^{1-\alpha} \tau_{sd} )^{-\theta}}{\sum_{n \in \boldsymbol{K}}{T_{n}  (P^M_n(t)^{\alpha} w_{n}(t)^{1-\alpha} \tau_{nd} )^{-\theta}}}  = \frac{T_{s}  (P^M_s(t)^{\alpha} w_{s}(t)^{1-\alpha} \tau_{sd} )^{-\theta}}{ P_s(t)^{-\theta} }    
\end{eqnarray*}

 Evaluating it at $\lambda_{ss}(t)$, noting that $\lambda^M_s(t) = \frac{M_s(t) (p^M_{ss}(t))^{1-\eta}}{(P^M_{s}(t))^{1-\eta}} = \frac{M_s(t) (\frac{1}{\alpha} P_s(t))^{1-\eta}}{(P^M_{s}(t))^{1-\eta}}$ and solving it for $P_s(t)$ allows me to write it linear in $M_s(t)$: 
\begin{eqnarray*}
    P_s(t) &=&  T_{s}^{-\frac{1}{\theta}}  \lambda^F_{ss}(t)^{\frac{1}{\theta}} ( P^M_s(t)^{\alpha} w_{s}(t)^{1-\alpha}  )  \\ 
    P_s(t) &=& \left(  \frac{ T_{s} }{   \lambda^F_{ss}(t) } \right) ^{- \frac{1}{\theta}} \left( \frac{M_s(t)}{\lambda^M_{ss}(t)}\right)^{\frac{\alpha}{1-\eta}} \cdot \alpha^{-\alpha} \cdot P_s(t)^{\alpha} (w_{s}(t)^{1-\alpha} ) \\
    P_s(t) &=&  \alpha^{-\frac{\alpha}{1-\alpha}} \cdot \left(  \frac{ T_{s} }{   \lambda^F_{ss}(t) } \right) ^{- \frac{1}{\theta(1-\alpha)}}\left( \frac{M_s(t)}{\lambda^M_{ss}(t)}\right)^{\frac{1}{1-\eta} \frac{\alpha}{1-\alpha} } \cdot w_{s}(t) \\
    P_s(t) &=& \alpha^{-\frac{\alpha}{1-\alpha}} \cdot \left(  \frac{ T_{s} }{   \lambda^F_{ss}(t) } \right) ^{- \frac{1}{\theta(1-\alpha)}} \frac{\lambda^M_{ss}(t)}{M_s(t)} \cdot w_{s}(t) \qquad \left( \because \frac{\alpha}{1-\alpha} = 1-\eta \right)
\end{eqnarray*}

\noindent which allows me to write real wages as a linear function of $M_s(t)$:

\begin{equation}
    \frac{w_s(t)}{P_s(t)} = M_s(t)  \times \alpha^{1-\eta} \times \left(  \frac{ T_{s} }{   \lambda^F_{ss}(t) } \right) ^{\frac{1}{\theta(1-\alpha)}} \times \left( \lambda^M_{ss}(t) \right)^{-1} \equiv M_s(t) \times \mathcal{R}^w_s(t)
\end{equation}

Similarly, I can write aggregate profits as a linear function of $M_s(t)$:

\begin{eqnarray*}
    \Pi_s(t) &=& \frac{\alpha}{\eta} \sum_{d \in \boldsymbol{K}} \lambda^M_{sd}(t) P_d(t) Y_d(t) \\
    \Pi_s(t) &=& \frac{\alpha}{\eta} \sum_{d \in \boldsymbol{K}} \frac{M_s(t) \left( \alpha^{-1} \tau_{sd} P_s(t) \right)^{1-\eta}}{(P_d^M (t) )^{1-\eta}} P_d(t) Y_d(t)
\end{eqnarray*}

Therefore: 

\begin{equation}
    \frac{\Pi_s(t)}{P_s(t)} = M_s(t)  \times  \frac{\alpha}{\eta} \sum_{d \in \boldsymbol{K}} \left( \frac{  \alpha^{-1} \tau_{sd} P_s(t) }{P_d^M (t) } \right)^{1-\eta} \frac{P_d(t) Y_d(t)}{P_s(t)} \equiv M_s(t)  \times \mathcal{R}^\pi_s(t)
\end{equation}

The budget constraint then is:

\begin{eqnarray}
    C_s(t) + I_s(t) &=& \frac{w_s(t)}{P_s(t)} L_s + \frac{\Pi_s(t)}{P_s(t)} \nonumber  \\
    &=& M_s(t) \times \left[ \alpha^{1-\eta} \left(  \frac{ T_{s} }{   \lambda^F_{ss}(t) } \right) ^{\frac{1}{\theta(1-\alpha)}}\left( \lambda^M_{ss}(t) \right)^{-1}L_s + \frac{\alpha}{\eta} \sum_{d \in \boldsymbol{K}} \left( \frac{  \alpha^{-1} \tau_{sd} P_s(t) }{P_d^M (t) } \right)^{1-\eta} \frac{P_d(t) Y_d(t)}{P_s(t)} \right] \nonumber \\  
    &=& M_s(t) \times \left[ \mathcal{R}^w_s(t) + \mathcal{R}^\pi_s(t) \right] = M_s(t) \times \mathcal{R}_s(t) 
\end{eqnarray}

\subsection{Balanced Growth Path}\label{appendix: bgp}

\paragraph{Autarky}

\paragraph{Proof of Proposition \eqref{prop: BGP-autarky}}

\begin{proof}
Without loss of generality, choose an arbitrary country $s \in \boldsymbol{K}$. Since this world economy is under autarky, evaluate \eqref{eq: euler-equation} replacing for the real interest rate using equations \eqref{eq: rents} and \eqref{eq: profits} and taking the limit $\tau_{sd} \to \infty (\forall s \neq d)$. By assumption \eqref{ass: trade-costs}, $\tau_{ss} = 1 (\forall s)$. Therefore, \eqref{eq: euler-equation} collapses to:

\begin{equation}
    g_s^{\text{autarky}} =  \frac{\alpha  \cdot \psi}{\eta} \cdot \frac{Y_s(t^*)}{M_s(t^*)}    - \rho
\end{equation}

\noindent for a BGP inclusive of each period $t \ge t^*$. 

The next step in the proof is to show that  $g_{M_s} = g_{Y_s} = g_{C_s} = g_{w_s} = g_{A_s}= g_s^{\text{autarky}}$. Since real wages, real profits, assets, and real output are linear functions of $M_s(t)$ in each period, it follows that $g_{M_s} = g_{Y_s} = g_{w_s} = g_{A_s}$. Since, with log preferences, consumption is a constant fraction of output, $g_{C_s} = g_{Y_s}$. Since the choice of $s$ was arbitrary, this holds for any $s \in \boldsymbol{K}$. 

To show uniqueness, one needs to solve for growth rate in terms of parameters. In order to do so, a few intermediate steps are necessary. First, note that one can express the demand for intermediates as:

\begin{equation*}
    \bar{x}_{ss}(t,\omega) \equiv x_{ss}(t,\omega,\nu) = \left[ \alpha z_s(\omega) p_{ss}(t,\omega) \right]^\frac{1}{1-\alpha} \cdot \ell_s(t,\omega)
\end{equation*}

\noindent which, in turn, implies that the optimal price of intermediate varieties is $p_{ss}(t,\omega,\nu) = \frac{1}{\alpha}$ and I can rewrite the production function of the final goods producer as:

\begin{eqnarray*}
    y_s(\omega) &=& z_s(\omega) \ell_s(t,\omega)^{1-\alpha} \left( \frac{1}{\alpha }\int_0^{M_s(t)} [\bar{x}_{ss}(t,\omega)]^{\alpha} d \nu \right) \\
                &=& z_s(\omega) \ell_s(t,\omega)^{1-\alpha} \left( \frac{1}{\alpha } \int_0^{M_s(t)} \left[ \left[ \alpha z_s(\omega) p_{ss}(t,\omega) \right]^\frac{1}{1-\alpha} \cdot \ell_s(t,\omega) \right ]^{\alpha} d \nu \right) \\
                &=& \left[ z_s(\omega) \right]^{\frac{1}{1-\alpha}} \cdot \left[ \alpha \cdot p_{ss}(t,\omega) \right]^{\frac{\alpha}{1-\alpha}} \cdot \ell_s(t,\omega) \cdot \frac{1}{\alpha} \cdot M_s(t) 
\end{eqnarray*}

Replacing for $p_{ss}(t,\omega)$ using the assumption of pricing under perfect competition:

\begin{eqnarray*}
    y_s(\omega) &=& \left[ z_s(\omega) \right]^{\frac{1}{1-\alpha}} \cdot \left[ \alpha  \frac{w_s(t)^{1-\alpha} \alpha^{-\alpha(1-\eta)} M_s^\alpha}{\alpha \cdot z_s(\omega)} \right]^{\frac{\alpha}{1-\alpha}} \cdot \ell_s(t,\omega) \cdot \frac{1}{\alpha} \cdot M_s(t) \\
     &=&  \alpha^{-(1-\alpha)} \cdot z_s(\omega)  \cdot w_s(t)^\alpha \cdot M_s(t)^{1-\alpha} \cdot \ell_s(t,\omega)
\end{eqnarray*}

By GDP expenditure clearing, total expenditure is equal wages plus profits:

\begin{equation*}
    Y_s(t) = w_s(t)L_s + \frac{\alpha}{\eta} Y_s(t) \implies \frac{1-\alpha}{\eta} Y_s(t) = w_s(t)L_s \implies  Y_s(t) = w_s(t)L_s
\end{equation*}

\noindent where the last equation states that, in the last equation, GDP is labor income because labor is the only factor of income in this economy. Hence, value added is equal to labor income.

Integrating the production function over $\omega$ and using the fact above gives us:

\begin{eqnarray*}
   Y_s(t) = \left[\int_0^1 z_s(\omega)  \ell_s(t,\omega)^{\frac{\sigma-1}{\sigma}} d\omega \right]^{\frac{\sigma}{\sigma-1}} \cdot \alpha^{-(1-\alpha)} \cdot  w_s(t)^\alpha \cdot M_s(t)^{1-\alpha} = L_s w_s(t)
\end{eqnarray*}

solving for $w_s(t)$:

\begin{equation*}
     w_s(t) = \left( \left[\int_0^\infty z  \ell_s(t,z)^{\frac{\sigma-1}{\sigma}} dF_s(z) \right]^{\frac{\sigma}{\sigma-1}} \right)^{\frac{1}{1-\alpha}} \cdot \alpha^{-1} \cdot M_s(t) L_s^{-\frac{1}{1-\alpha}} 
\end{equation*}

The term in the integral denotes the joint product of productivity and labor allocation across firms. In aggregate terms, since both the distribution of productivity and the population are fixed for every $t$; and relative wages are fixed along the BGP, this term will be constant.

Following \textcite{alvarez_general_2007}, note that all goods enter symmetrically in the definition of the aggregate final good and they differ only by their productivity level. Therefore, one can express the BGP growth rate of the economy fully in terms of exogenous objects:

\begin{equation}
    g_s^{\text{autarky}} =  \frac{\psi}{\eta} \cdot \left( \left[\int_0^\infty z  \ell_s(t,z)^{\frac{\sigma-1}{\sigma}} dF_s(z) \right]^{\frac{\sigma}{\sigma-1}} \right)^{\frac{1}{1-\alpha}} \cdot   L_s^{-\frac{\alpha}{1-\alpha}}     - \rho
\end{equation}

Since neither the productivity distribution $F_s(z)$ nor the demand functions $\ell_s(t^*,z)$ will change along the BGP and all other terms in the growth rate are parameters, this pins down the uniqueness of the BGP under autarky, which completes the proof.
\end{proof}

\paragraph{Zero gravity}

\paragraph{Proof of Proposition \eqref{prop: bgp-zero-gravity}}

\begin{proof}
Without loss of generality, choose an arbitrary country $s \in \boldsymbol{K}$. Since this world economy is under zero gravity, evaluate \eqref{eq: euler-equation} replacing for the real interest rate using equations \eqref{eq: rents} and \eqref{eq: profits} and evaluating $\tau_{sd} =1 (\forall s,d)$.Therefore, \eqref{eq: euler-equation} collapses to:

\begin{equation}
    g_s^{\text{zero gravity}} =  \left[ \frac{\alpha \cdot \psi}{\eta \cdot P_s(t^*)}  \cdot       \frac{ \sum_{k \in \boldsymbol{K}} Y_{k}(t^*) }{\sum_{k \in \boldsymbol{K}} M_{k}(t^*) }   - \rho \right]
\end{equation}

\noindent for a BGP inclusive of each period $t \ge t^*$. Since there are no trade costs, the law of one price holds, and $P_s(t^*) = P_d(t^*) \equiv P(t^*)$ for every $s, d \in \boldsymbol{K}$. Choosing $P(t^*)$ to be num\'eraire of this economy shows that the growth rate will follow the stated equation. 

Since the choice of the $s$ of arbitrary and the expression in the right-hand side of the equation is equal for every $s \in \boldsymbol{K}$, it follows that the $g_s^{\text{zero gravity}} = g^{\text{zero gravity}}$ for all $s \in \boldsymbol{K}$, which shows that the growth rate must be common across all countries. Furthermore, since $Y_k(t^*) = M_k(t^*) \mathcal{R}_k(t^*)$ and the fact that $g_s$ must be constant along a BGP, $\frac{ \sum_{k \in \boldsymbol{K}} Y_{k}(t^*) }{\sum_{k \in \boldsymbol{K}} M_{k}(t^*) }$ will only be homogeneous of degree zero in $[M_n(t^*)]_{n \in \boldsymbol{K}}$ if $\mathcal{R}_k(t^*) = \mathcal{R}(t^*)$.

With log preferences, households will consume a constant fraction $(1-\rho)$ of their income and invest a fraction $\rho$. The non-arbitrage condition shows that real interest rate and returns to R\&D equalize globally along the BGP:

\begin{equation*}
    \frac{r_s(t^*)}{P_s(t^*)} = \frac{\psi \pi_s(t^*,\nu)}{P_s(t^*)} = \frac{\psi \Pi_s(t^*)}{M_s(t^*) P_s(t^*)} = \frac{\psi }{M_s(t^*)} M_s(t^*) \times  \mathcal{R}^{\pi}_s(t^*) =  \psi \rho \mathcal{R}(t^*)
\end{equation*}

The next step in the proof is to show that  $g_{M_s} = g_{Y_s} = g_{C_s} = g_{w_s} = g_{A_s}= g_s^{\text{zero gravity}}$. Since real wages, real profits, assets, and real output are linear functions of $M_s(t)$ in each period, it follows that $g_{M_s} = g_{Y_s} = g_{w_s} = g_{A_s}$. Since, with log preferences, consumption is a constant fraction of output, $g_{C_s} = g_{Y_s}$. Since the choice of $s$ was arbitrary, this holds for any $s \in \boldsymbol{K}$. 

 For uniqueness, one needs to show that the cross-sectional equilibrium is unique. Start from equation \eqref{eq: expenditure-market-clearing}. Evaluating it under zero gravity and noting that prices of final goods and intermediate goods equalize in that situation results in:

 \begin{equation*}
     P_s(t^*) Y_s(t^*) = \sum_{d \in \boldsymbol{K}} \left[ (1-\alpha) \frac{T_s w_s(t^*)^{-(1-\alpha)\theta}}{\sum_{k \in \boldsymbol{K}} T_k w_k(t^*)^{-(1-\alpha)\theta}} + \frac{\alpha}{\eta} \frac{M_s(t^*)}{ \sum_{k \in \boldsymbol{K}} M_k(t^*)} \right]  P_d(t^*) Y_d(t^*) 
 \end{equation*}

 Recall that $P_s(t^*) Y_s(t^*) = w_s(t^*) L_s + \Pi_s(t^*)$ and note that, under zero gravity, $\Pi_s(t^*) = \frac{\alpha}{\eta} \frac{M_s(t^*)}{ \sum_{k \in \boldsymbol{K}} M_k(t^*)} P_d(t^*) Y_d(t^*)$. So, given $M_s(t^*)$ the expenditure determination system becomes a simple system in wages: 

  \begin{equation*}
     w_s(t^*) L_s  = \sum_{d \in \boldsymbol{K}} \left[ (1-\alpha) \frac{T_s w_s(t^*)^{-(1-\alpha)\theta}}{\sum_{k \in \boldsymbol{K}} T_k w_k(t^*)^{-(1-\alpha)\theta}} \right]  w_d(t^*) L_d 
 \end{equation*}

 Define the excess demand function:

 \begin{equation*}
     Z_s(\textbf{w},t) \equiv \frac{1}{w_s(t^*)}\left( \sum_{d \in \boldsymbol{K}} \left[ (1-\alpha) \frac{T_s w_s(t^*)^{-(1-\alpha)\theta}}{\sum_{k \in \boldsymbol{K}} T_k w_k(t^*)^{-(1-\alpha)\theta}} \right]  w_d(t^*) L_d  - w_s(t^*) L_s \right)
 \end{equation*}

 and note:

 \begin{equation*}
     \frac{\partial Z_s(\textbf{w},t^*)}{\partial w_d(t^*)} = \frac{1}{w_s(t^*)} (1-\alpha) \lambda^F_{sd}(t^*) \left( L_d + \frac{\lambda^F_{dd}(t^*)}{w_{d}(t^*)} \right) > 0
 \end{equation*}

\noindent which shows that it satisfies the gross substitution property and the cross-section equilibrium is unique. Therefore, the BGP under zero gravity will be unique.

 \end{proof}

\paragraph{General case}

\paragraph{Proof of Proposition \eqref{prop: bgp-general-case}}

\begin{proof}
Without loss of generality, choose an arbitrary country $s \in \boldsymbol{K}$. From \eqref{eq: gdp-linear}, real GDP is a linear function of $M_s(t)$:

\begin{equation*}
    C_s(t) + I_s(t) = \frac{w_s(t)}{P_s(t)} L_s + \frac{\Pi_s(t)}{P_s(t)} = M_s(t) \times \mathcal{R}_s(t)
\end{equation*}
     
Over the BGP, with log preferences, consumption is a constant fraction of GDP: $C_s(t^*) =  (1-\rho)M_s(t) \times  \mathcal{R}_s(t).$ From the Poisson arrival process, $g_{M_s} = \frac{\dot{M}_s(t^*)}{M_s(t)} = \psi \rho \frac{I_s(t^*)}{M_s(t^*)}$. Since trade is balanced, $I_s(t^*) = \frac{\rho}{1-\rho} C_s(t^*)$ and varieties grow at the following rate:

\begin{eqnarray}
    g_{M_s} &=& \psi \rho  \left[ \alpha^{1-\eta} \left(  \frac{ T_{s} }{   \lambda^F_{ss}(t^*) } \right) ^{\frac{1}{\theta(1-\alpha)}}\left( \lambda^M_{ss}(t^*) \right)^{-1}L_s + \frac{\alpha}{\eta} \sum_{d \in \boldsymbol{K}} \lambda_{sd}^M(t^*) \frac{P_d(t^*) Y_d(t^*)}{P_s(t^*)M_s(t^*)} \right] \nonumber
\end{eqnarray}

The following statements are true:

\begin{enumerate}
    \item $\lambda^F_{ss}(t^*), \lambda^M_{ss}(t^*)$ are homogeneous of degree zero in $\{M_n(t^*)\}$;
    \item $\left( \frac{  P^M_s(t^*) }{P_d^M (t^*) } \right), \left( \frac{P_d(t^*) }{P_s(t^*)} \right)$ are homogeneous of degree zero in $\{M_n(t^*)\}$;
    \item $\frac{Y_d(t^*)}{M_s(t^*)} = \frac{ M_s(t^*) \times \mathcal{R}_s(t^*) }{M_s(t^*)}$ is homogeneous of degree zero in $\{M_n(t^*)\}$ if and only if $\mathcal{R}_s(t^*)$ is homogeneous of degree zero in $\{M_n(t)\}$ for all $s \in \boldsymbol{K}$.
\end{enumerate}

Therefore, for $g_{M_s}$ to be consistent with a BGP it must also be homogeneous of degree zero in $[ M_n(t^*) ]_{n \in \boldsymbol{K}}$. As a result, if $g_{M_s}$ is consistent with a BGP, $\mathcal{R}_s(t^*)$ must be homogeneous of degree zero in $[M_n(t^*)]_{n \in \boldsymbol{K}}$ for all $s \in \boldsymbol{K}$. As a result, it must be that varieties grow at the same rate across countries, which implies that $\mathcal{R}_s(t^*) = \mathcal{R}(t^*)$.

With log preferences, households will consume a constant fraction $(1-\rho)$ of their income and invest a fraction $\rho$. The non-arbitrage condition shows that real interest rate and returns to R\&D equalize globally along the BGP:

\begin{equation*}
    \frac{r_s(t^*)}{P_s(t^*)} = \frac{\psi \pi_s(t^*,\nu)}{P_s(t^*)} = \frac{\psi \Pi_s(t^*)}{M_s(t^*) P_s(t^*)} = \frac{\psi }{M_s(t^*)} M_s(t^*) \times  \mathcal{R}^{\pi}_s(t^*) =  \psi \rho \mathcal{R}(t^*)
\end{equation*}

The next step in the proof is to show that  $g_{M_s} = g_{Y_s} = g_{C_s} = g_{w_s} = g_{A_s}= g_s$. Since real wages, real profits, assets, and real output are linear functions of $M_s(t)$ in each period, it follows that $g_{M_s} = g_{Y_s} = g_{w_s} = g_{A_s}$. Since, with log preferences, consumption is a constant fraction of output, $g_{C_s} = g_{Y_s}$. Since the choice of $s$ was arbitrary, this holds for any $s \in \boldsymbol{K}$. 
    
\end{proof}

\paragraph{Changes in trade costs}

\paragraph{Proof of Proposition \ref{prop: comparative-statics-long-run}} 

\begin{proof}

The equilibrium growth rate of varieties:

\begin{eqnarray}
    g_{M_s} &=& \psi \rho  \left[ \left(  \frac{ T_{s} }{   \lambda^F_{ss}(t^*) } \right) ^{\frac{1}{\theta(1-\alpha)}}\left( \lambda^M_{ss}(t^*) \right)^{-1}L_s + \frac{\alpha}{\eta} \sum_{d \in \boldsymbol{K}} \lambda_{sd}^M(t^*) \frac{P_d(t^*) Y_d(t^*)}{P_s(t^*)M_s(t^*)} \right] \nonumber
\end{eqnarray}

Recall that:
\begin{eqnarray*}
    \sum_{d \in \boldsymbol{K}} \lambda_{sd}^M(t^*) &=&   \sum_{d \in \boldsymbol{K}} \frac{M_s (\tau_{sd} P_s(t^*))^{1-\eta}}{\sum_{k' \in \boldsymbol{K}} M_{k'} (\tau_{k' d} P_{k'}(t^*))^{1-\eta}} \frac{P_d(t^*) Y_d(t^*)}{P_s(t^*) M_d(t^*)} 
\end{eqnarray*}

Since these economies are symmetric, then: $P_s(t^*) = P_{s'}(t^*)$, $w_s(t^*)= w_{s'}(t^*)$, $M_s(t^*)= M_{s'}(t^*)$ for every $s, s'$ and $\tau_{sd} = \tau$ for every ${sd}$. Evaluated with symmetric economies, the expression above becomes:

\begin{eqnarray*}
    \sum_{d \in \boldsymbol{K}} \lambda_{sd}^M(t^*) &=&   \frac{(N-1)\tau^{1-\eta}}{[1+(N-1)\tau^{1-\eta}]}  + \frac{1}{[1+(N-1)\tau^{1-\eta}]} = 1 
\end{eqnarray*}

Therefore, denoting $P_s(t^*) = P(t^*)$, $M_s(t^*) = M(t^*)$ and noting that $Y_d(t^*) = M_d(t^*) \times \mathcal{R}(t^*)$ the growth rate becomes to:

\begin{eqnarray}
    g^* &=& \psi \rho  \left[ \left(  T_{s} \right) ^{\frac{1}{\theta(1-\alpha)}} \left( \frac{1}{1+(N-1)\tau^{1-\theta}} \right) ^{-\frac{1}{\theta(1-\alpha)}} \left( \frac{1}{1+(N-1)\tau^{1-\eta}} \right) ^{-1} L_s + \frac{\alpha}{\eta} \mathcal{R} \right] \nonumber
\end{eqnarray}

Then, take the derivative of $g^*$ wrt $\tau$:

\begin{eqnarray*}
    \frac{\partial  g^*}{\partial \tau} &=& \psi \rho   \left(  T_{s} \right) ^{\frac{1}{\theta(1-\alpha)}} \left( \frac{1}{1+(N-1)\tau^{1-\theta}} \right) ^{-\frac{1}{\theta(1-\alpha)}} \left( \frac{1}{1+(N-1)\tau^{1-\eta}} \right) ^{-1} L_s \times  \\
    & & \Bigg(\frac{(1-\eta) \tau^{-\eta}}{1+(N-1)\tau^{1-\eta}} -  \frac{\theta \tau^{-\theta-1}}{1+(N-1)\tau^{-\theta}}  \Bigg) < 0
\end{eqnarray*}

which is negative because $(1-\eta) < 0$ and every other term in the parenthesis is positive.

\end{proof}

\subsection{Welfare}\label{appendix: welfare}

 Recall that $C_s(t^*)$ can be expressed as a constant fraction of total lifetime wealth:

    \begin{equation*}
    C_s(t^*) = \rho \left[ A_s(t^*)   +  \int_{t^*}^\infty  \frac{w_s(\tau)}{P_s(\tau)} L_s \cdot\exp \left\{  - \bar{r}_s(\tau) \cdot \tau \right\} d\tau   \right]  \nonumber
    \end{equation*}

    where $\bar{r}_s = \frac{1}{\tau} \int_{t^*}^\tau r_s(t) dt$ is the average interest rate between $t^*$ and $\tau$. Since this holds along the BGP, $\frac{w_s(\tau)}{P_s(\tau)} = \frac{\exp\{ (\tau - t^*) g_{w_{s}}\}w_s(t^*)}{P_s(t^*)}$. Furthermore, since $\frac{r_s(t^*)}{P_s(t^*)}$ is constant along the BGP, $\bar{r}_s(\tau) = \frac{r_s(t^*)}{P_s(t^*)}$ for all $\tau \ge t^*$. Replacing those above results in:

    \begin{eqnarray*}
    C_s(t^*) &=& \rho \left[ A_s(t^*)   +  \frac{w_s(t^*)}{P_s(t^*)} L_s \int_{t^*}^\infty   \cdot\exp \left\{  - \left( \frac{r_s(t^*)}{P_s(t^*)} - g_{w_s}  \right) \cdot (\tau - t^*)\right\} d\tau   \right]  \\
    &=& \rho \left[ A_s(t^*)   +  \frac{w_s(t^*)}{P_s(t^*)}  \frac{L_s}{\frac{r_s(t^*)}{P_s(t^*)} - g_{w_{s}}} \right]  \\
    &=& \rho A_s(t^*)   +  \rho \frac{w_s(t^*)}{P_s(t^*)}\frac{L_s}{\frac{r_s(t^*)}{P_s(t^*)} - g_{w_s}} 
    \end{eqnarray*}

    Since $g_{w_s} = g_{C_s}$ and $g_{C_s} = \frac{r_s(t^*)}{P_s(t^*)} - \rho$, $\frac{r_s(t^*)}{P_s(t^*)} - g_{w_s} = \rho$. Hence, over the BGP, real consumption is a fraction of assets plus real labor income:

    \begin{equation*}
        C_s(t^*) = \rho A_s(t^*)   +   \frac{w_s(t^*)L_s}{P_s(t^*)}  
    \end{equation*}

    Welfare over the BGP is:

    \begin{eqnarray*}
        \int_{t^*}^\infty \exp \{-\rho (t-t^*)\}   \log \left(\exp \{g^* t\} C_s(t^*) \right)  dt &=& \int_{t^*}^\infty \exp \{-\rho (t-t^*)\}   \log \left( C_s(t^*) \right) dt   \\
        &+& \int_{t^*}^\infty \exp \{-\rho (t-t^*)\} g^* t dt \\
        &=& \frac{\log \left( C_s(t^*) \right)}{\rho} +  \frac{g^*}{\rho^2} \\
        &=& \log \left( A_s(t^*) \right) + \frac{1}{\rho} \log \left( \frac{w_s(t^*)L_s}{P_s(t^*)} \right) + \frac{g^*}{\rho^2}
    \end{eqnarray*}

    Finally, using the fact that $\psi A_s(t^*) = M_s(t^*)$, I can write:

    \begin{equation*}
        \int_{t^*}^\infty \exp \{-\rho (t-t^*)\}   \log \left(\exp \{g^* t\} C_s(t^*) \right)  dt =  \log \left( \frac{1}{\psi} M_s(t^*) \right) + \frac{1}{\rho} \log \left( \frac{w_s(t^*)L_s}{P_s(t^*)} \right) + \frac{g^*}{\rho^2}
    \end{equation*}

\paragraph{Static welfare} For real labor income, start from equation ~\eqref{eq: gravity} evaluated at $s=d$ and use the fact that, as shown in equation \eqref{eq: appendix-price-level} of Appendix \ref{appendix: ek-trade}, 
\begin{equation*}
    P_s(t) = \gamma \cdot \left[ \sum_{n\in \boldsymbol{K}}{T_{n}  (w_{n}(t)^{1-\alpha} P_{n}^M(t)^{\alpha} \tau_{nd} )^{-\theta}} \right]^{-\frac{1}{\theta}}
\end{equation*}

\noindent where $\gamma \equiv \Gamma \left( \frac{\theta + 1 - \sigma}{\theta} \right)^{\frac{1}{1-\sigma}}$. Then, own trade share in a given country can be represented by:

\begin{equation*}
    \lambda^F_{dd}(t) = \gamma^{\theta} \cdot \frac{T_{d}  (w_{d}(t)^{1-\alpha} (P_d(t)^M)^{\alpha} )^{-\theta}}{[P_{d}(t)]^{-\theta}}
\end{equation*}

Solving for $\frac{w_d(t)}{P_d(t)}$ delivers:

\begin{eqnarray*}
    \frac{w_{d}(t)}{P_d(t)} &=& \gamma^{\frac{1}{1-\alpha}} \lambda_{dd}(t)^{-\frac{1}{(1-\alpha)\theta}} T_d^{\frac{1}{(1-\alpha)\theta}} \left( \frac{P_d^M(t)}{P_d(t)} \right)^{-\alpha}
\end{eqnarray*}

Replacing for the definition of $P^M_d(t) = \left[ \sum_{k \in \boldsymbol{K}} M_k \left( \frac{\tau_{kd} P_k(t)}{\alpha} \right)^{-\frac{\alpha}{1-\alpha}} \right]^{-\frac{1-\alpha}{\alpha}}$ results in:

\begin{eqnarray*}
    \frac{w_{d}(t)}{P_d(t)} &=& \gamma^{\frac{1}{1-\alpha}} \lambda_{dd}(t)^{-\frac{1}{(1-\alpha)\theta}} T_d^{\frac{1}{(1-\alpha)\theta}} \left[ \sum_{k \in \boldsymbol{K}} M_k \left( \frac{\tau_{kd} P_k(t)}{\alpha P_d(t)} \right)^{-\frac{\alpha}{1-\alpha}} \right]^{1-\alpha}
\end{eqnarray*}
Consider what happens to welfare after a change in trade costs from $\boldsymbol{\tau}$ to $\boldsymbol{\tau} + d\boldsymbol{\tau}$, as in \textcite{arkolakis_new_2012}. In this dynamic setting, to compare the static component of welfare, I need to compare what happens across the two BGPs, comparing the two initial equilibria. Suppose $t^*$ is the initial period of the original BGP while $t^{**}$ is the first period of the final BGP. To fit this framework to the general trade literature, I will compare the static component of these BGP as if they happened in the same period, and compound the difference over time.

Let $\hat{x} \equiv x(t^**)/x(t^*)$. Then cumulative changes in static welfare are:

\begin{equation*}
    \frac{1}{\rho} \log \left( \widehat{\frac{w_s(t^{**})}{P_s(t^{**})}} \right) =  \frac{1}{\rho}  \log \left( \widehat{\lambda_{dd}^F}(t^{**})^{-\frac{1}{(1-\alpha)\theta}} \right) + \frac{1}{\rho\eta} \log \left(\sum_{k \in \boldsymbol{K}}  \mu_{k}(t^*)  \widehat{M}_{k}(t^{**}) \cdot  \left( \frac{\widehat{\tau}_{kd}  \widehat{P}_{k}(t^{**})}{\widehat{P}_{d}(t^{**})} \right)^{1-\eta}  \right) 
\end{equation*}

\noindent where $\mu_{k}(t) \equiv \frac{ M_{k}(t) \cdot  \left( \frac{\tau_{kd}  P_{k}(t)}{ P_{d}(t)} \right)^{1-\eta} }{\sum_{k \in \boldsymbol{K}} M_{k}(t) \cdot  \left( \frac{\tau_{kd}  P_{k}(t)}{P_{d}(t)} \right)^{1-\eta} }$

\newpage

\subsection{Nesting of Romer and Eaton-Kortum} \label{appendix: nesting}

In this subsection, I will briefly describe how to recover the canonical \textcite{romer_endogenous_1990} and \textcite{eaton_technology_2002} models from the framework described above. 

\paragraph{Eaton-Kortum} Setting $\alpha = 0$ implies that the value of new varieties is zero since the demand for and profits of varieties is also zero. Therefore, $I_s(t) = 0$ and $A_s(t) = 0$ for all $t$ and $s$. While the Eaton-Kortum model is a static model, here it can be thought of as an infinite sequence of static models with no intertemporal decision, since there are no longer asset markets that permit households to save:

\begin{eqnarray*}
    \max_{   C_{s}(t), c_s(t,\omega)_{\omega \in [0,1]} } & & \int_0^{\infty} \exp\{-\rho t\} \log \left( C_{s}(t)\right) dt \\
 s.t. \quad P_{s}(t) C_{s}(t)  &=& w_s(t) L_{s}   \\
    C_{s}(t) &=&  \Big[ \int_0^1 c_{s}(t,\omega)^{\frac{\sigma-1}{\sigma}}  d\omega \Big] ^{\frac{\sigma}{\sigma-1}} \\
    P_{s}(t) C_{s}(t) &=& \int_0^1 p_{s}(t,\omega) c_{s}(t,\omega)   d\omega
\end{eqnarray*}

Furthermore, since $\alpha=0$, the intermediate and research and development sectors disappear. The problem of the final goods producer becomes:

\begin{equation*}
 \max_{\ell_{s}(t, \omega)}  p_{ss}(t, \omega) \cdot z_{s}(t, \omega) \cdot \ell_{s}(t, \omega)   - \ell_{s}(t, \omega) w_{s}(t) 
\end{equation*}

\noindent which is identical to the one in the standard Eaton-Kortum model. Equilibrium will take the form of a system of labor market determination equations that solve for $N$ wages using trade expenditure shares.

\paragraph{Romer} Setting $\tau_{sd} \to \infty$ for $s \neq d$ implies trade costs are prohibitively high internationally, such that varieties of both final goods and intermediate goods become sold only locally. Normalizing the price of the domestic final good to be the num\'eraire in each country, I write the dynamic household problem as:
\begin{eqnarray*}
    \max_{   C_{s}(t), c_s(t,\omega)_{\omega \in [0,1]} } & & \int_0^{\infty} \exp\{-\rho t\} \log \left( C_{s}(t)\right) dt \\
 s.t. \quad I_s(t) = \dot{A}(t) &=& r_s(t) A_s(t) + w_s(t) L_{s} - C_{s}(t)  \\
    C_{s}(t) &=&  \Big[ \int_0^1 c_{s}(t,\omega)^{\frac{\sigma-1}{\sigma}}  d\omega \Big] ^{\frac{\sigma}{\sigma-1}} 
\end{eqnarray*}

Furthermore, redefine assumption \eqref{ass: productivity} in the following terms:

\begin{assumption}[Productivity draws to recover Romer] \label{ass: productivity-romer}To recover the Romer model as a special case of the general model, I need to specify productivity terms $z_s(\omega)$ which are homogeneous across firms in each country. In order to do so, redefine the cumulative distribution function $F_s(t)(z)$ of the baseline case to be one of a degenerate random variable with a point mass concentrated at a certain scalar for each country. Formally:

\begin{equation*}
    F_s(t)(z) = \begin{cases}
                    0 \text{ for } z < T_s \\
                    1 \text{ for } z \ge T_s
                \end{cases}
\end{equation*}     
\end{assumption}

Using the symmetry assumption above, the num\'eraire normalization and the unavailability of foreign intermediate goods in the domestic market, the final goods assembler technology becomes:

\begin{equation*}
    y_{s}(t, \omega) = T_s  [\ell_{s}(t, \omega)]^{1-\alpha} \left( \frac{1}{\alpha} \int_{0}^{M_{s}(t)} [x_{ss}(t, \omega,\nu)]^{\alpha} d\nu  \right)    
\end{equation*}

\noindent which is identical to the single-country Romer model. Profits and demand per variety $\nu \in [0,M_s(t)]$ will be constant and growth will be driven by the domestic R\&D sector. Equilibrium will take the following form: labor markets will clear; total final goods produced being equal to total final goods used for consumption, intermediate production; and R\&D production; and optimized household optimal dynamics will be described by an Euler equation and an asset/measure accumulation equation.

\newpage

\section{Qualitative evidence: life among product innovators}

As an initial exploratory part of this research, I conducted a qualitative survey of managers in firms of New Member States. I first collected a list of notable firms from publicly available sources, restricted the sample to those who were active for at least two years before the time their respective countries joined the European Union, and then crawled through their English-language websites to collect the publicly available contact information. I sent the questionnaire below to 221 firms.

My goal was to assess if the description of the world that macro theorists set forth aligns with the practical intuitions of entrepreneurs. And it turns out that, at among the group of managers that responded to my email, they do. I will highlight two illustrative cases in the text.

For instance, the dynamic mechanism that propels growth, as I have described in the theory section is that increased access to foreign markets increases expected profits, thereby increasing the incentive to invest in research and development. This is entirely consistent with the description of the facts by one Czech biotech entrepreneur:

    \begin{quote}
        ``Once we joined the EU [...] \textcolor{cyan}{this allowed us to increase our exports and fund our own genetic programmes}.'' \textemdash \textcolor{blue}{\textit{CEO of a Czech Biotech company}}
    \end{quote}

In their comments, they went on to specify the importance of having access of not only to the European market itself, but also third party markets. They mentioned that after the Czech Republic joined the EU, his firm had immediate access to the standards for labeling and certification in existing trade agreements between the EU and third parties, which facilitated their firm's exports. These kinds of non-tariff barriers are typically considered part of trade costs $\tau$ in most trade models.

In this firm's particular case, product innovation came through the invention of breeding of new varieties of farm animals, that were then commercialized. But we see a similar story in a very different market: alcoholic beverages:

    \begin{quote}
        ``In 2004, we first started producing the ultra-luxury variation of our signature vodka, \textcolor{cyan}{which became a popular export product [...] and later started production of 18 new products.}'' \textemdash \textcolor{blue}{\textit{Spokesperson of a Latvian liquor manufacturer}}
    \end{quote}

In this case, the firm reported having used the European market's exports as a platform for global expansion. For context, 2004 marks the year Latvia accessed the EU \textemdash and also the year that this manufacturer decided to expand its product line by introducing the ultra-luxury versions of its signature product, which they claimed was adequate to the Western European market.

Once again, this is qualitatively consistent with the theoretical mechanism proposed in the model, with market access likely inducing product innovation. Of course, these individual experiences are not necessarily representative of a large universe of firms, which is why in the next two sections I will perform a detailed quantitative exploration of the data, first detailing some stylized facts, then going into causal inference. Nonetheless, the type of qualitative evidence presented here is useful to show to that the big picture is consistent with the individual experiences.

\subsection{Qualitative Questionnaire}

\begin{enumerate}
\item After your country joined the European Union, did your company:
\begin{itemize}
    \item start producing more products/services or varieties;
    \item start producing fewer products/services or varieties; or
    \item keep producing about the same number of products/services or varieties?
\end{itemize}
\item If your company changed the number of products/services or varieties after EU accession, how was the change implemented and what were the results? Please include any important information or relevant anecdotes.
\item If your company changed the number of products/services or product/service
varieties after EU accession, was the decision primarily motivated by access to
new technologies/imports, access to new markets/exports, or both? Explain.

\item After your country joined the European Union, did your company:
\begin{itemize}
    \item stay in the same industry;
    \item expanded to another industry; or
    \item move completely to a new industry?
\end{itemize}

\item If your company expanded to another industry or moved to a new industry. Please explain whether the change was related to your country’s EU accession.

\end{enumerate}

\newpage

\section{Data and Empirical Appendix} \label{appendix: data}
\renewcommand{\theequation}{D.\arabic{equation}}
\setcounter{equation}{0}

\subsection{Extensive Description of the Data}

\paragraph{Production data} Production data comes from Eurostat's Prodcom (\textit{Production Communautaire}), which is an annual full coverage survey of the European mining, quarry and manufacturing sectors, reporting the value of production of 4,000+ different product-lines of EU members and candidate countries. Prodcom reports, for each product line, country, and year, the value (in euros) and volume (in kg, $m^2$, number of items, etc.) of production. Product lines follow the Statistical Classification of Products by Activity in the EU (CPA). 

The target population of the full coverage sample is every enterprise that manufactures some good in the Prodcom List. Data quality is good for member countries since European Law\footnote{``PRODCOM statistics are compiled under the legal basis provided by Council Regulation (EEC) NO 3924/1991 of 19 December 1991 and by Commission Regulation (EC) No 0912/2004 of 29 April 2004 implementing the Council Regulation (EEC) No 3924/91 on the establishment of a Community survey of industrial production. Additionally, a Commission Regulation updating the PRODCOM classification is available annually since 2003.''} mandates National Statistical Institutes to collect enterprise-level information on the value and volume of production covering at least 90\% of national production in each NACE class, defined as the first four digits of each product code. In practice, reporting goes beyond this minimum threshold and, according to Eurostat, the coverage error is estimated to be below 10\%. 

Let $n,i,p,t$ index countries, sectors, products, and periods, respectively; and denote $Y_{inpt}$ as the market value of production of product $p$\footnote{To construct sector codes, I use Eurostat concordances to map Prodcom product codes to Harmonized System (HS) product codes. I then used the respective HS-2 division codes as sector codes.}. The set of varieties produced in each sector is $\mathcal{M}_{nit} = \{ k : Y_{nikt} > 0 \}$. The measure of varieties is simply the cardinality of the set of produced varieties $M_{nit} = |\mathcal{M}_{nit}| = \sum_k \mathbb{1}_{\{ k : Y_{nikt} > 0 \}}$. The overall measure over varieties produced in a country is, then: $M_{nt} = \sum_{i} M_{nit}$. These measures can be directly calculated from Prodcom's table.

Oftentimes, the value of production is labeled as confidential information by the National Statistical Institute, particularly in cases in which production is concentrated on a few enterprises. In those cases, while the value and volume are not publicly available, Eurostat reports this number as \textit{confidential}, which still allows one to infer that $Y_{nikt} > 0$ for that particular variety $k$, implying that the variety is produced.

Typically, production information at the variety level is not available, which pushed researchers to use product-level trade data instead. Some exceptions include  \textcite{goldberg_imported_2010} and \textcite{rachapalli_learning_2021}, who use firm-product links from the Indian Survey of Manufacturers; \textcite{bernard_multiproduct_2011}, who use US Manufacturing Censuses firm-product data.

\paragraph{Tariff and trade flow data} Bilateral tariff data come from WITS (World Integrated Trade Solution Trade Stats). It consolidates tariff data from the UNCTAD's Trade Analysis Information System (TRAINS) as well as from the WTO. 

To construct effective tariff rates, one starts from baseline tables of most favored nation tariffs at the source-country $\times$ destination-country $\times$ HS6-code $\times$. Then, one superimposes every bilateral product level preferential tariff available in the WITS database on each of these tables. Furthermore, whenever there are gaps between two identical bilateral preferential tariffs, one fills in those gaps. The result is a dataset of effectively applied tariff rates. 

Bilateral trade flow data comes from UNCOMTRADE. These data, which are widely used in research, come natively in a source-country $\times$ destination-country $\times$ HS-6 product-code $\times$ year format, which makes it readily compatible with the tariff data mentioned above.

Let $s,d,i,p,t$ index source countries, destination countries, sectors, products, and periods, respectively; and denote $X_{sdipt}$ as the market value of bilateral trade of product $p$.

The set of traded varieties in each sector is $\mathcal{X}_{nit} = \{ k : X_{sdikt} > 0 \}$. Analogously as with production, one can observe the total number of traded varieties $\sum_k \mathbb{1}_{\{ k : X_{nikt} > 0 \}}$. To make sure these are comparable to PRODCOM's codes, whenever possible, I used concordances and restricted the set of goods to create a dataset that matched both trade and production.

\paragraph{Other data} I also collected data on (a) the dates of accession of new member states to the European Union; (b) trade agreements existent and entered into force between the European Union and third parties before 2004; and (c) expenditure in private research \& development expenditures per capita. The first two come from hand collecting documents and tables from the European Commission's official websites while the latter comes from Eurostat.

\subsection{Formal Description of the Callaway \& Sant'Anna Estimator}

Formally, let a ``treatment'' group $g$ be defined as being treated for all periods $t \ge g$. Note that, since the EU enlargement happened simultaneously for more than one country, there is more than one country $n$ for each $g_n = g$. If some country cluster is in group $g$, then $G_{nt} = g$ $(\forall t)$. If it is never treated, it is in the control group, and then $G_{nt} = \infty $ $(\forall t)$.

The parameter of interest is the average treatment on the treated for a given treatment group $g$ and horizon $t$, i.e.:

\begin{equation}
    ATT (g, t) = \mathbb{E} [M_{nt} (g) - M_{nt} (0) |  G_{nt} = g]
\end{equation}

\noindent where $M_{nt} (g)$ is the potential outcome of country $n$ at period $t$ if treated at period $g$; $M_{nt} (0)$ is the potential outcome country $n$ at period $t$ if untreated; $X_{n {g-1}}$ are pre-treatment time-invariant covariates; and $G_{nt} = g$ is a group indicator.

Note that the $ATT(g,t)$ is group and period-specific. It can be recovered under assumptions similar to the standard difference-in-differences framework: parallel trends and no-anticipation\footnote{Formally, parallel trends is the assumption that potential outcomes evolve almost surely equally to the untreated group: $\mathbb{E}[M_{nt}(0) - M_{n {t-1}}(0)| G_{nt} = g] = \mathbb{E}[M_{nt}(0) - M_{n {t-1}}(0)| G_{nt} > g]$ for all $t \ge g$. No anticipation means that potential outcomes for a treated group are equal to the untreated group for any date before the treatment \textemdash i.e., for all $t < g$, $\mathbb{E}[M_{nt}(g)| G_{nt} = 1] = \mathbb{E}[M_{nt}(0)| G_{nt} = 1]$ almost surely.}. The next step is to summarize the ATT across groups by appropriately weighting the results as:

\begin{equation}\label{eq: cs-theta}
    \theta(t) = \sum_{g} \mathbb{1}\{g \le t\} w_{gt} ATT(g, t)
\end{equation}

\noindent for some weights $w_{gt}$. \textcite{callaway_difference--differences_2021} propose the weights $w_{gt} = P(G_{nt} = g|G_{nt} \le t)$, which is the share of country clusters from group $g \ge t$ out of all country clusters being treated at time $t$.

\subsection{Further Details on Causal Inference}

\begin{figure}[htp!]
    \centering
    \includegraphics[width=\textwidth]{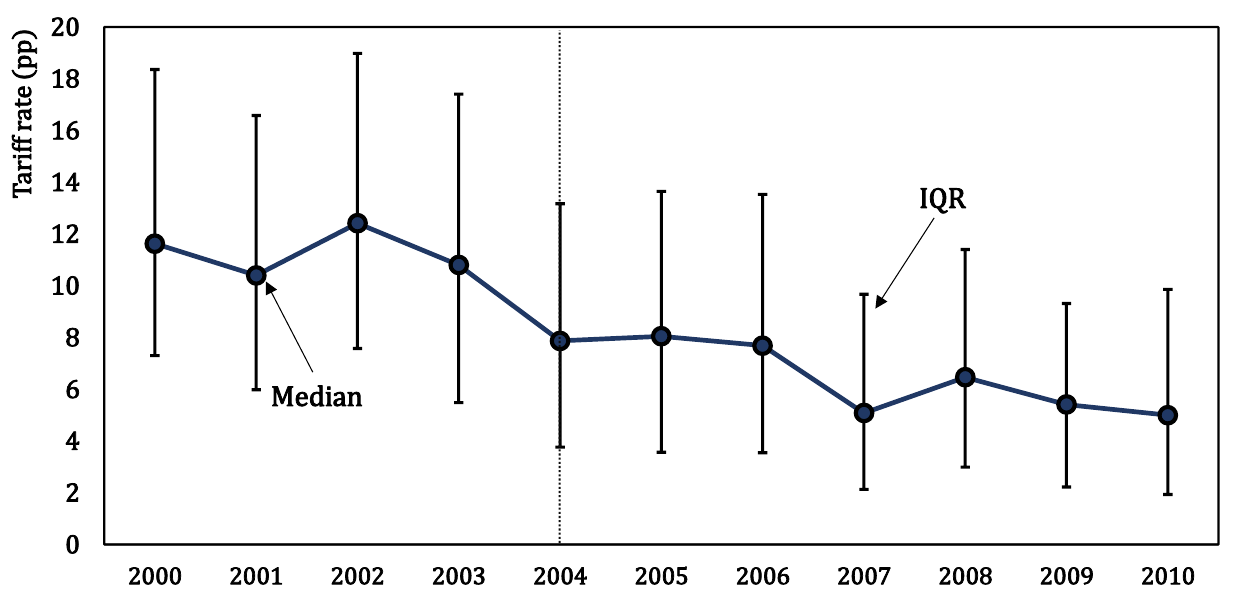}
    \caption{\textbf{Distribution of Tariff Changes Over Time}: \footnotesize{Interquartile Range Bilateral HS6-Product-Level Tariff Rates Between New Member States (2004 EU Enlargement) and Set of Countries that Concluded Trade Agreements with EU prior to 2004. Data were constructed from WITS Preferential and MFN databases. }}
     \label{fig:event-tariff-distribution}
\end{figure}

Since the largest wave of enlargement was in 2004, in this analysis I will focus exclusively on that wave. The source of variation is at the source-country $\times$ destination-country $\times$ HS-code product level. In each year, there are about 300 thousand observations. Figure \ref{fig:event-tariff-distribution} shows the interquartile range of bilateral product-level tariff rates between NMS and the set of countries that had concluded trade agreements with the EU prior to 2004.

It shows that there is not much change in tariffs leading up to membership and then a median drop of about 2.5 percentage points between 2003 and 2004. In the years immediately after membership, there is also not a large change in the distribution of bilateral tariff rates. There are some changes after 2007, possibly because some future provisions in trade agreements kick in.

\begin{figure}[htp!]
    \centering
    \includegraphics[width=0.55\textwidth]{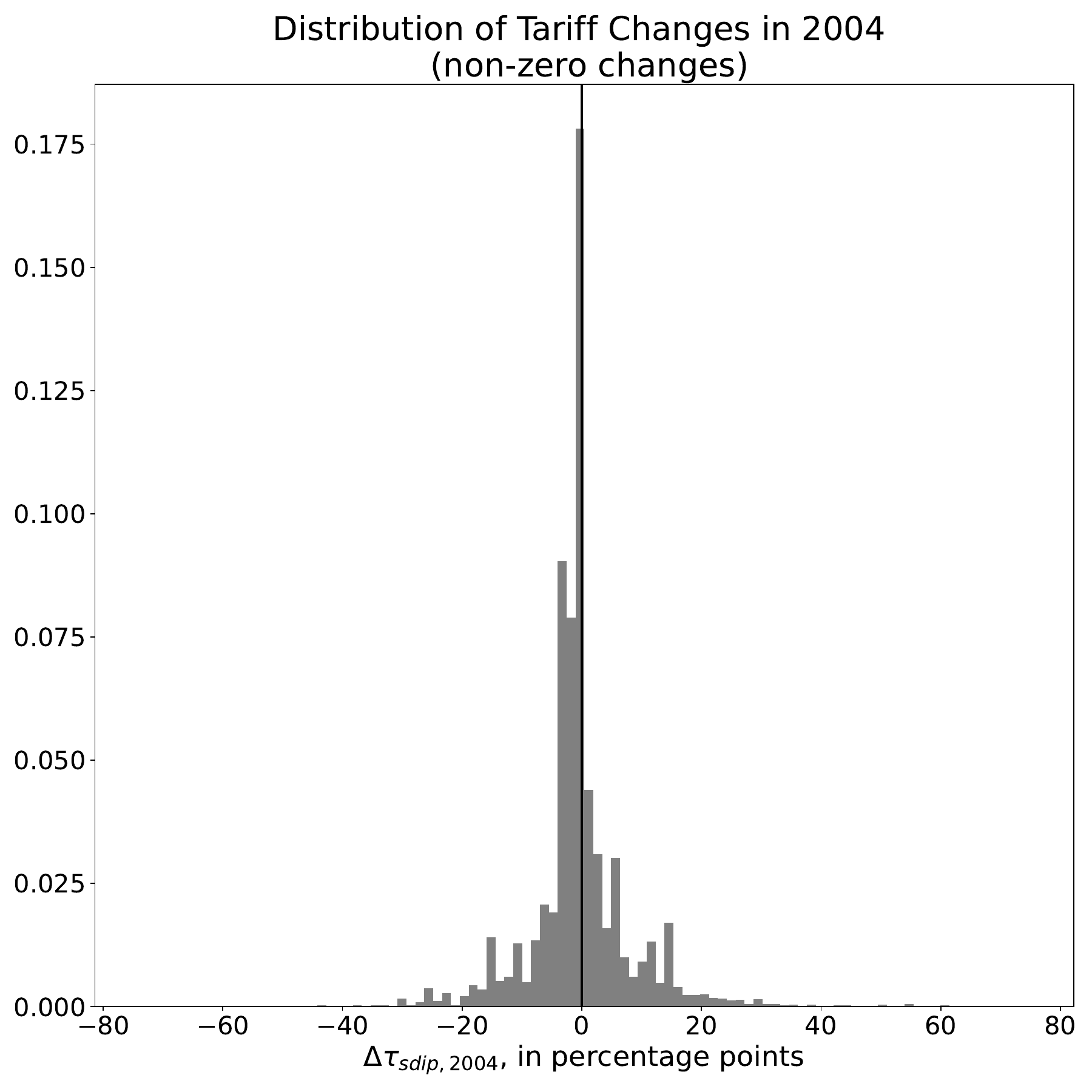}
    \caption{\textbf{Tariff Shock:} \footnotesize{Distribution of the (non-zero) observations of the changes in Bilateral HS6-Product-Level Tariff Rates Between New Member States (2004 EU Enlargement) and Set of Countries that Concluded Trade Agreements with EU prior to 2004. Data were constructed from WITS Preferential and MFN databases.}}
    \label{fig:event-delta-tau}
\end{figure}

The metric of the tariff shock change is simply $\Delta \tau_{sdip,2004} \equiv (\tau_{sdip,2004} - \tau_{sdip,2003})$, which is the change in the level of effectively applied bilateral tariffs at the product level between 2003 and 2004. Figure \ref{fig:event-delta-tau} plots the distribution of $\Delta \tau_{sdip,2004}$, excluding the zero-valued observations. The average $\Delta \tau_{sdip,2004}$ is $2.14\%$ and the standard deviation is $12\%$.

I estimate a sequence of cross-sectional local-projection linear probability models, which estimate what is the marginal effect of an \textit{increase} in the tariffs on exports of a given product $p$, \textit{conditional on that country $s$ not producing that particular product before joining the EU in 2003}. The fact the data is highly granular permits me to exploit within $industry \times source \times destination \times horizon$ (across
product) variation. 

Formally, I estimate the following equation:
\begin{eqnarray}
    P\left(X_{sdip,h} > 0 \middle| Y_{s\cdot ip,2003} = 0 \right) &=& 
    \alpha_h + \beta_h \cdot \Delta \tau_{sdip,2004} + \gamma_{sdi,h} + \nu_{sdip,h} \\
    & & \text{for } h \in \{2000, \cdots, 2010\} \nonumber
\end{eqnarray}

\noindent where $X_{sdip,h}$ is the market value of exports between country $s$ and country $d$ of product $p$ of industry $i$ at horizon $h$; $Y_{s\cdot ip,2003}$ is the market value of production in country $s$ of product $p$ of industry $i$ in $2003$; $\alpha_h$ are horizon (time) fixed-effects; $\gamma_{sdi,h}$ are $source \times destination \times industry$ interactions fixed-effects for each $h$. 

Note that, since these are local projections, the right-hand side coefficients, the regressor $\tau_{sdip,2004}$ is fixed for all horizons, and the coefficients $\beta_h$ change. As initially argued by \textcite{chodorow-reich_regional_2020} and later formalized by \textcite{dube_local_2023}, these types of cross-sectional event studies with local projections can be interpreted as differences in differences with continuous treatments. If consistently estimated, the estimated coefficients $\beta_h$, then, are simply the average treatment on treated compared to the potential outcomes of not being treated, normalized to a treatment of intensity of one unit.

This strategy takes the assertion in \textcite{baier_free_2007} (henceforth B\&B) that countries engage endogenously in free trade agreements (FTAs) and one needs to look for a plausibly exogenous source of variation to check whether or not FTA ``actually increase members' international trade'' seriously. Here, I rely on their strategy of running dynamic panels with fixed effects to control for unobserved heterogeneity.

Importantly, while they estimate their models at the aggregate country level with $source \times destination \times period$ fixed effects, I have enough variability and data availability to estimate it at the product-level adding $industry \times source \times destination \times period$ fixed effects. Hence, this approach adds granularity to B\&B's strategy, thereby controlling for more unobserved heterogeneity.

The identification assumption is that conditional on the very saturated fixed effects that this model includes, the unobserved components $\nu_{sdip,h}$ are uncorrelated with the change in tariffs $\Delta \tau_{sdip,2004}$. Intuitively, the identification is robust to a NMS (say, Poland's) policymakers endogenously targeting EU accession to have preferential access to a third-party's (say, Mexico's) car industry (relative to other industries and countries), but not if they want to have preferential access to compact cars relative to SUVs in Mexico.

The identification strategy is plausible. In general, neither lobbyists of industry trade groups nor trade negotiations work in such a disaggregated product-level setting. Typically, lobbyists consolidate the interests of the producers of many products under the same umbrella and try to influence negotiations. Similarly, even when governments are negotiating tariffs schedule changes \textemdash which was not the case in this particular case \textemdash these negotiations typically also happen in blocs, with governments exchanging positions in some products for others. Hence, the fact that this is a highly disaggregated dataset at the product level adds a lot of strength to the identification strategy. 

As shown in Figure \ref{fig:event-treatment-entry}, an increase in market access by 1 percentage point increases the probability of starting to produce and export a given product by about 1 percent by 2010. To benchmark this result, it is about one-third of the conditional mean $\mathbb{E}[X_{sdip,h} > 0 | X_{s\cdot ip,h} > 0, h > 2003] = 2.9\%$. There are no signs of a pre-existing trend before 2004: both the magnitude of the coefficients and the standard errors are very small before the treatment date.

The related set of continuation regressions, is very similar to the model estimated in equation \eqref{eq: entry-regressions}, except that now it conditions in initial production being active:

\begin{eqnarray}\label{eq: continuation-regressions}
    P\left(X_{sdip,h} > 0 \middle| Y_{s\cdot ip,2003} = 1 \right) &=& 
    \alpha_h + \beta_h \cdot \Delta \tau_{sdip,2004} + \gamma_{sdi,h} + \nu_{sdip,h} \\
    & & \text{for } h \in \{2000, \cdots, 2010\} \nonumber
\end{eqnarray}

In this case, there are no effects observed on the extensive margin. When countries already have the ability to produce a given product, additional market access produces very noisy results in the extensive margin. The coefficients are large and bounce between positive and negative and the confidence bands are even larger. One potential explanation is that the countries possibly already had market access before 2004, as illustrated by the positive (albeit insignificant results) for 2000-03, since they already had the production capacity. It is possible that most of the effects concentrate on the intensive margin, something that futures iteration of this paper would need to check.

\begin{figure}[htp!]
    \centering
    \includegraphics[width=0.95\textwidth]{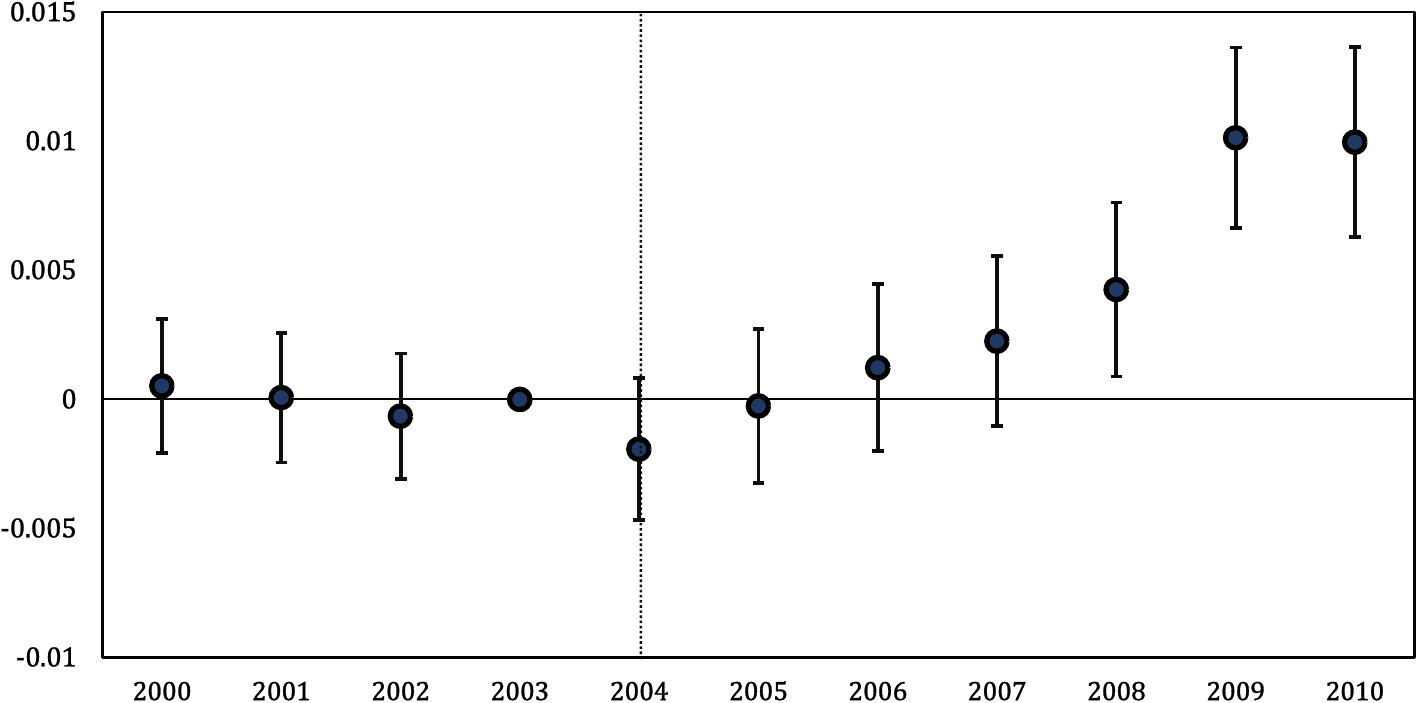}
    \caption{\textbf{Entry Regressions.} \footnotesize{This plot shows the coefficients $\beta_h$ of the local projection linear probability models specified in equation \eqref{eq: entry-regressions}. Each year is a different cross-sectional regression with approximately 300 thousand observations. The whiskers show 95\% confidence intervals with robust standard errors clustered at the source-destination-industry level.}}
\end{figure}

\begin{figure}[htp!]
    \centering
    \includegraphics[width=0.95\textwidth]{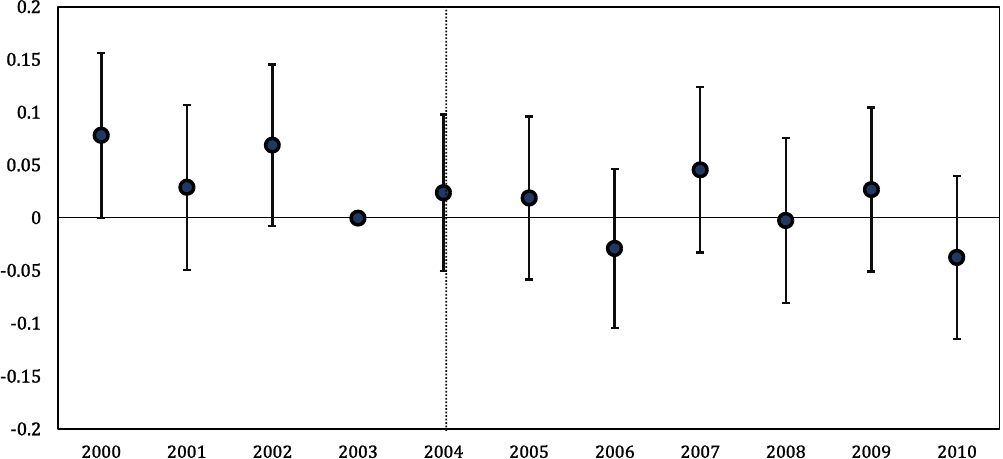}
    \caption{\textbf{Continuation Regressions.} \footnotesize{This plot shows the coefficients $\beta_h$ of the local projection linear probability models specified in equation \eqref{eq: continuation-regressions}. Each year is a different cross-sectional regression with approximately 300 thousand observations. The whiskers show 95\% confidence intervals with robust standard errors clustered at the source-destination-industry level.}}
    \label{fig:event-treatment-continuation}
\end{figure}

\newpage

\section{Computational Appendix} \label{appendix: computational}
\renewcommand{\theequation}{F.\arabic{equation}}
\setcounter{equation}{0}

This computation appendix explains how I solve for the BGP growth rate.

\begin{enumerate}
    \item \textbf{Inner loop (Prices of Final Goods)}. Given parameters $\{ \theta, \psi, \alpha, \textbf{L}, \textbf{T}, \boldsymbol{\tau} \}$ and guesses for wages $\boldsymbol{w}$, measures of varieties $\boldsymbol{M}$ and some common return $\mathcal{R}$, use the input-output structure of the model to solve for the prices of the final goods. 

    \begin{eqnarray*}
        P_s(t) &=& \gamma \cdot \left [ \sum_{n \in \boldsymbol{K}} T_n \left( P_n^M(t)^{\alpha} w_n(t)^{1-\alpha} \tau_{ns} \right)^{-\theta} \right]^{-\frac{1}{\theta}} \\
        P_s(t)  &=& \gamma \cdot \left [  \sum_{n \in \boldsymbol{K}} T_n\left( w_n(t)^{1-\alpha} \tau_{ns} \right)^{-\theta} \left( \sum_{k \in \boldsymbol{K} } M_k(t) \left( \frac{\tau_{kn} P_{k}(t) }{\alpha} \right)^{- \frac{\alpha}{1-\alpha}} \right)^{\theta(1-\alpha)} \right]^{-\frac{1}{\theta}}
    \end{eqnarray*}

    \noindent with $\gamma \equiv \Gamma\left(\frac{\theta+1-\sigma}{\theta}\right)^{\frac{1}{1-\sigma}}$. The last equation makes it explicit that, given parameters, wages, and measures of varieties, this is a system of $|N|$ equations and $|N|$ unknowns in final goods prices. A simple grid search algorithm finds a fixed point for final goods prices.

    \item \textbf{Intermediate loop}. Given parameters $\{ \theta, \psi, \alpha, \textbf{L}, \textbf{T}, \boldsymbol{\tau} \}$ and guesses the measures of varieties $\boldsymbol{M}$, some common return $\mathcal{R}$, and the prices from the following step, use the expenditure determination equation to solve for final demand.

    \begin{equation*}
         P_s(t) Y_s(t) = \sum_{d \in \boldsymbol{K}}  \left[  (1-\alpha) \lambda_{sd}^F(t) + \frac{\alpha }{\eta } \lambda_{sd}^M(t) \right]   P_d(t) Y_d(t)
    \end{equation*}

    Update the guess for usages using a constant fraction $(1-\rho)$ of income over the BGP and taking advantage of returns $\mathcal{R}^g$ and measures $M^g_k$:

    \begin{equation*}
        w_s(t) L_s(t) = (1-\alpha) \sum_{k \in \boldsymbol{K}} \lambda_{sk}(t)^F (1-\rho) \mathcal{R}^g M_k(t)^g
    \end{equation*}

    Re-normalize $w_s(t) = \frac{w_s(t)}{L_s \cdot \sum_{k \in \boldsymbol{K}} w_k(t) L_k}$ to ensure it always maps onto a compact space, it is an operator and converges according to the contraction mapping theorem.

\item \textbf{Outer loop (Growth rates)}. Given parameters $\{ \theta, \psi, \alpha, \textbf{L}, \textbf{T}, \boldsymbol{\tau} \}$, prices, wages, and trade shares calculated in the previous steps, update the guesses for $M_s^g$ using:

\begin{equation*}
    M_s^{g'} = \left( \frac{\lambda_{ss}^F(t^*)}{T_s} \right)^{-\frac{1}{\theta (1-\alpha)}} \left( \lambda_{ss}^M(t^*)  \right)^{-1} L_s \frac{M_s^g}{\mathcal{R}^g} + \frac{\alpha}{\eta} \sum_{k \in \boldsymbol{K}} \left( \lambda_{sk}^M(t^*) \frac{ \sum_{l \in \boldsymbol{K}} \lambda_{kl}^F(t^*) (1-\rho) M_l^g  }{P_s(t^*) M_s^g} \right) 
\end{equation*}

    Again, to make sure it always maps onto a compact space, it is an operator and converges according to the contraction mapping theorem, renormalize the measure of varieties: $M_s^{g'} = \frac{M_s^{g'}}{P_s(t^*) \cdot \sum_{k \in \boldsymbol{K}} M_s^{g'} P_s(t^*)}$

    And update the guesses for the global return rates:

    \begin{equation*}
        \mathcal{R}^{g'} = \frac{1}{\sum_{k \in \boldsymbol{K}} M_s^{g'} P_s(t^*)}
    \end{equation*}
\end{enumerate}

A test of this algorithm is, starting from a random guess, knowing that a group of symmetric countries will eventually converge towards the same measure of varieties within some tolerance criterion $<\varepsilon$. One numerical illustration of this convergence is the Figure below, for a group of 4 symmetric countries, starting for a random guess, that eventually converge to $0.25$ (the sum of the measure of varieties is normalized to sum to 1).

\begin{figure}[htp!]
    \centering
    \includegraphics[width=0.5\textwidth]{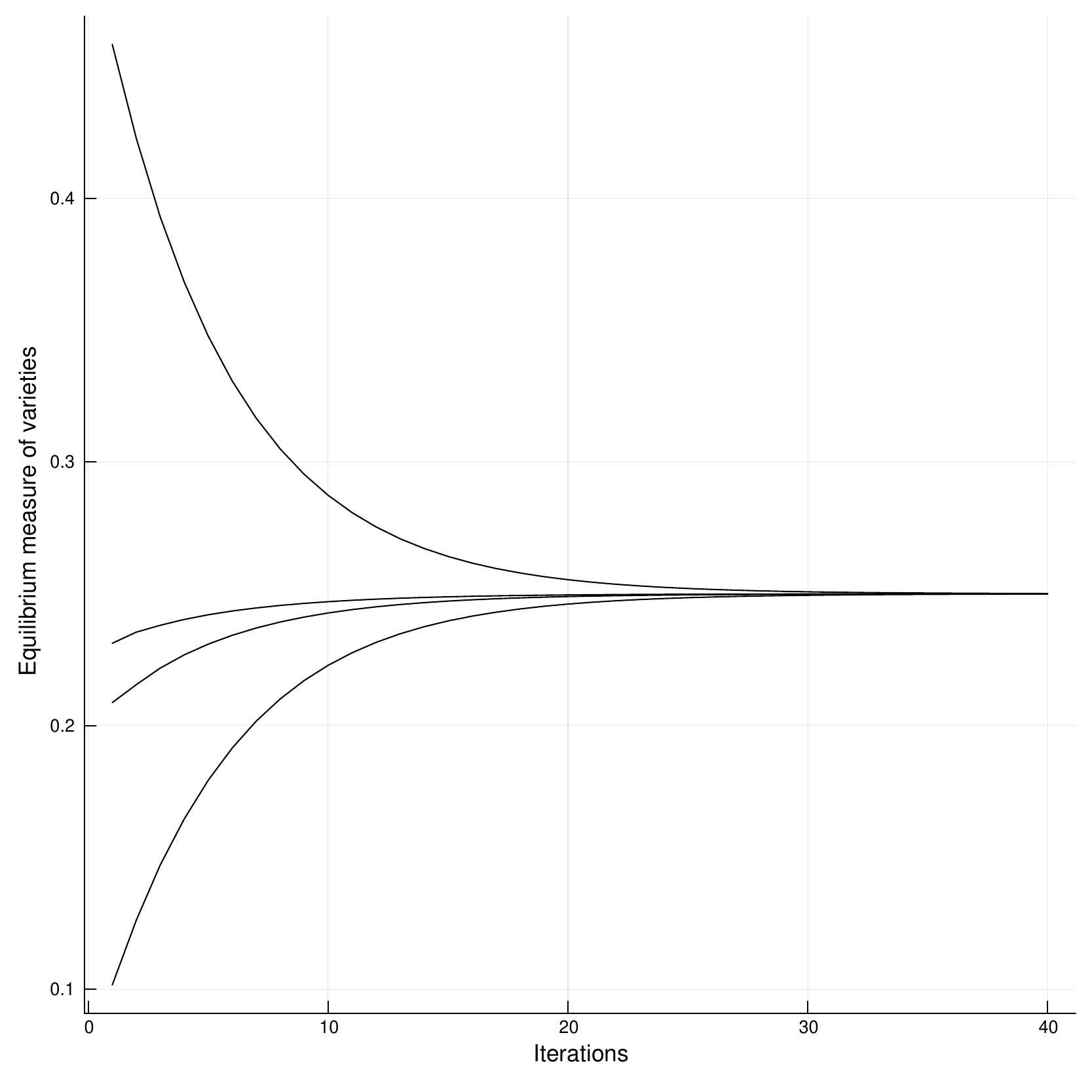}
\end{figure}
\newpage

\subsection{Calibration of Trade Shares}

I use observed trade flows to infer trade costs. The strategy goes back to \textcite{head_increasing_2001}. According to the handbook chapter by \textcite{head_gravity_2014}, the index is called the Head-Ries Index (HRI) since 2011 (when the working paper version of \textcite{eaton_trade_2016} was published). 

Expenditure in final goods is defined as:

\begin{equation*}
    E^F_{sd}(t) = \lambda_{sd}^F(t) P_d(t) Y_d(t) = \frac{T_{s} \left( \tilde{M}_{s}(t)^{1-\alpha} \right)^{\theta}  (w_{s}(t)^{1-\alpha} \tau_{sd} )^{-\theta}}{\sum_{n=1}^{N}{T_{n} \left( \tilde{M}_{n}(t)^{1-\alpha} \right)^{\theta} (w_{n}(t)^{1-\alpha} \tau_{nd} )^{-\theta}}} \cdot P_d(t) Y_d(t)
\end{equation*}

The ratio between $E^F_{sd}(t)$ and $E^F_{dd}(t)$ is, then:

\begin{equation*}
    \frac{E^F_{sd}(t)}{E^F_{dd}(t)} = \frac{T_{s} \left( \tilde{M}_{s}(t)^{1-\alpha} \right)^{\theta}  (w_{s}(t)^{1-\alpha} \tau_{sd} )^{-\theta} }{ T_{d} \left( \tilde{M}_{d}(t)^{1-\alpha} \right)^{\theta}  (w_{d}(t)^{1-\alpha} \tau_{dd} )^{-\theta} }
\end{equation*}

Analogously, the ratio between $E^F_{ds}(t)$ and $E^F_{ss}(t)$ is:

\begin{equation*}
    \frac{E^F_{ds}(t)}{E^F_{ss}(t)} = \frac{T_{s} \left( \tilde{M}_{d}(t)^{1-\alpha} \right)^{\theta}  (w_{d}(t)^{1-\alpha} \tau_{ds} )^{-\theta} }{ T_{d} \left( \tilde{M}_{s}(t)^{1-\alpha} \right)^{\theta}  (w_{s}(t)^{1-\alpha} \tau_{ss} )^{-\theta} }
\end{equation*}

Therefore:

\begin{equation*}
    \frac{E^F_{sd}(t)}{E^F_{dd}(t)} \cdot \frac{E^F_{ds}(t)}{E^F_{ss}(t)} = \left( \frac{\tau_{sd} \tau_{ds} }{ \tau_{ss} \tau_{dd} } \right)^{-(1-\alpha)\theta}
\end{equation*}

Using Assumption \eqref{ass: trade-costs}, $\tau_{ss} = \tau_{dd} = 1$ and $\tau_{sd} = \tau_{ds}$. Hence, I can express the trade cost $\tau_{sd}$ as:

\begin{equation}
    \tau_{sd} = \left(  \frac{E^F_{sd}(t)}{E^F_{dd}(t)} \cdot \frac{E^F_{ds}(t)}{E^F_{ss}(t)}  \right)^{- \frac{1}{2 \theta (1-\alpha)}}
\end{equation}

\end{document}